\documentclass[reprint, onecolumn,  amsmath, amssymb, aip ]{revtex4-1}   	
\pdfoutput=1
\usepackage{geometry}                		
\usepackage{graphicx}				
\usepackage[cm]{fullpage}
\usepackage{amssymb}
\usepackage{amsmath,amssymb}
\usepackage{color} 
\usepackage{caption}
\usepackage{subcaption}
\usepackage{cleveref}
\usepackage{multirow}

\usepackage{fancyhdr}
\newcommand{\pd}[2]{\frac{\partial#1}{\partial#2}}     	

\newcommand\etal{{\em et al.}}

\newcommand{\pth}[1]{\left( #1 \right)}				

\newcommand{\divg}{\nabla\cdot}
\newcommand{\divergence}{\divg}
\newcommand{\grad}{\nabla}

\def\bgk#1{\mbox{\boldmath $#1$}}

\usepackage[stable]{footmisc}
\long\def\symbolfootnote[#1]#2{\begingroup%
\def\thefootnote{\fnsymbol{footnote}}\footnote[#1]{#2}\endgroup} 

\newcommand{\apri}{\emph{a priori}~}
\newcommand{\apos}{\emph{a posteriori}~}

\newcommand{\ThreeFig}[9]{
\begin{figure}
        \centering
        \begin{subfigure}{2.55in}
                \includegraphics[height=1.3\textwidth,trim=.25in .3in .1in 0, clip=true]{#1}
                \caption{{#2}}
                \label{{#8}a}
        \end{subfigure}%
        \hspace{-.2in}
        \begin{subfigure}{2.55in}
                \includegraphics[height=1.3\textwidth,trim={#9} .3in .1in 0, clip=true]{#3}
                \caption{#4}
                \label{{#8}b}
        \end{subfigure}%
         \hspace{-.3in}
        \begin{subfigure}{2.55in}
                \includegraphics[height=1.3\textwidth,trim={#9} .3in .1in 0, clip=true]{#5}
                \caption{#6}
                \label{{#8}c}
        \end{subfigure}%
	\caption{\label{#8}
	#7
	}
\end{figure}
}

\newcommand{\barr}[1]{\langle #1\rangle}

\begin{document}
\title{Large eddy simulation requirements for the Richtmyer-Meshkov Instability }
\author{Britton J. Olson}
\email[B. Olson:  ]{olson45@llnl.gov}
\affiliation{Lawrence Livermore National Laboratory\\
Livermore, CA 94550}

\author{Jeff Greenough}
\affiliation{Lawrence Livermore National Laboratory\\
Livermore, CA 94550}

\lfoot[LLNL-JRNL-647614-DRAFT]{LLNL-JRNL-647614-DRAFT}
\rfoot[Submitted to Physics of Fluids]{Submitted to Physics of Fluids}

\fancyhead{}
\renewcommand{\headrulewidth}{0pt}

\begin{abstract}

	The shock induced mixing of two gases separated by a perturbed interface is investigated through Large Eddy Simulation (LES) and Direct Numerical Simulation (DNS).  In a simulation, physical dissipation of the velocity field and species mass fraction often compete with numerical dissipation arising from the errors of the numerical method.  In a DNS the computational mesh resolves all physical gradients of the flow and the relative effect of numerical dissipation is small.  In LES, unresolved scales are present and numerical dissipation can have a large impact on the flow, depending on the computational mesh.
A suite of simulations explores the space between these two extremes by studying the effects of grid resolution, Reynolds number and numerical method on the mixing process.  Results from a DNS are shown using two different codes, which use a high- and low-order numerical method and show convergence in the temporal and spectral dependent quantities associated with mixing.  Data from an unresolved, high Reynolds number LES are also presented and include a grid convergence study.  A model for an effective viscosity is proposed which allows for an \emph{a posteriori} analysis of the simulation database that is agnostic to the LES model, numerics, and the physical Reynolds number of the simulation.  An analogous approximation for an effective species diffusivity is also presented.  This framework is then used to estimate the effective Reynolds number and Schmidt number of future simulations, elucidate the impact of numerical dissipation on the mixing process for an arbitrary numerical method, and provide guidance for resolution requirements of future calculations.  
	
\end{abstract}

\maketitle

\section{Introduction}

The mixing of fluids is enhanced in the presence of fully developed turbulent flow.  The turbulent cascade transports entrained fluid from the large scale eddies to the small scale eddies, increasing the net interfacial surface area and the speed at which the fluids molecularly diffuse.  This fluid dynamics process is of importance in numerous applications in engineering and nature.  For example, Inertial Confinement Fusion (ICF) capsules are known to be Rayleigh-Taylor unstable during the late compression phase of the ignition process.  If this instability transitions to turbulence, the rate at which the ablator mixes with the fuel rapidly increases, potentially impacting capsule performance.  Given the extreme conditions of ICF, physicists must rely heavily on computational and theoretical models to elucidate the actual state of the mixing.  

Large eddy simulation (LES) is a powerful simulation tool for capturing the large scale dynamics of unsteady fluid flow.  Of LES, W.C. Reynolds~\cite{reynolds:90} wrote, 
``The objective of large eddy simulations is to compute the three-dimensional time-dependent details of the largest scales of  motion (those responsible for the primary transport) using a simple model for the smaller scales.  LES is intended to be useful in the study of turbulence physics at high Reynolds number, in the development of turbulence models, and for predicting flows of technical interest in demanding complex situations where simpler model approaches (e.g. Reynolds stress transport) are inadequate''.  

Traditional LES approaches use high-order numerics and explicit sub-grid scale (SGS) models to account for the unresolved scales of motion at or below the grid cut-off frequency.  Although a complete overview of existing SGS models is not given here, a review of general SGS model development and scale invariance is given by Meneveau and Katz~\cite{meneveau:2000} with select analysis of popular SGS models.  The low numerical dissipation associated with high-order schemes is a desired attribute in LES as it allows for a broader range of length scales to be captured on the computational mesh.  Indeed, Kravchenko and Moin~\cite{kravchenko:1997} found that errors in the SGS model and numerical truncation were reduced when high-order methods, with lower numerical dissipation and a broader range of resolved scales, were used. 
Since the fidelity of an LES calculation is proportional to the percentage of energy explicitly captured on the mesh, using a scheme with less numerical dissipation will generally produce more accurate results.  

In all LES approaches, dissipation works to inhibit and damp out energy in the fine scales.  Dissipation is introduced into the simulation by the numerical method, the physical viscosity or the SGS model viscosity, if one is used.
In the absence of an explicit SGS model viscosity, the method must rely on the underlying numerical discretization scheme to supply the ``implied'' SGS viscosity through numerical dissipation.
Schemes which have no SGS model and no physical transport properties are classified as Implicit Large Eddy Simulation (ILES) methods.  A complete development of various ILES methods is given by Grinstein, Margolin, and Rider~\cite{grinstein:book} with additional development of the general ILES approach given by Boris~\cite{boris:1992}.  In the present work, we have considered methods that include the Navier-Stokes properties with and with out explicit SGS terms.    

For a given LES calculation with unresolved scales of motion, the effects of the three sources of dissipation are difficult to segregate.
Although the physical transport coefficients are directly known, or not included in the case of ILES, their relative effect on the solution will depend on the amount of model and numerical dissipation.  SGS dissipation and numerical dissipation will effectively vanish relative to the physical dissipation as the grid resolution increases and the DNS limit is approached.  However, DNS solutions are often not computationally feasible for high Reynolds number flows.  Furthermore, if physical viscous terms are not included, as is typical with ILES, the notion of a DNS limit is nonexistent.   


Efforts to quantify the dissipation of LES methods through an effective viscosity have been previously made. Aspden \etal~\cite{aspden:2008} derived an effective viscosity model for an incompressible fluid which was verified by a suite of viscous calculations for sustained isotropic turbulence.  Aspden's model was particularly instructive in that it was applied \emph{a posteriori} to data and allowed for an effective Reynolds number to be computed for viscous and inviscid simulations.  
Grinstein~\etal~\cite{grinstein:2011} briefly showed for a Taylor-Green vortex that there exists a connection between under-resolved LES calculations at high Reynolds number and resolved DNS calculations at a much lower Reynolds number, implying an effective Reynolds number for the LES calculations.  
Thornber~\etal~\cite{thornber:2007} examined the numerical viscosity of decaying isotropic turbulence using high-order methods in ILES calculations.
Zhou~\etal~\cite{zhou:2014} provided a method for estimating effective viscosities for ILES calculations where approximations of the dissipation rate are made to derive an effective viscosity.  


LES studies of the Richtmyer-Meshkov instability have led to significant scientific insight and have been done using the gamut of LES methodologies.  
Hill~\etal\cite{hill:2006} used the stretched vortex SGS method and a hybrid WENO scheme to simulate the effect of shock Mach number on RM growth with reshock.  Thornber~\etal~\cite{thornber:2010} used ILES and a high-order Gudonov-type finite volume method to investigate the dependence of initial conditions on RM induced mixing.  Shankar and Lele~\cite{shankar:2012} used a 6$^\text{th}$-order compact finite difference scheme and an explicit hyper-viscosity model~\cite{kawai:2008,cook:2007:POF} to perform LES studies of recent experiments \cite{balakumar:2008} of the RM instability.
Although a full review of all the LES studies of the RM instability is outside the scope of this paper, the aforementioned examples serve to illustrate the diversity of numerical methods and models used in LES of the RM instability.
Results from independent studies which employ different LES methodologies and different numerical methods will be subject to some degree of variability.  With the exception of DNS, results from different methods should not be expected to be identical.

To elucidate uncertainties in the LES approach, a comparison study of two Large Eddy Simulation methodologies is made by simulating the Richtmyer-Meshkov instability.  The range between the viscous (DNS) and inviscid (Euler) limits is explored by variation of the physical viscosity.  A grid convergence study in conducted at each Reynolds number for both LES approaches.  The resulting database of simulation data allows the various sources of dissipation to be explored and is unprecedented for three-dimensional RM instability.  A new {\it a posteriori} analysis is proposed which treats all methods, resolutions and Reynolds numbers in one common framework, which includes a formulation for both an effective viscosity and an effective species diffusivity.

The paper is divided into five subsequent sections.  Section~\ref{sec:methods} gives an overview of the equations of motion of multi-component flow which are being solved in the LES/DNS calculations.  The numerical methods of the two codes and LES models are outlined as well.  Section~\ref{sec:LESresults} includes results for the high- and low-Reynolds number cases, showing diagnostics for mixing and scale dependent energy.  Dependence of the results on grid resolution and numerical method are discussed.  In Section~\ref{sec:MUeff} a framework for comparing results of LES calculations at different Reynolds numbers, grid resolutions and using different numerics is given.  An effective viscosity and diffusivity are proposed which collapse the data and provide an estimate for an effective Reynolds number, P\'eclet number, and Schmidt number for the flow.  Additional discussion and suggestions for predicting the requirements for LES/DNS calculations is given in Section~\ref{sec:discuss} and a summary of the present work is given in the Conclusion in Section~\ref{sec:conclusion}.

%
%
%
%
%
%
%
%
%

\section{LES Methodology}
\label{sec:methods}

Variation of the numerical method is achieved by using two different LES codes for simulating the RM instability.  Both codes, Ares and Miranda, are developed at Lawrence Livermore National Laboratory and are capable of solving the compressible Navier-Stokes equations in three spatial dimensions.  In this section, an overview of the equations of motion is given.  A brief summary of each LES solver is provided including discussion of the numerical method and the LES model, if any.

\subsection{Equations of Motion}

The compressible multi-component Navier-Stokes equations for $N$ fluids can be written in strong conservation form as:

\begin{align}
	\pd{\rho Y_i}{t}  &+  \divergence \pth{\rho Y_i{\bf u} + {\bf J}_i} = 0, \ \ \ \text{for i=1,2,..,N} \label{eq:NS1} \\
	\pd{\rho {\bf u}}{t} &+ \divergence \pth{\rho{\bf u u^T} + \underline{\bf\delta} p - \underline{\bf\tau}} = 0 \label{eq:NS2} \\
	\pd{E}{t}  &+ \divergence \pth{ {\bf u} \pth{ E + p }  + {\bf q}-  {\bf u \cdot\underline{ \tau }}    } = 0	\label{eq:NS3}
\end{align}
where $\rho$ is the density, $Y_i$ is the mass fraction of species $i$, $\bf u$ is the velocity vector, $E=\rho\pth{ e+ {\bf u}^2/2}$ is the total energy of the mixture, $T$ is the temperature of the gas, $e$ is the internal energy, $p$ is the pressure, and $\underline{\bf\delta}$ is the Kronecker delta tensor.  The diffusive mass flux ${\bf J}_i$, viscous stress tensor $\underline{\tau}$, and energy flux $\bf q$ are given by

\begin{align}
	{\bf J_i} &= -\rho \pth{ D_i \grad Y_i      -Y_i \sum_{j=1}^N D_j \grad Y_j } , \label{eq:mass_diff_J}  \\
	\underline{\tau} &= 2 \mu  \underline{S} + \pth{\beta - \frac{2}{3}\mu} \pth{\divergence{\bf u}}  \underline{\delta} , \label{eq:tau} \\
	{\bf q} &= {\bf q}_T + {\bf q}_E
\end{align}
where the strain rate tensor $\underline{S}$, the conductive heat flux ${\bf q}_T$ and the interdiffusional enthalpy flux ${\bf q}_E$ are written as

\begin{align}
	\underline{S} &= \frac{1}{2}\pth{\grad {\bf u} + \pth{\grad {\bf u}}^T} \\
	{\bf q}_T &= - \kappa \grad T  \label{eq:heat_flux} \\
	{\bf q}_E &= \sum_i^N h_i {\bf J}_i
\end{align}
and where $h_i$ is the individual species enthalpy~\cite{cook:2009}.
\subsubsection{Mixture equation of state}
The Navier-Stokes terms in \cref{eq:NS1,eq:NS2,eq:NS3} contain the physical transport coefficients $\mu$, $\beta$, $\kappa$ and $D_i$; which are the shear viscosity, bulk viscosity, thermal conductivity, and species diffusivity, respectively.
For low-Mach number flow, the temperature dependence of the species diffusivities is small.  Once the shock wave has passed, the mean turbulent Mach number of the mixing layer remains below 0.05 for all time.  Therefore, for problem simplification, a constant physical viscosity is prescribed through a Reynolds number, species diffusivity through a constant Schmidt number and thermal conductivity through a constant Prandtl number as follows;

\begin{equation}
  \mu_i = \frac{ \rho_{0,i} V_0 \lambda_0 }{ Re_{\lambda_0,i} },
\label{eq:muf}
\end{equation}

\begin{equation}
         D_{f} = \frac{\mu_i }{ \rho_{0,i} Sc_i}
\label{eq:Diff}
\end{equation}

\begin{equation}
  \kappa_f = \frac{c_p \mu_f }{ Pr}
\label{eq:cond}
\end{equation}
where $V_0$ is the post-shock velocity, $\lambda_0$ is the fastest growing perturbed wave length (eq. \ref{eq:init}) and $c_p$ is the specific heat capacity at constant pressure.  For the present calculations, the initial Reynolds numbers ($Re_{\lambda_0}$) in the pre-shocked air and SF$_6$ are 30,000 and 180,000, respectively.  The Schmidt numbers ($Sc$) are 1.11 and 0.18 in the the Air and SF$_6$, respectively, and give a constant diffusivity, $D_f$.  The Prandtl number ($Pr$) is 1.0 and is based on the mixture viscosity, $\mu_f$, which is given as,

\begin{equation}
	\mu_f  = \pth{\sum_{i=1}^2 \frac{Y_i}{\mu_i}  }^{-1},
\end{equation}
where the species index $i$ refers to Air ($i=1$) and SF$_6$ ($i=2$).  The constant thermodynamic and species transport properties are summarized in table~\ref{tab:props}.  The ideal gas law is assumed, giving temperature and pressure as

\begin{align}
	T &= \frac{(\gamma_f-1)e}{R_f}, \\
	p &= (\gamma_f-1)\rho e.
\end{align}
The mixture ratio of specific heats ($\gamma_f$) and mixture specific gas constant ($R_f$) are given as

\begin{align}
	\gamma_f &= \frac{c_p}{c_v}, \\
	R_f &= R_{\text{univ}} \sum_{i=1}^2 \frac{Y_i}{M_{w,i}}
\end{align}
with 

\begin{align}
	c_p &= R_{\text{univ}} \sum_{i=1}^2 \frac{Y_i \gamma_i }{M_{w,i} \pth{\gamma_i-1}} ,\\
	c_v &= R_{\text{univ}} \sum_{i=1}^2 \frac{Y_i }{M_{w,i} \pth{\gamma_i-1}} 
\end{align}
and where $R_{\text{univ}} = 8.314\times 10^7$ [erg/K/mol] is the universal gas constant.

\begin{table}
\begin{center}
	\begin{tabular}{ |c|  c  |c|  c  |c|  c  |r|}
	\hline 
	Gas$_i$& $\gamma_i $ & $\mu_i$ [g/cm$\cdot$s] & $D_f$ [cm$^2$/s]& $M_{w,i}$ [g/mol] & Re$_{\lambda_0,i}$ & Sc$_{0,i}$ \\
	\hline 
	Air ($i=1$)& 1.4 & $18.26\times 10^{-5}$ &  $15.05 \times 10^{-2}$  & 28.8 & $30 \times 10^3$ & 1.11 \\
	SF$_6$ ($i=2$) & 1.1 & $14.75\times 10^{-5}$ &  $15.05 \times 10^{-2}$  & 146.0544 & $180\times 10^3$ & 0.18\\
	\hline
	\end{tabular}
\end{center}
	\caption{Constant thermodynamics and molecular transport properties for the present study.  
	\label{tab:props}
	}
\end{table}

\subsection{The Miranda code}

The Miranda code has been used extensively for simulating turbulent flows with high Reynolds numbers and multi-component mixing~\cite{cook:2004:JFM,cabot:2006:nature,olson:2007,olson:2011}.  Miranda uses a 10$^\text{th}$-order compact differencing scheme for spatial differentiation and a 5-stage, 4$^\text{th}$-order Runge-Kutta scheme for temporal integration.  Full details of the numerical method are given by Cook\cite{cook:2007:POF}.  For numerical regularization of the sharp, unresolved gradients in the flow, artificial fluid properties are used to locally damp structures which exist on the length scales of the computational mesh.  
	
In this approach, artificial diffusion terms are added to the physical ones which appear in Eqs. \ref{eq:mass_diff_J}, \ref{eq:tau} and \ref{eq:heat_flux}. 
This method of AFLES was originally proposed by Cook~\cite{cook:2007:POF} and has been altered by computing the artificial bulk viscosity term using $\divg\bgk{u}$ rather than $S$ (magnitude of the strain rate tensor).
Mani \etal \cite{mani:2009} showed that this modification substantially decreased the dissipation error of the method.  
The artificial transport coefficients are computed by taking higher derivatives of the resolved fields.  The explicit form for the terms and various test problems for validating the method are given in the references~\cite{cook:2007:POF,johnsen:2010,olson:2013:JCP}.

\subsection{ARES}
	
ARES is an Arbitrary Lagrange Eulerian (ALE) code developed at Lawrence Livermore National Laboratory (LLNL). The Lagrange time step uses a second order predictor-corrector method.  The Gauss divergence theorem is applied to solve the discrete finite difference equations~\cite{wilkins:1963} of the compressible multi-component Navier-Stokes equations (eqs.~\ref{eq:NS1}-\ref{eq:NS3}).  The spatial derivatives are approximated using a second-order finite difference method.  Artificial viscosity~\cite{kolev:2009} is applied to damp out spurious, high frequency oscillations which are generated near shocks and contact discontinuities.  

Velocities are defined as nodal quantities, while density and internal energy are defined at zone centers using piecewise constant profiles.
After each Lagrangian time step, a second order remap is applied to all variables (nodal and cell centered) to a new mesh, in keeping with the general ALE methodology.  For the simulations of this study, a fixed Eulerian mesh is used.

Although the ARES code includes an adaptive mesh refinement (AMR) capability~\cite{berger:1984,berger:1989}, it was not exercised in this study to facilitate a direct comparison with Miranda.
No explicit sub-grid scale model is applied to the equations of motion for all simulations presented in this study.

\section{Richtmyer-Meshkov Instability}
\label{sec:LESresults}

\subsection{Problem Setup} 

To focus the scope of the present study, only the single shock RM problem is considered.  
In this case, dependence on the initial conditions is strong, therefore a particular realization of initial conditions is used for both codes at all resolutions and Reynolds numbers.  The problem is solved in the post shock interface frame of reference, such that after the shock passes through the interface, it remains motionless in one dimension (1D).  This motion was analytically prescribed and verified numerically in 1D.

\begin{figure}[!h]
	\centering
	\includegraphics[width=7in,trim=0in 3.25in 0in 1.0in,clip=true]{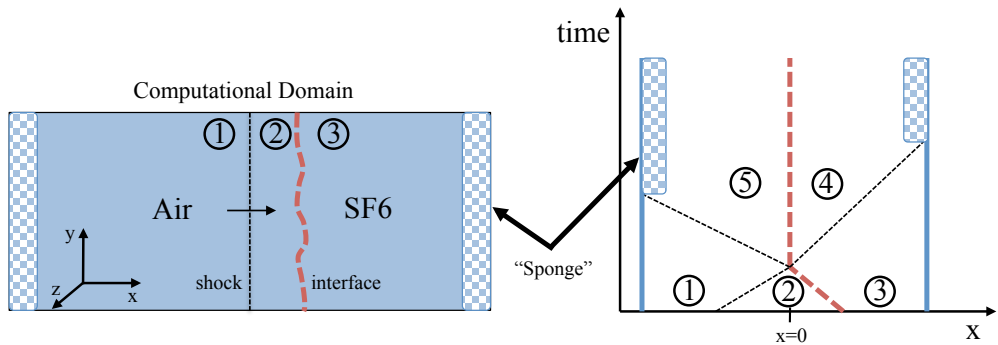}
	\caption{
	\label{fig:RM_setup}
	Schematic setup of the Richtmyer-Meshkov instability showing the initial conditions (left) and the 1D evolution of the shock waves and interface locations on the $x-t$ diagram (right).  The ``sponge'' boundary conditions are used to absorb the outward moving shock wave with minimal spurious reflection.  The states from region 1-5 are given in Table~\ref{tab:init}. 
	}
\end{figure}

A Mach 1.18 shock wave is initialized in air, ahead of a perturbed interface of sulfur-hexaflouride (SF$_6$).  The shock wave is initialized at $x_s$ such that the two discontinuities intersect at $x=0$ (see Fig.~\ref{fig:RM_setup}).  The shock wave satisfies the Rankine-Hugoniot jump conditions, which are used to prescribe the conditions ahead of and behind the shock wave.  The states in regions 1-5 of Figure~\ref{fig:RM_setup} are explicitly given in Table~\ref{tab:init} below.  The three dimensional domain extents are 16cm$\times$8cm$\times$8cm in the $x$,$y$ and $z$ directions, respectively.  

\begin{table}[!h]
\begin{center}
	\begin{tabular}{|c   |c   |c   |c  |c|  }
	\hline 
	Region & $p ~\mathrm{[g/(cm\cdot s^2)]}$ &  $\rho~\mathrm{[g/cm^3]}$ & $u_x~\mathrm{[cm/s]}$  & Species \\
	\hline 
	1 & 1.36e6 & 1.42e-3 & 3.33e3     & Air\\
	2 & 0.931e6 & 1.08e-3 & -6.33e3    & Air\\
	3 & 0.931e6 & 5.50e-3 & -6.33e3    & SF$_6$ \\
	4 & 1.53e6 & 8.66e-3 & 0.0     & SF$_6$ \\
	5 & 1.53e6 & 1.55e-3 & 0.0     & Air\\
	\hline
	\end{tabular}
\end{center}
	\caption{Values for initial flow field in the post shock air (region 1), pre shock air (region 2) and pre shock SF$_6$ (region 3).  The final states after the transmission and reflection of the shock wave are given for the SF$_6$ and Air, regions 4 and 5, respectively.  
	\label{tab:init}
	}
\end{table}

\subsubsection{Perturbed initial interface}
The perturbation of the initial interface is necessary to generate baroclinic vorticity, instability growth and eventual transition to turbulence.  The perturbation is defined in Fourier space as a power spectrum which is a function of the two-dimensional wave number.  In this study, the general form for the power spectrum suggested by Thornber~\etal~\cite{thornber:2010} is assumed as

\begin{equation}
	P(k) = \begin{cases}
			Ck^m, & k_\text{min}< k < k_\text{max}, \\
			0,  & \text{otherwise},
		\end{cases}
	\label{eq:init}
\end{equation}  
where $k=\sqrt{k_y^2+k_z^2}$ is the two-dimensional wave number.  For the present study, $C=\lambda_0/10$, $\lambda_0 = 1/k_\text{max}$ , $m$ is set to -2, ($k_\text{min}$,$k_\text{max}$) is set to (4,16) and the random phase shifts used to construct the Fourier modes were determined \apri and used to initialize all calculations.  Since $k_\text{max}$ is less than the Nyquist wave number on the coarsest mesh, all initial fields are spectrally exact.  
		  
The interface height is therefore given as 
\begin{equation}
	\eta(y,z) = \sum_j^{k_y} \sum_k^{k_z} P(k) \cos (k_y y + \phi_{y,j} )  \sin (k_z z + \phi_{z,k}), 
\end{equation}
where $\phi$ are the set of random numbers used for all initializations.  The mass fraction fields are diffusely initialized over a finite width using a hyperbolic tangent function as
\begin{align}
	Y_{SF_6}(x,y,z,\tau=0) &= \frac{1}{2}\pth{1 + \tanh\pth{ \frac{x - \eta(y,z)}{\delta_p}}  } , \\
	Y_{Air}(x,y,z,\tau=0) &= 1 - Y_{SF_6}(x,y,z,\tau=0) 
\end{align}
where $\delta_p$ is the initial interface thickness and is set to $\lambda_0 / 4$ for all calculations and where the non-dimensional time is given as $\tau = t V_0/\lambda_0$.  Prior to first shock the two fluids have a constant ambient temperature of 297 K, which is implicitly given the values of Table~\ref{tab:props} and~\ref{tab:init} and the ideal gas equation of state.  

\begin{figure}
	\centering
	\includegraphics[width=.75\linewidth, angle=0]{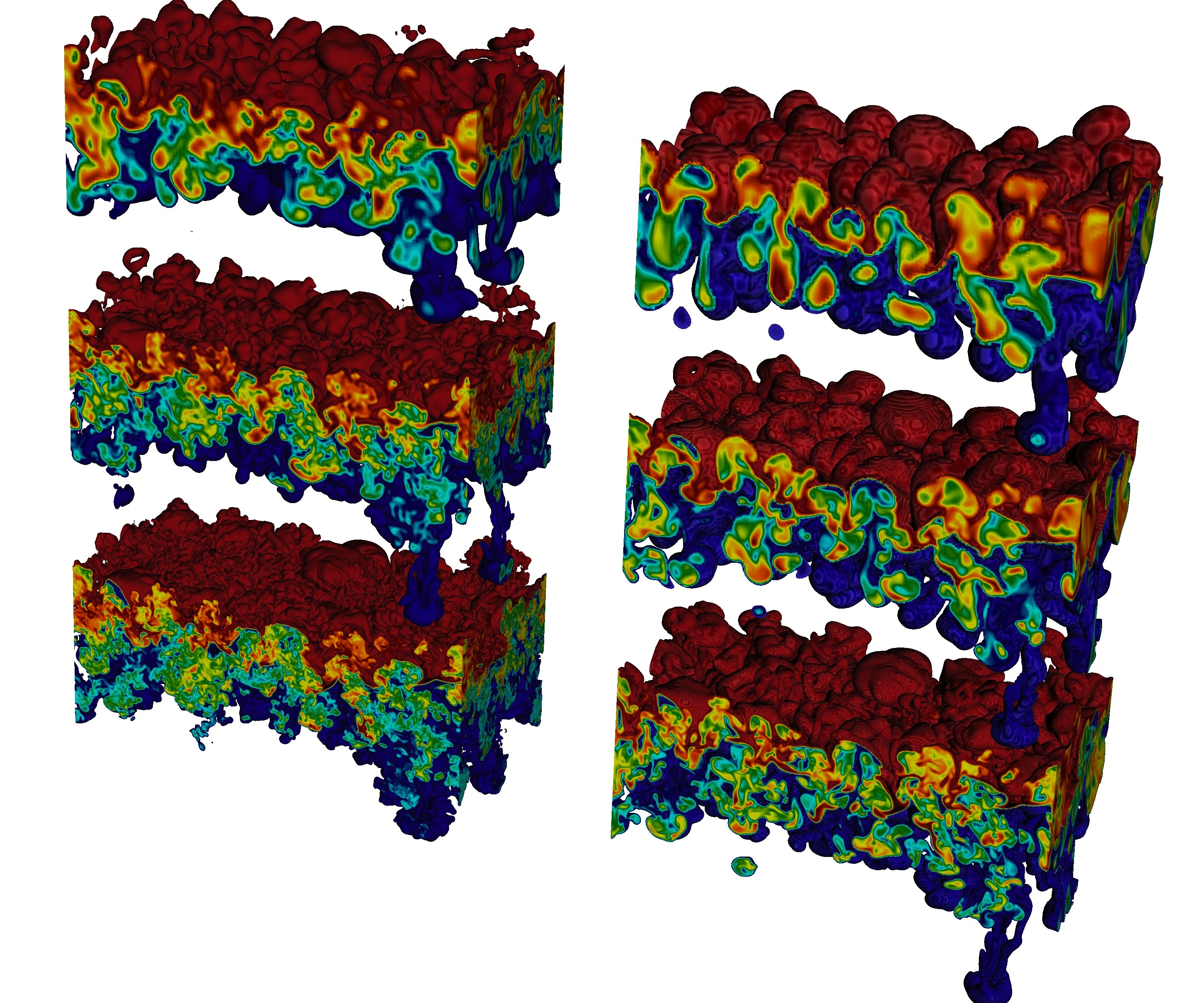}
	\caption{\label{fig:3Dscales}
	Iso-volume of the mass fraction of $SF_6$ between .1 and .9 for cases B, C, and D (top to bottom) from Miranda (left) and Ares (right) calculations at the nominal Reynolds number at $t V_0 / \lambda_0=35$.  Data from mesh D show the existence of a broad range of length scales in the mixing layer.  
	}
\end{figure}
		  
\subsection{Low Reynolds number DNS}
\label{sec:DNS}
		To establish a baseline for convergence to the resolved scales, a grid refinement study was conducted at a Reynolds number 1/25$^\text{th}$ the nominal value.  This reduction in Reynolds number (and subsequent reductions) was achieved by multiplying the species diffusivity and viscosities by the relevant factor, thereby maintaining a constant Schmidt number.  At this Reynolds number it was possible to approach the DNS limit for the given high-resolution grid spacing selected.  Table~\ref{tab:DNS} shows the various resolutions selected and the resulting number of total grid points.
	
	\begin{table}[!h]

\begin{center}
	\begin{tabular}{|c |c |c |c |c|}
	\hline 
	Mesh & $N_{x}$ & $N_{y}$ & $N_{z}$  & Total Pts.\\
	\hline 
	A & 128 & 64 &                   64              & 0.5 M \\
	B & 256 & 128 &               128             & 4.2 M \\
	C & 512 & 256 &               256             & 33.5 M \\
	D & 1024 & 512 &             512            & 268.4 M \\
	\hline
	\end{tabular}
\end{center}
	\caption{Computational mesh parameters for various levels of refinement and the resulting number of total grid points.  Grids were uniformly spaced in all three coordinate directions.
	\label{tab:DNS}
	}
\end{table}

\subsubsection{Mixing region growth}
\label{sec:MRG}

Several integral measures of the mixing region are compared here.  These global measures show the time dependent mixing state and are typically used for experimental comparison where only gross mixing measures are available.  The mixing width is defined as

\begin{equation}
	W  = 4\int_{-\infty}^{\infty} \langle Y_\text{SF6} \rangle \langle Y_\text{Air}\rangle dx,
\end{equation}
where the $\langle\cdot\rangle$ operator denotes planar averages taken in the $y$-$z$ plane and is defined as

\begin{align}
	\langle f \rangle (x,t) &= \frac{1}{A} \int  f(x,y,z,t) dy dz \text{  ,  where } \\
	A &= \int dy dz.
\end{align}

Another measure of mixing is the ``mixedness'', which is the ratio of mixed fluid to entrained fluid defined as

\begin{equation}
	\Theta  = \frac{\int_{-\infty}^{\infty} \langle Y_\text{SF6}  Y_\text{Air}  \rangle dx}{ \int_{-\infty}^{\infty} \langle Y_\text{SF6} \rangle \langle Y_\text{Air}\rangle dx} .
\end{equation}
For fully developed three-dimensional mixing, this quantity approaches~\cite{cabot:2006:nature} $\approx 0.8$.  $\Theta$ represents the $2^\text{nd}$ statistical moment of mixing and can be identically related to the variance of the mass fractions as

\begin{equation}
	\Theta  = 1 + 4\frac{\int_{-\infty}^{\infty} \langle Y_\text{SF6}^\prime  Y_\text{Air}^\prime  \rangle dx}{ W } 
	\label{eq:theta}
\end{equation}
where the primed values are defined as $F^\prime = F-\langle F\rangle$.  Therefore, where $W$ is an integral measure of the mean of species mass fraction, $\Theta$ is an integral measure of its variance or fluctuations.  

\ThreeFig
{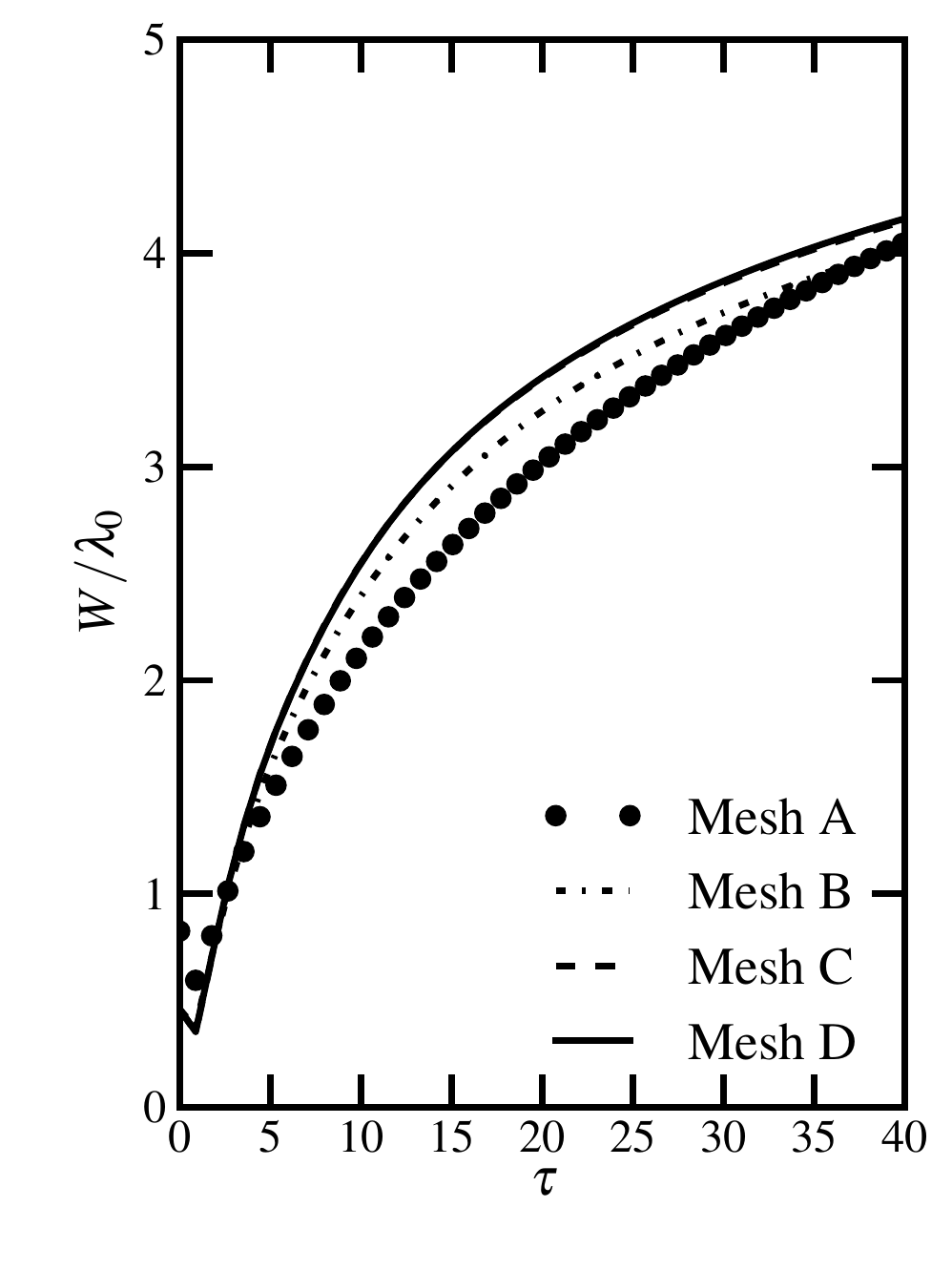}{Miranda}
{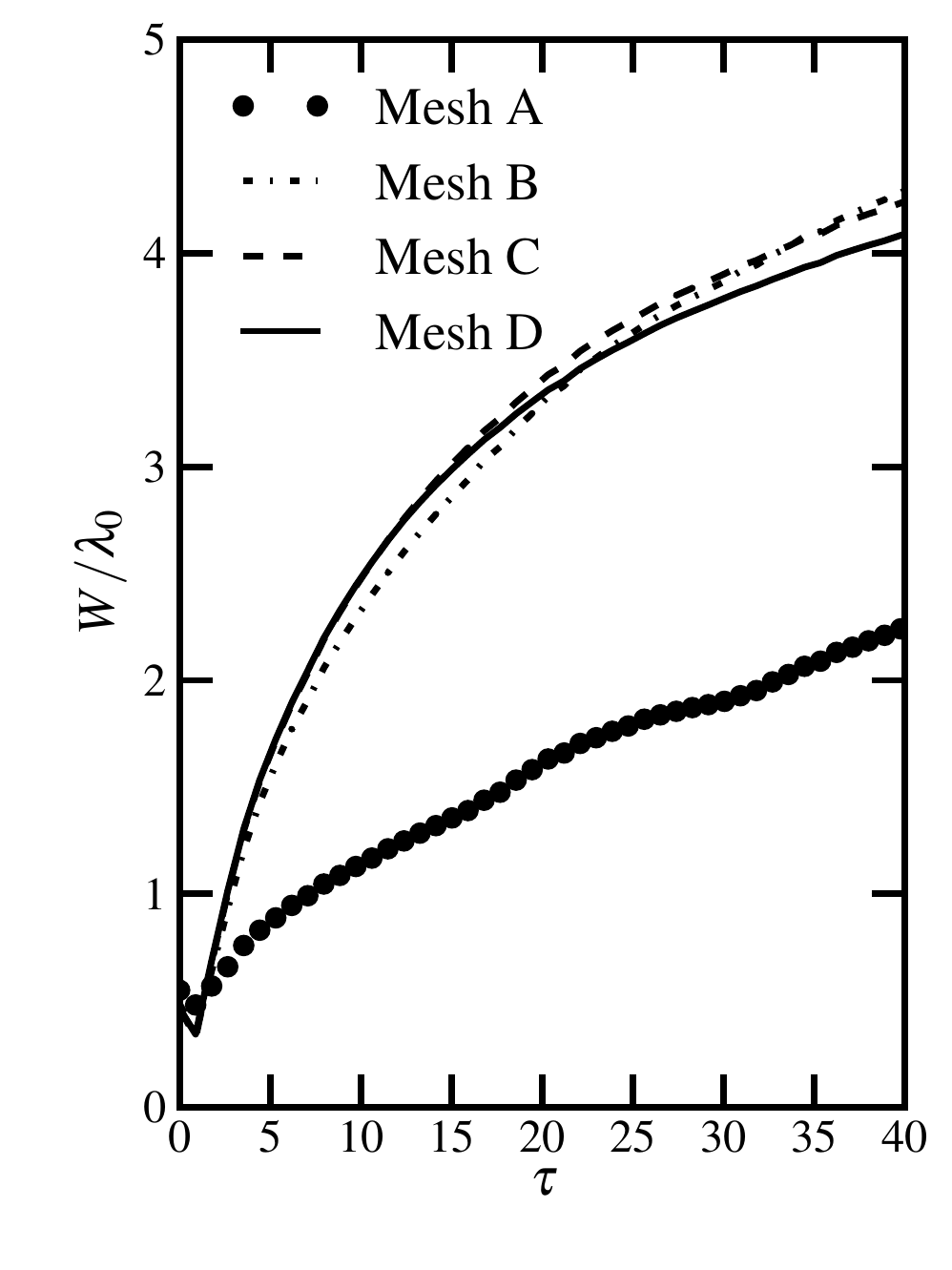}{Ares}
{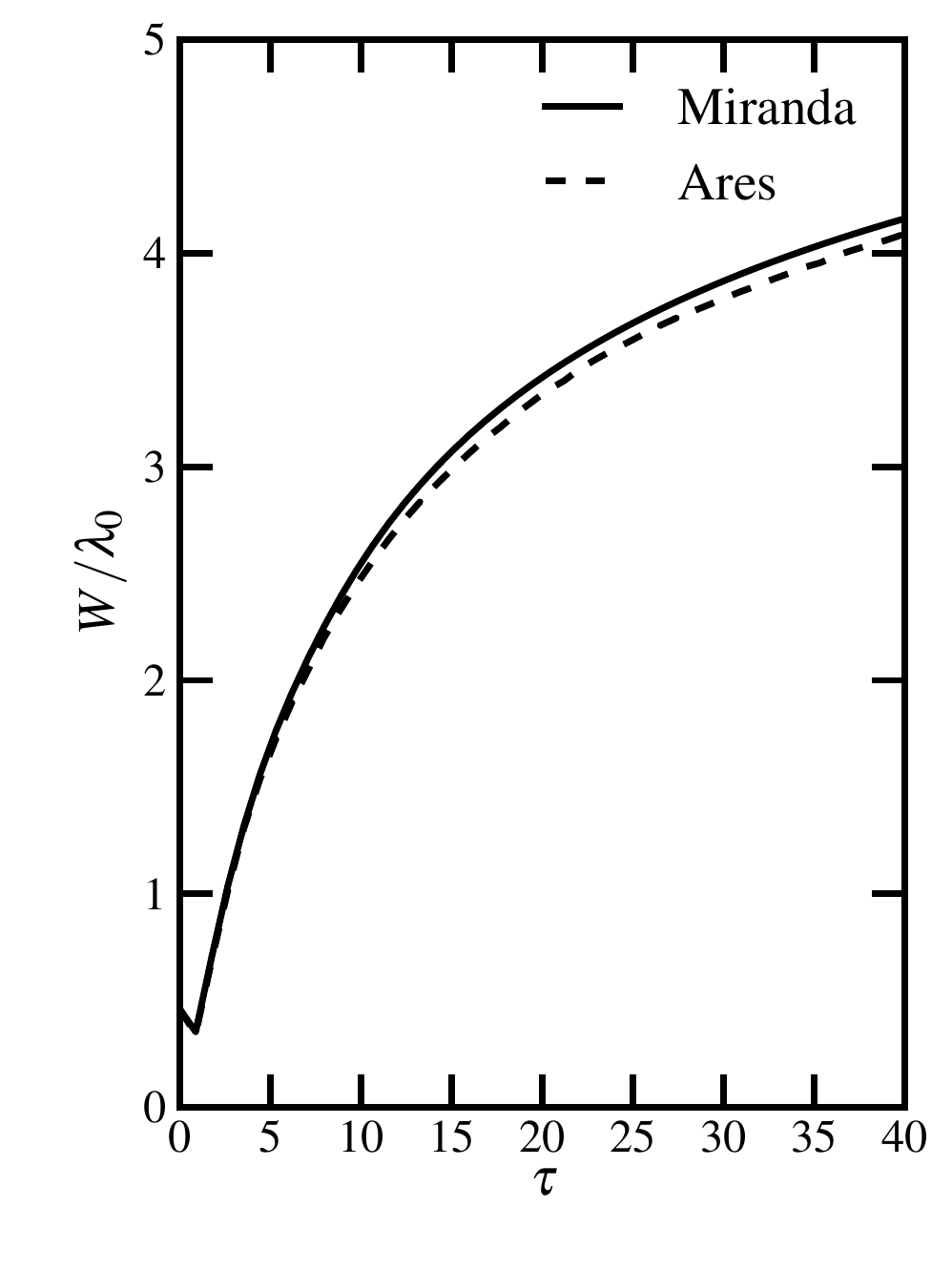}{Mesh D for Miranda \& Ares}
{Non-dimensional mixing width vs. time for meshes A-D at the reduced Reynolds number from Miranda (a) and Ares (b).  Data between codes at the finest resolution are plotted in (c) and show that agreement worsens with time and is at most 2\% different at $\tau=40$.}
{fig:Wdns}{.55in}

\ThreeFig
{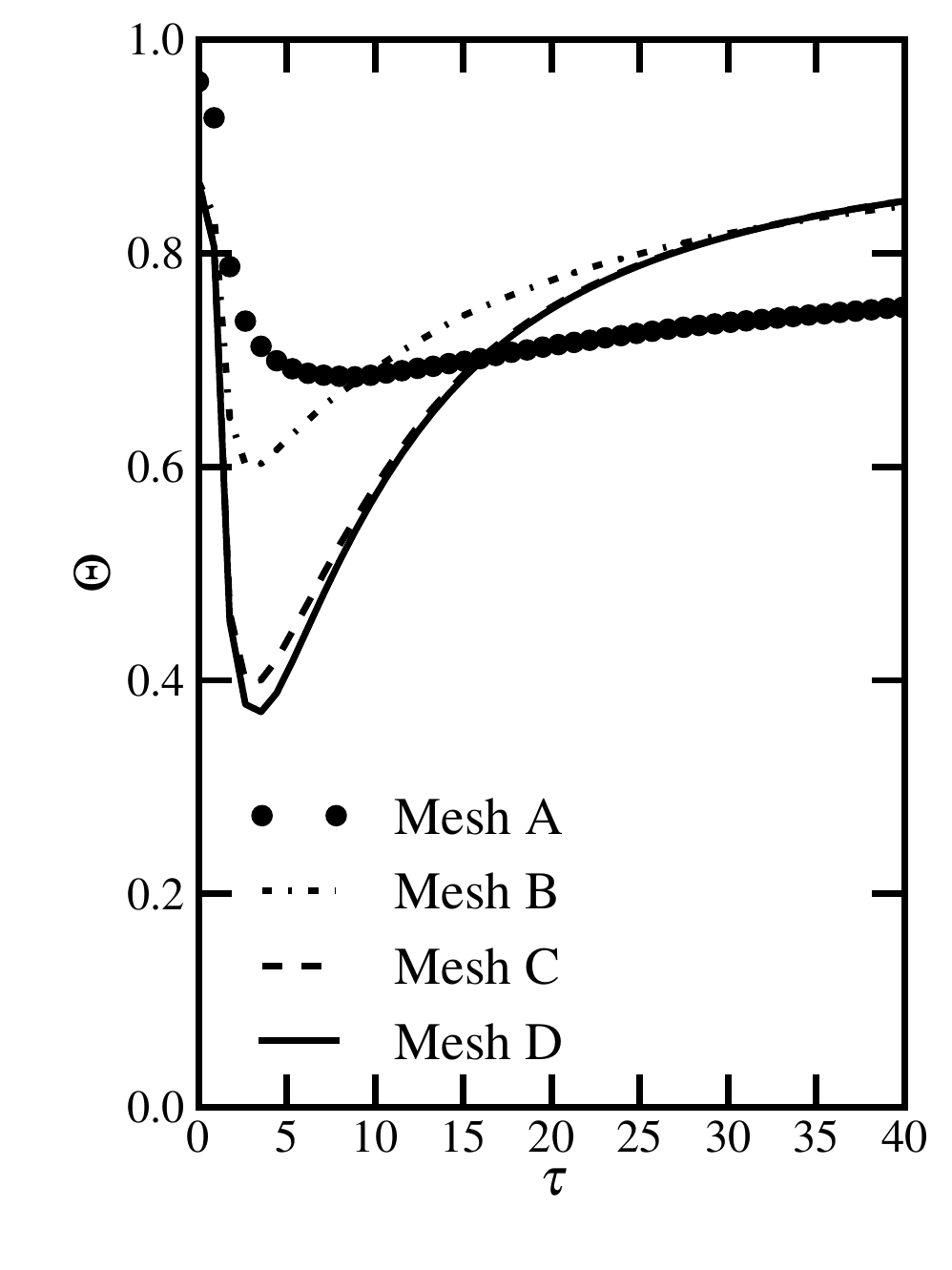}{Miranda}
{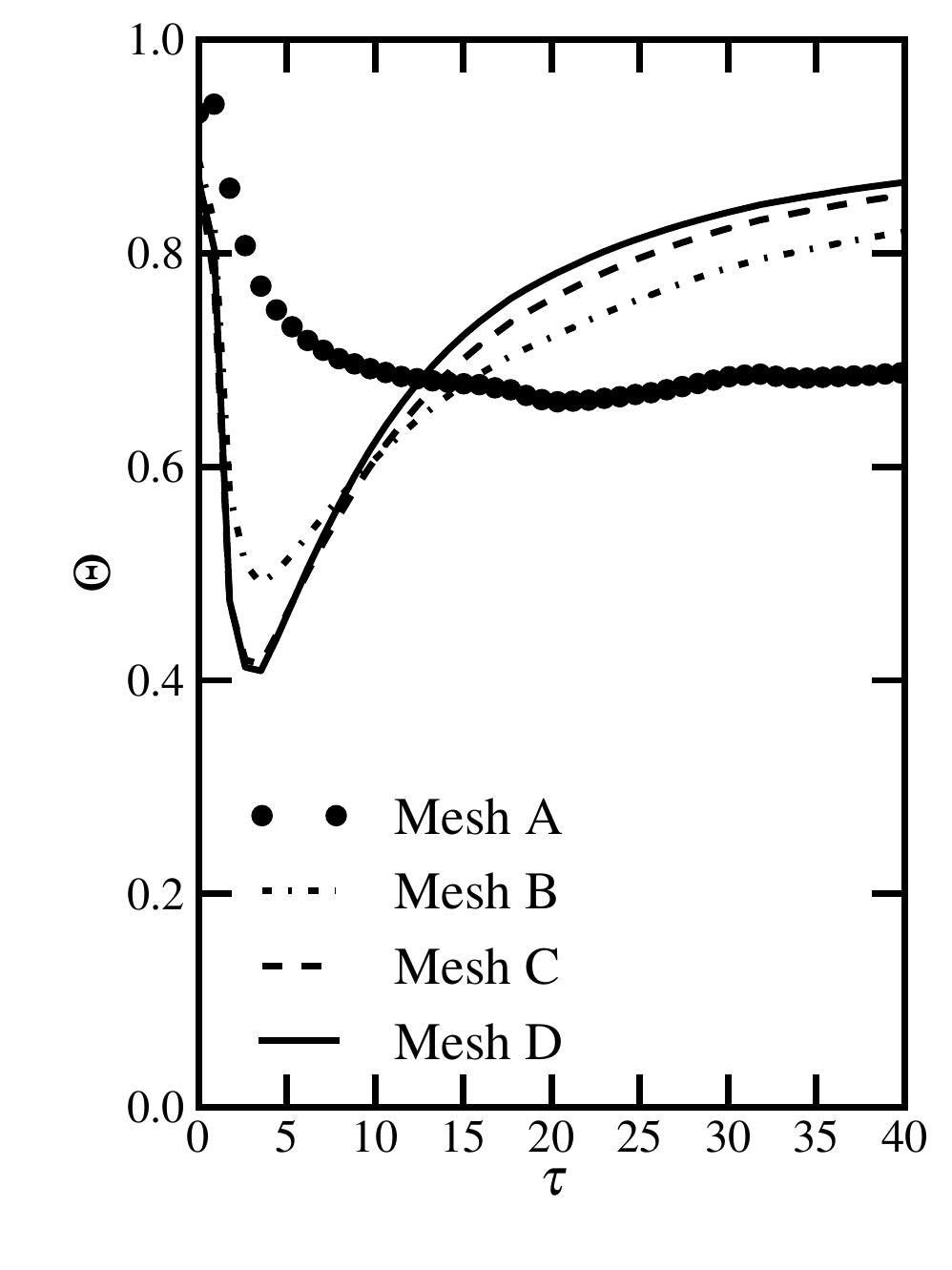}{Ares}
{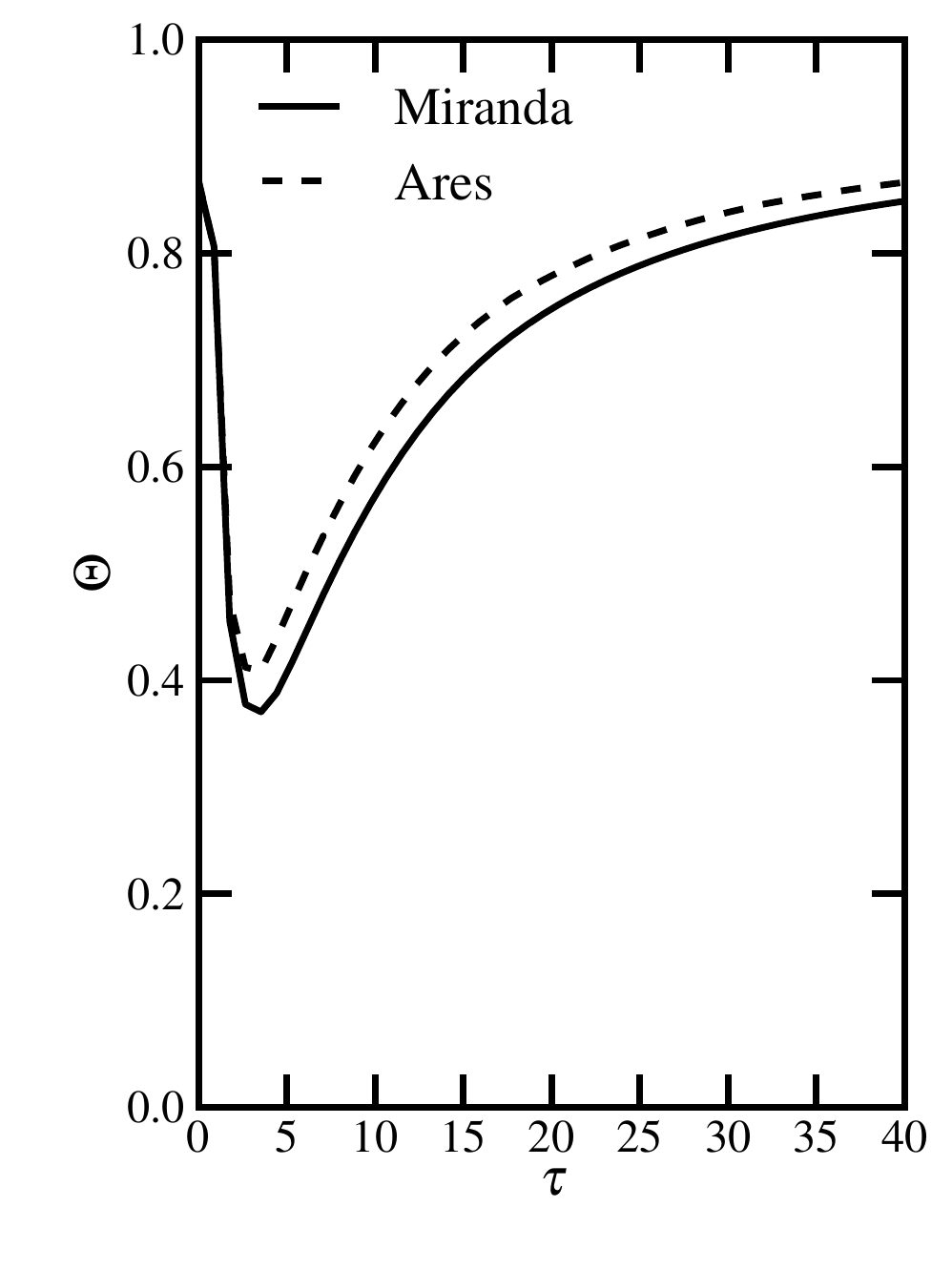}{Mesh D for Miranda \& Ares}
{Mixedness ($\Theta$) vs. time for meshes A-D at the reduced Reynolds number from Miranda (a) and Ares (b).  Data between codes at the finest resolution are plotted in (c) and show that differences arise early on and remain constant over the observed time.}
{fig:Phidns}{.50in}

Figures~\ref{fig:Wdns} and~\ref{fig:Phidns} show the time history of $W/\lambda_0$ and $\Theta$ at the various resolutions for the two codes.  Curves for resolutions C and D are nearly indistinguishable for $W/\lambda_0$ through $\tau = 40$ in Miranda (Fig.~\ref{fig:Wdns}a) and up to $\tau=15$ in Ares (Fig.~\ref{fig:Wdns}b) .  The comparison of the fine mesh calculation between codes (Fig.~\ref{fig:Wdns}c) shows the solutions differ more as time progresses, reaching approximately 2\% difference at $\tau=40$.   

Figure~\ref{fig:Phidns}a shows values for $\Theta$ are converged for $\tau>10$ in Miranda.  For Ares in Figure~\ref{fig:Phidns}b, the opposite occurs, where convergence is most pronounced for $\tau < 10$.  At $\tau=5$, $W$ is constant and $\Theta$ differs between the two codes (Figure~\ref{fig:Phidns}c) which (by enforcing equation~\ref{eq:theta}) implicates a larger mass fraction variance in Miranda at $\tau=5$ and indeed, for all time.  This statement is confirmed by differences in the power spectra of the mass fraction given below.


\subsubsection{Spectra}

Evaluating the wave number dependence of the fluctuating turbulent quantities can elucidate characteristics of the flow physics as well as the numerical errors associated with the particular LES approach.  In LES comparisons, directly measuring the energy of high wave numbers will indicate the range of scales which are resolved on the LES mesh.
Spectra are computed at each $y$-$z$ plane where $4\langle Y_\text{SF6} \rangle \langle Y_\text{Air}\rangle > 0.7$.  The two-dimensional Fourier transforms from these $N$ planes are then averaged, binned into annuli and plotted as a function of two-dimensional wave number, $k$.  This procedure is applied to the fluctuating mass fraction field as well as the velocity field, which are plotted in Figure~\ref{fig:SpecUdns} and~\ref{fig:SpecYdns}, which shows the spectra for both Ares and Miranda at all resolutions at $\tau=35$.

\ThreeFig
{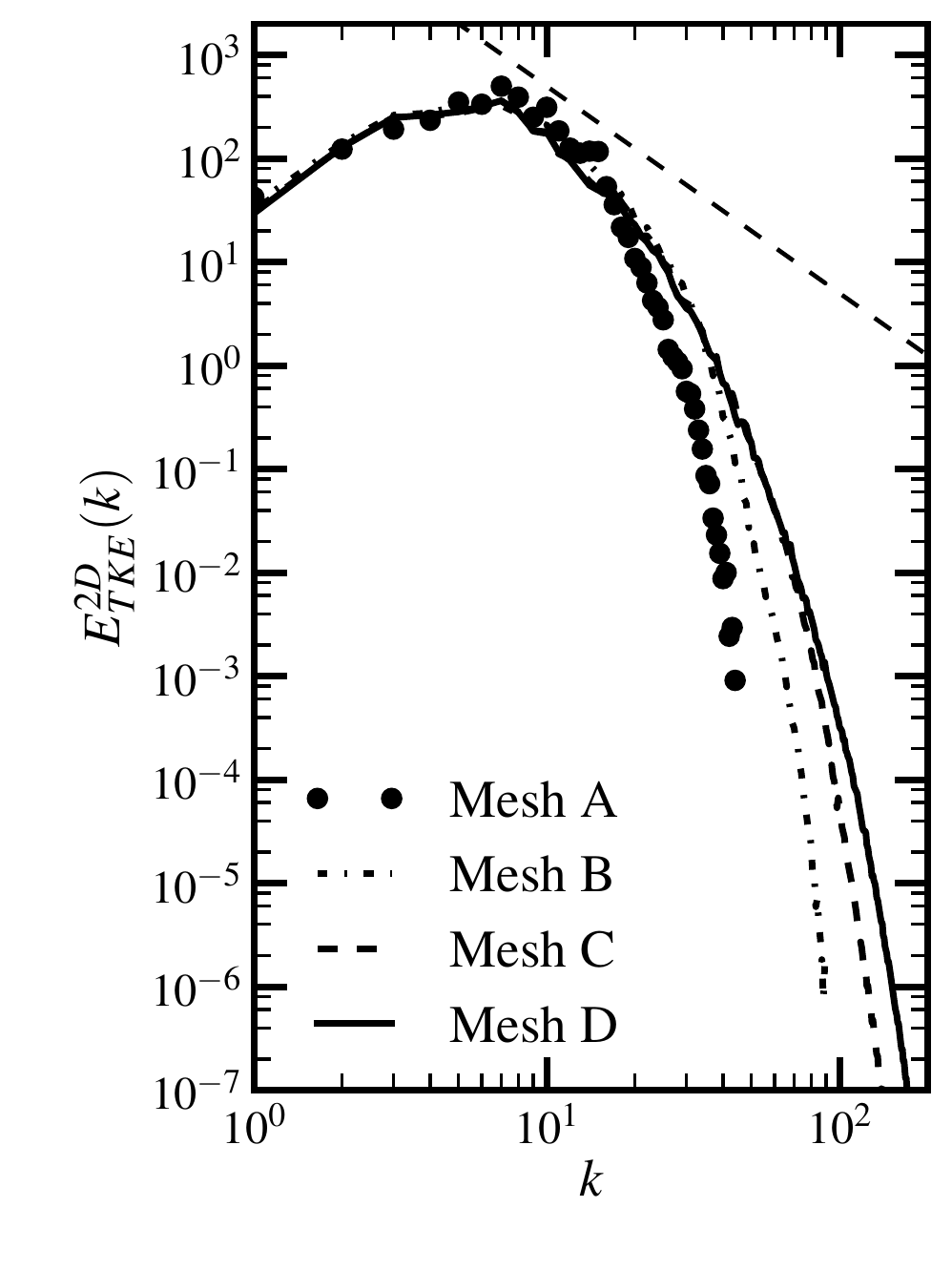}{Miranda}
{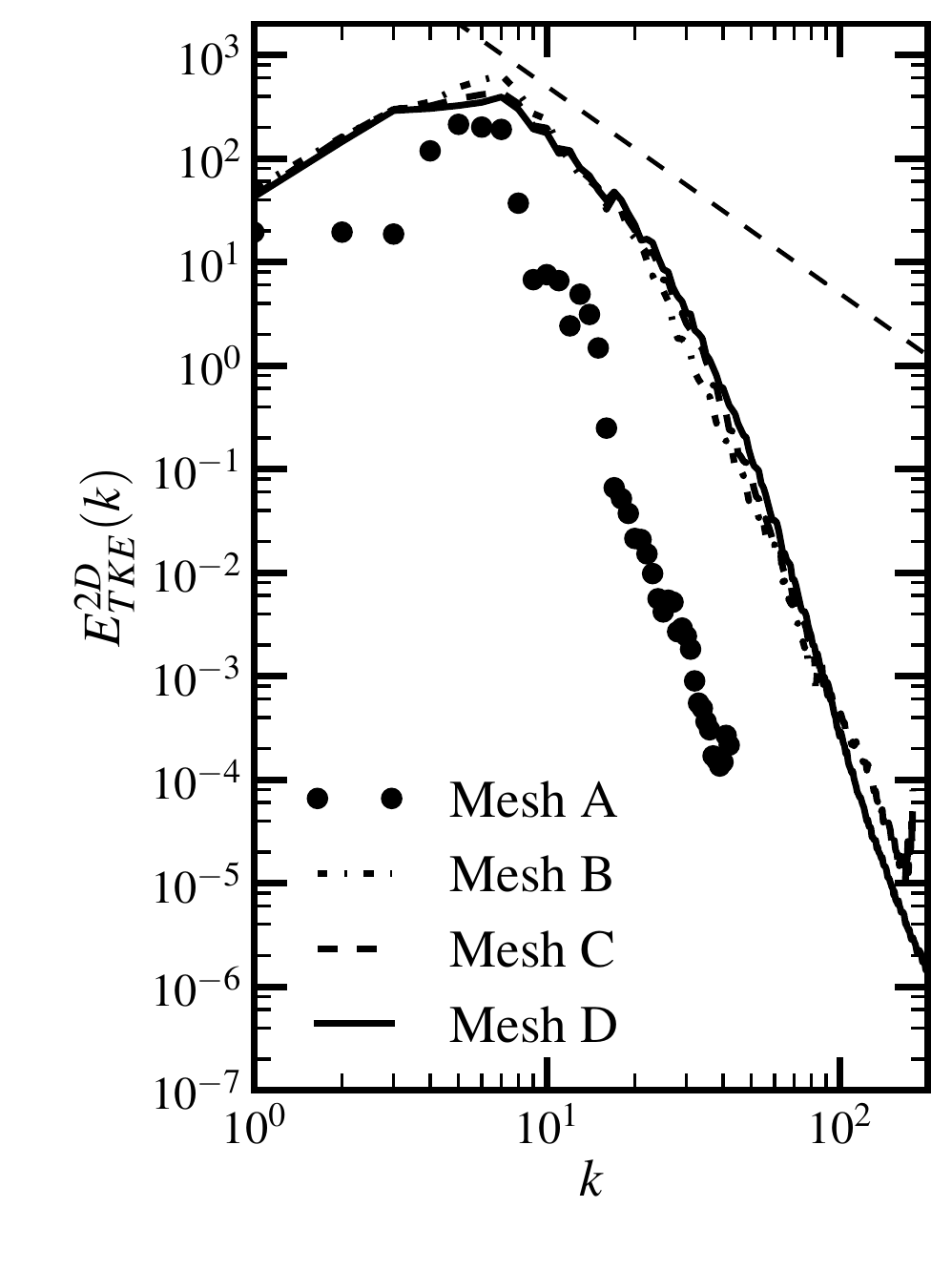}{Ares}
{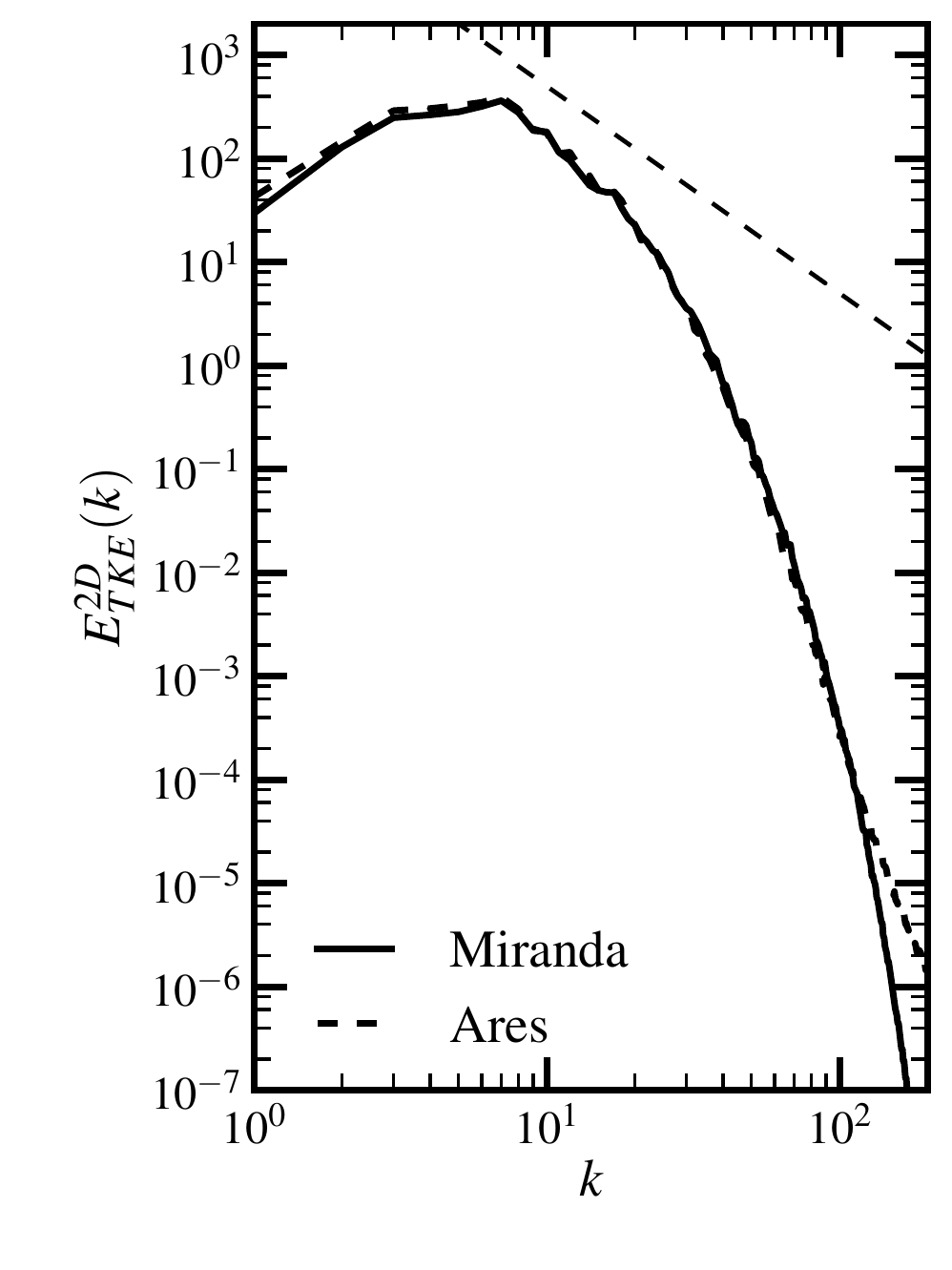}{Mesh D for Miranda \& Ares}
{Power spectra of the fluctuating velocity at $\tau=35$ for Miranda (a) and Ares (b) for meshes A-D at the reduced Reynolds number.  Convergence for wave numbers less than 70 is observed in both codes.  The difference of the spectra for mesh D between the codes (c) is negligible up to wave number 100.  The $k^{-5/3}$ fiducial is plotted (dashed) and shows a lack of an inertial subrange.}
{fig:SpecUdns}{.57in}

\ThreeFig
{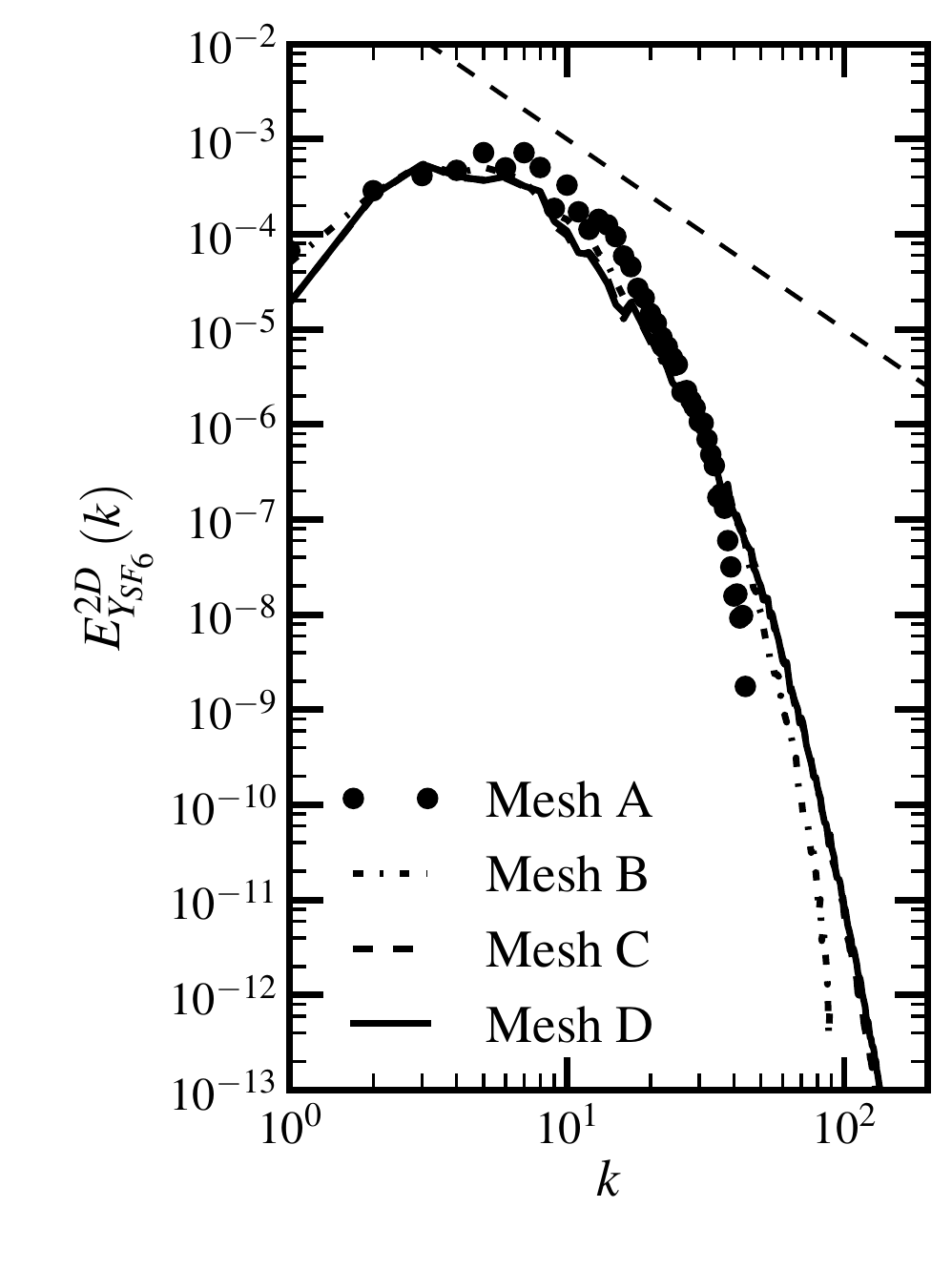}{Miranda}
{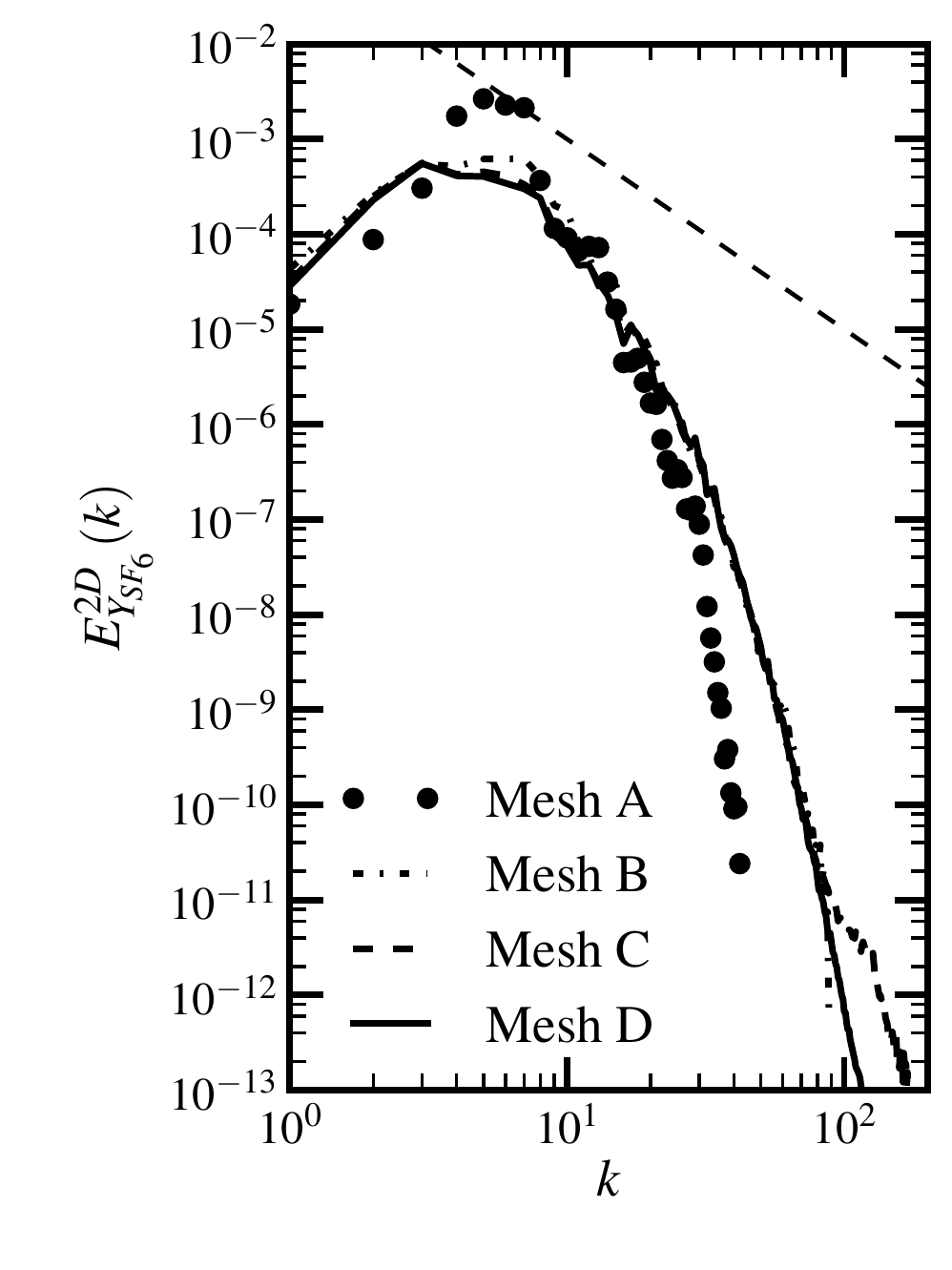}{Ares}
{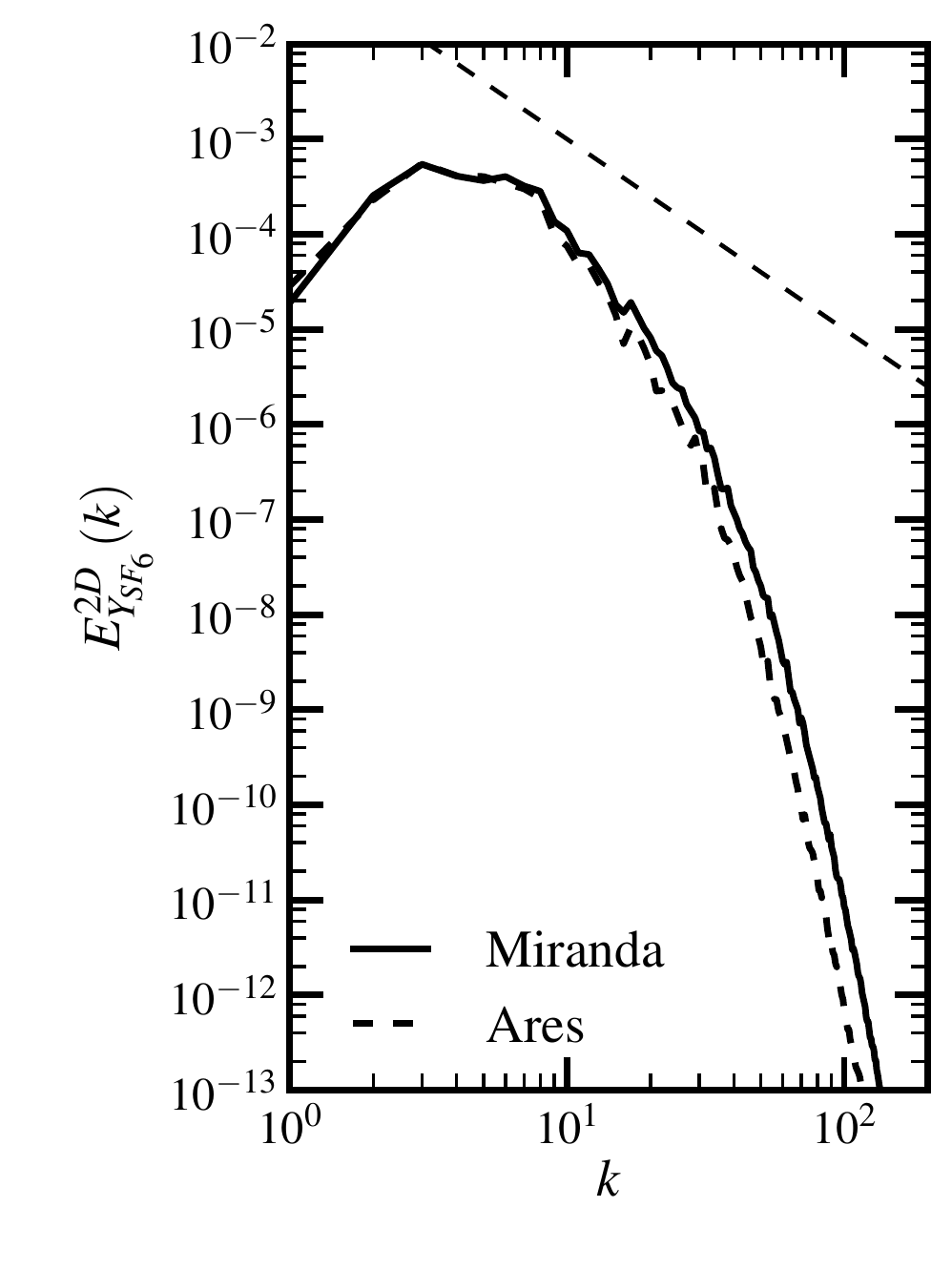}{Mesh D for Miranda \& Ares}
{Power spectra of the mass fraction of SF$_6$ at $\tau=35$ for Miranda (a) and Ares (b) for meshes A-D at the reduced Reynolds number.  Convergence for wave numbers less than 100 is observed in both codes.  The difference of the spectra for mesh D between the codes (c) is noticeable for wave numbers larger than 15.  The $k^{-5/3}$ fiducial is plotted (dashed) and shows a lack of an inertial subrange.}
{fig:SpecYdns}{.65in}

The spectra show excellent convergence at the low wave numbers and a strong trend toward convergence at the higher wave numbers.  
For the velocity spectra, data from the finest resolution grid of Ares and Miranda (Figure~\ref{fig:SpecUdns}c) are nearly indistinguishable for all wave numbers below $k=100$.  The spectra of the mass fraction indicate that nearly all scales are captured for both Miranda (Figure~\ref{fig:SpecYdns}a) and Ares (Figure~\ref{fig:SpecYdns}b), as no differences are observed between spectra from meshes C and D.  However, Ares and Miranda data are converging to different solutions in the high wave numbers, indicating a dependence on the numerical method.  Since the integral of the power spectral density is proportional to the variance, Figure~\ref{fig:SpecYdns}c supports the previous assertion made in Section~\ref{sec:MRG}, that Miranda has a larger mass fraction variance.

\subsubsection{Dissipation Measures}

Numerical dissipation is most active on the fine scales which are unresolved on the computational grid.  Quantities which are more dependent on the small scales, therefore, will exhibit larger sensitivity to both grid resolution and numerical method.  To explore these sensitivities, the domain integrated enstrophy and normalized scalar dissipation rate are computed to explore the high wave number behavior of dissipation of the velocity field as well as the scalar field.  
Enstrophy is given by 
\begin{equation}
	\Omega (t) = \int_V \rho \| \mathbf{\omega} \| ^2 \ \  {dxdydz}
	\label{eq:enst}
\end{equation}
where $\mathbf{\omega} = \nabla\times\mathbf{u}$.  The scalar (mass fraction) dissipation rate is defined as

\begin{equation}
	\chi (t) = \int_V D_{SF_6} \nabla Y_{SF_6} \cdot \nabla Y_{SF_6} \  {dx dy dz}.
	\label{eq:tmr}
\end{equation}
Given that the simplified equation of state produces a constant value for diffusivity in the mixing layer, the diffusivity may be pulled outside the integral.  Therefore, comparing $\chi/D_{SF_6}$ allows data from LES calculation of various Schmidt and Reynolds numbers to be compared directly.  

Differences between the codes and resolutions are largest at the temporal maxima of the enstrophy and scalar dissipation rate curves in Figures~\ref{fig:Enstdns} and \ref{fig:TMRdns}.  This maximum occurs at around $\tau=12$ for enstrophy and at $\tau=8$ for scalar dissipation and is indicative of the time when the flow is becoming damped by the dissipation scales.  Therefore, energy coupling to higher modes has taken place and the flow is beginning to transition to broadband turbulence.  Note here that the ``turbulence'' referred to is in the diffusive/dissipative regime as the flow is relaxing and decaying and is not being driven.  For the scalar field, this dissipation threshold occurs slightly before that of the velocity field.

Convergence is significantly slower for these global measures of dissipation as compared to the mean mixing measures.  In both Miranda and Ares (Figure~\ref{fig:Enstdns}a-b), enstrophy values are getting closer under grid refinement but have not fully converged by mesh D.  The difference between mesh C and D in Miranda is smaller than that in Ares.  The difference between the enstrophy at the finest mesh (Figure~\ref{fig:Enstdns}c) between the two codes is large, at nearly 10\%.

The measure of scalar dissipation also exhibits slow convergence, with the mesh D solution differing from that of mesh C by approximately 20\% for both Miranda and Ares (Figure~\ref{fig:TMRdns}a-b).  The difference between codes in $\chi/D_{SF_6}$ at the mesh D resolution (Figure~\ref{fig:TMRdns}c) is even larger than the differences in the enstrophy as might be expected, given that the power spectra of the species mass fraction differ more than the power spectra of the velocity.

\ThreeFig
{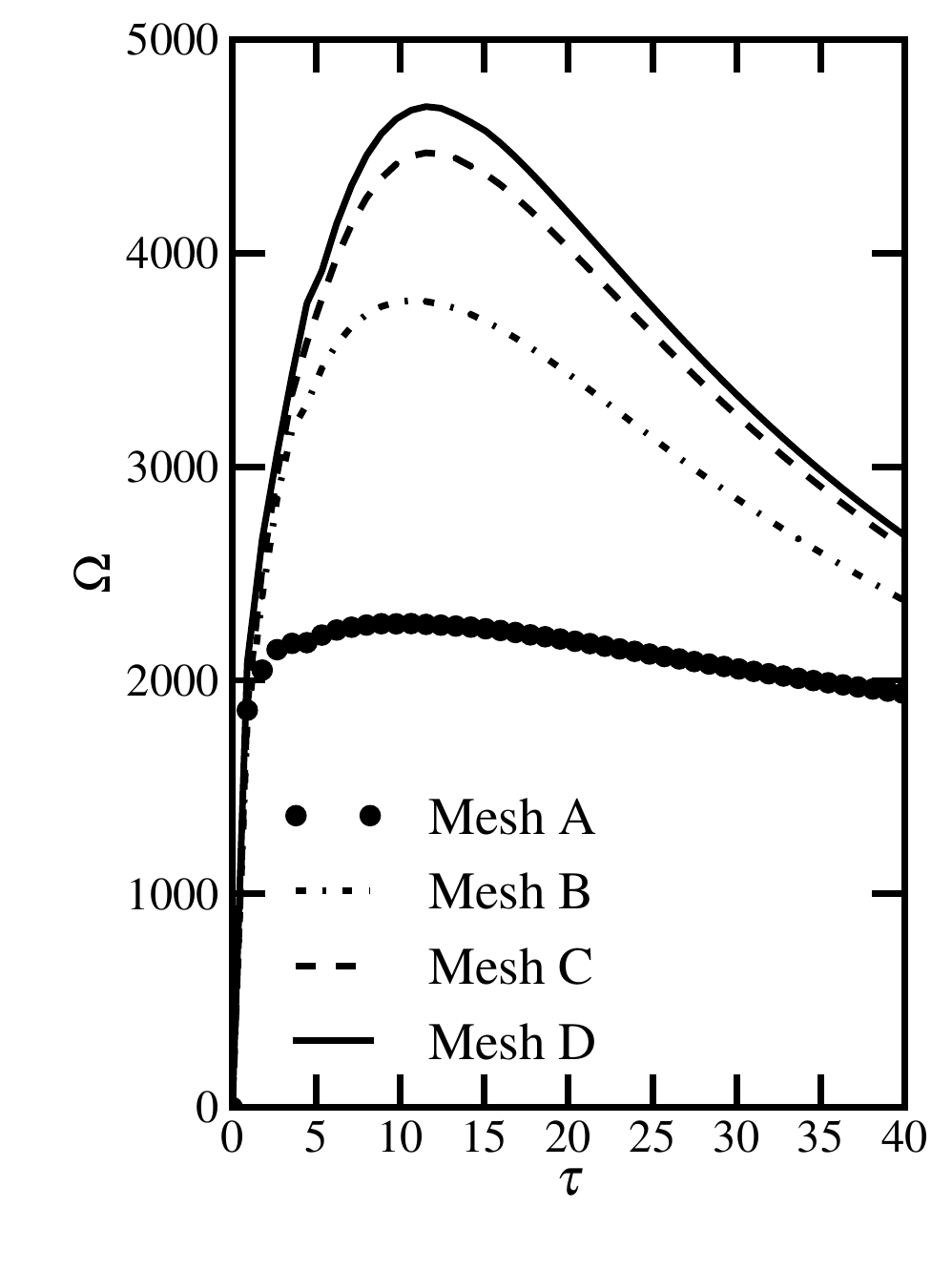}{Miranda}
{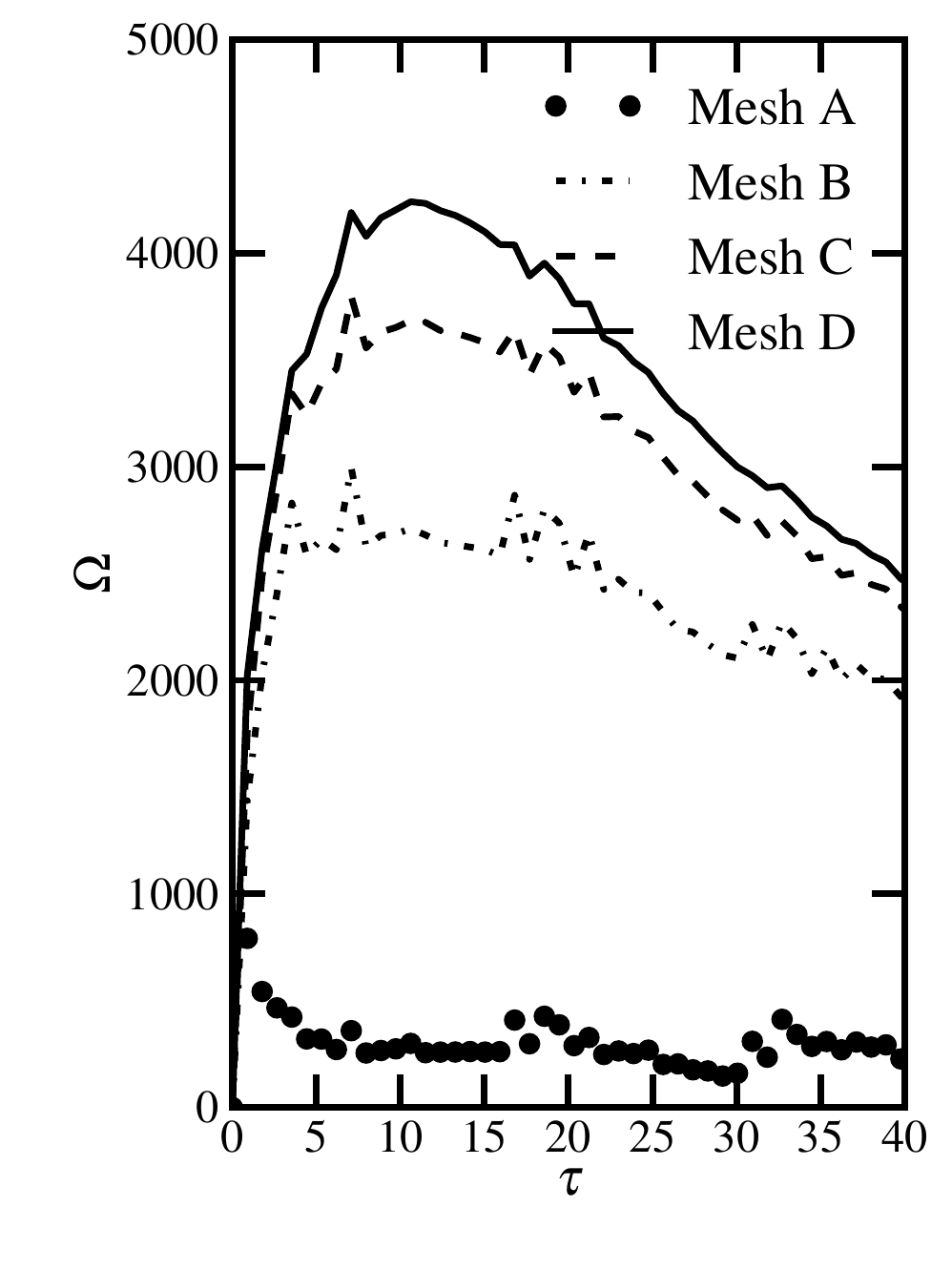}{Ares}
{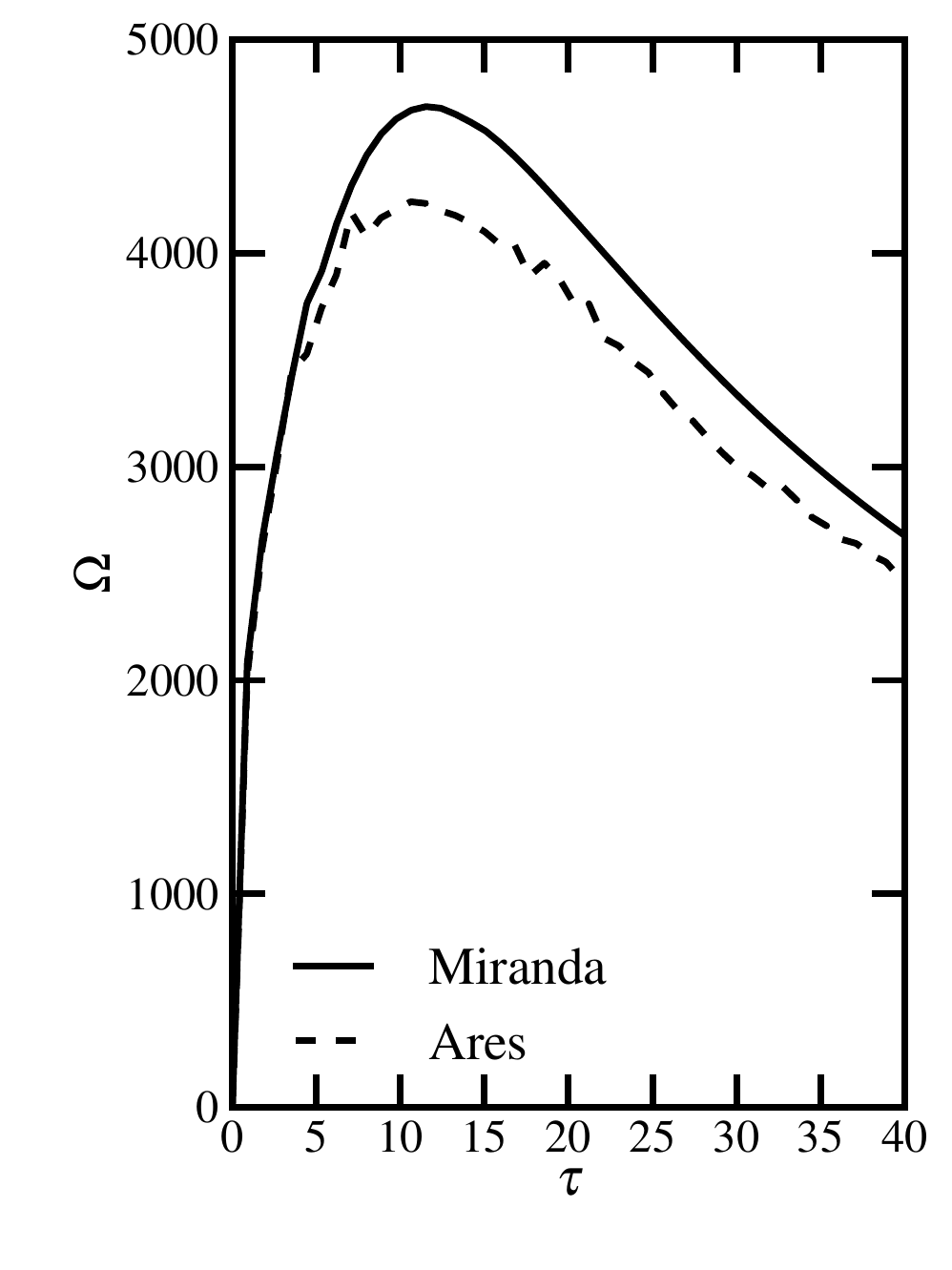}{Mesh D for Miranda \& Ares}
{Time history of the domain integrated enstrophy ($\Omega$, eq.~\ref{eq:enst}) for meshes A-D at the reduced Reynolds number from Miranda (a) and Ares (b).  Data between codes at the finest resolution are plotted in (c).}
{fig:Enstdns}{.5in}

\ThreeFig
{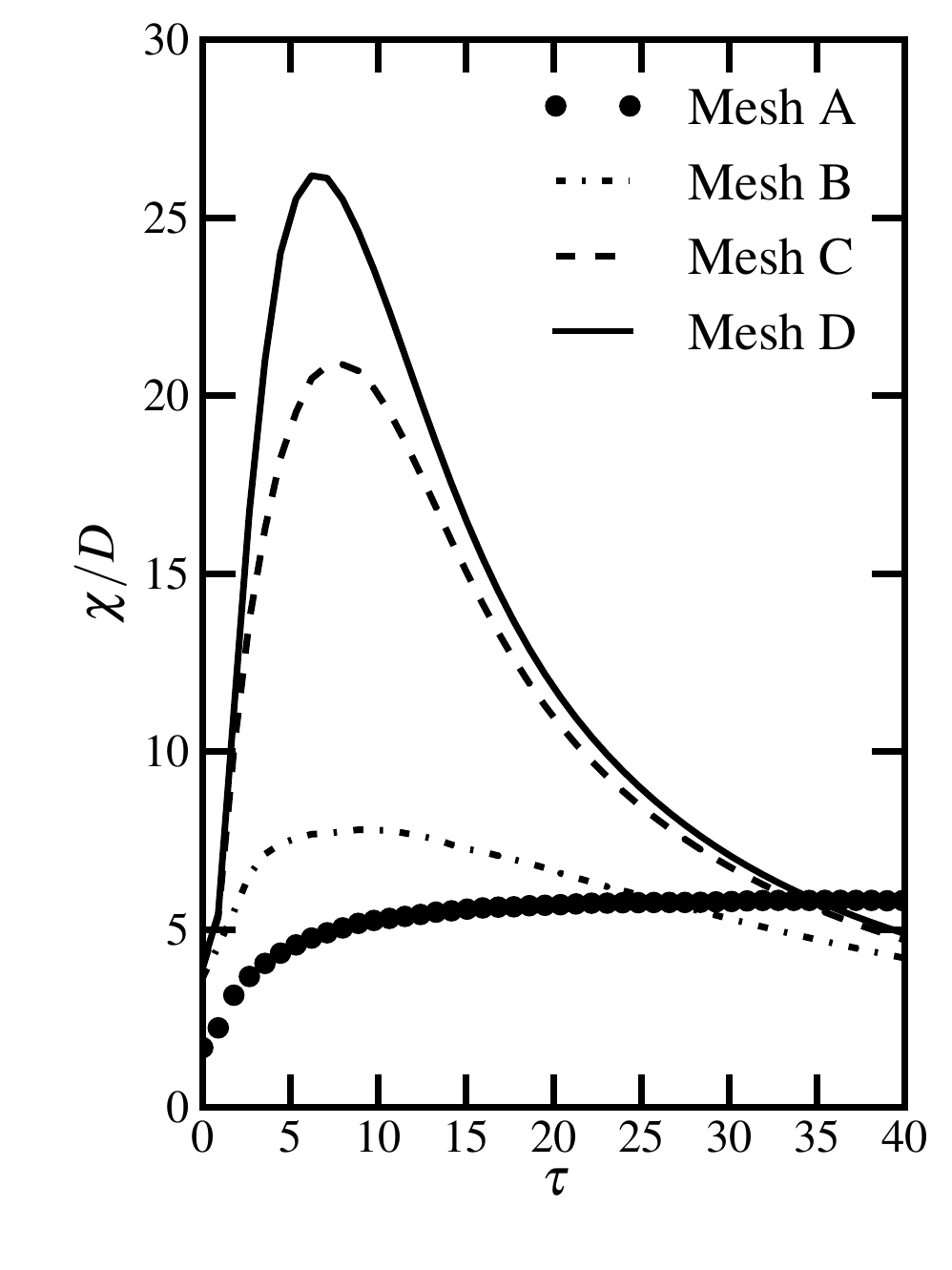}{Miranda}
{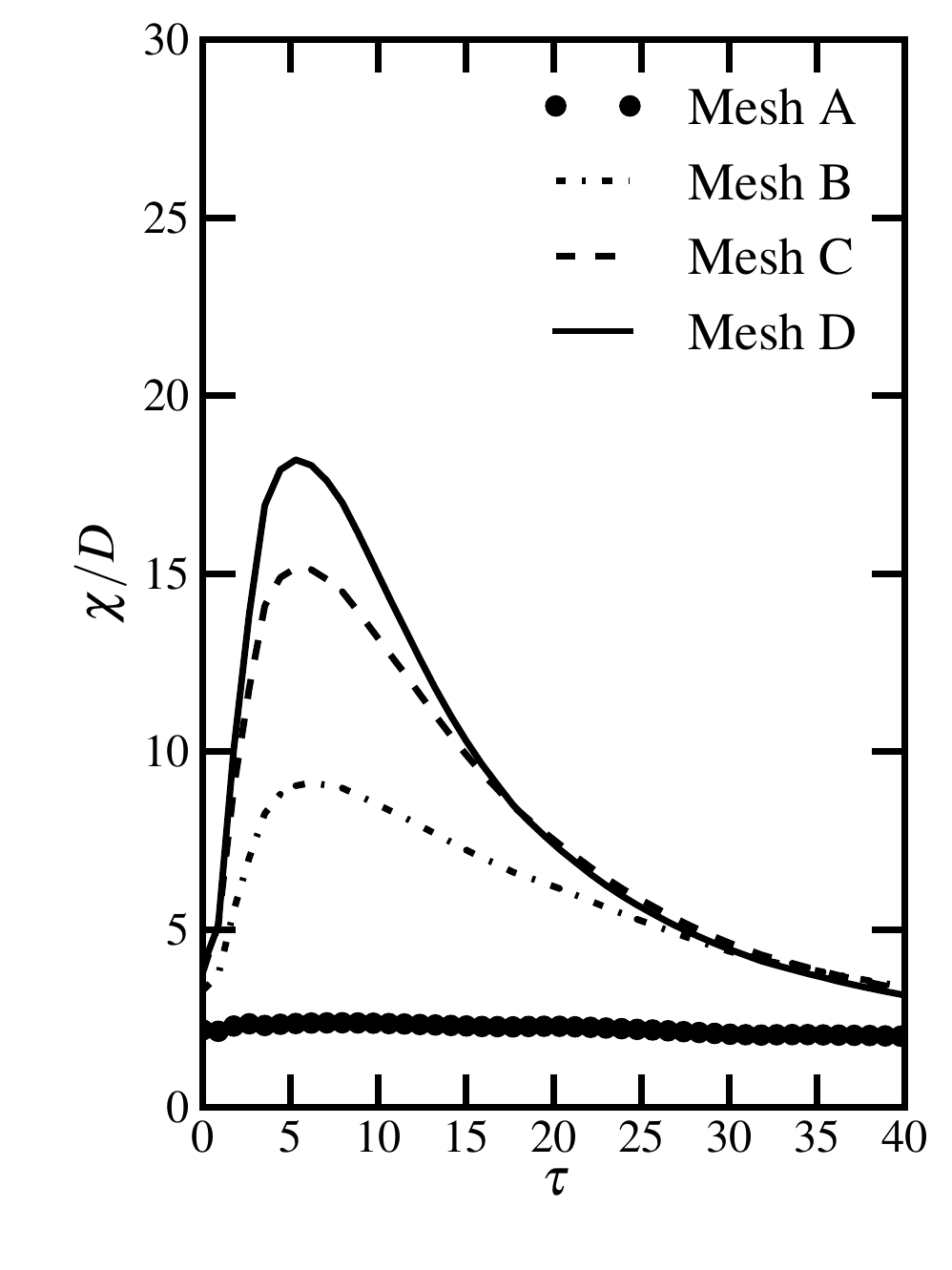}{Ares}
{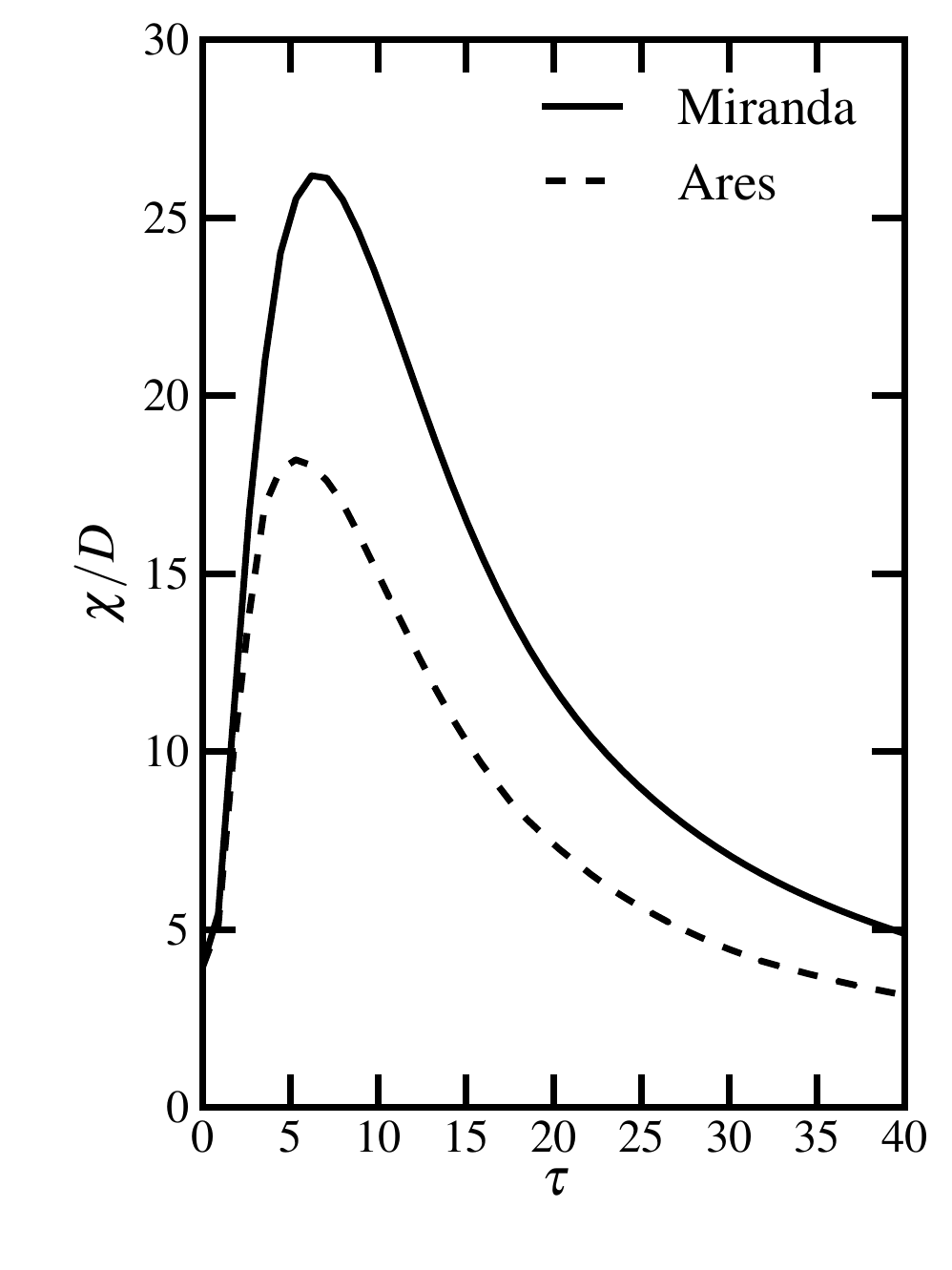}{Mesh D for Miranda \& Ares}
{
Time history of the domain integrated scalar dissipation rate ($\chi/D_{SF_6}$, eq.~\ref{eq:tmr}) for meshes A-D at the reduced Reynolds number from Miranda (a) and Ares (b).  Data between codes at the finest resolution are plotted in (c).
}
{fig:TMRdns}{.55in}

As will be shown later, the behavior of the mean flow at this reduced Reynolds number is largely dependent on the Reynolds number.  With no inertial subrange, the smallest viscous scales will directly impact the large scales and alter the energy containing scales.  As the Reynolds number gets sufficiently large and the inertial range forms and broadens, this dependence will gradually subside.  Indeed, Grinstein~\etal~\cite{grinstein:2011} have suggested that DNS at low Reynolds number can resemble poorly resolved LES calculations at infinite Reynolds numbers, loosely linking the notion of grid dependence with Reynolds number dependence.  This will be explored further in Section~\ref{sec:MUeff}.

\clearpage
	
\subsection{High Reynolds number LES}
\label{sec:LES}
The second set of calculations were conducted at the Reynolds number given in Table~\ref{tab:props}, which is close to the experimental conditions of previous studies~\cite{jacobs:2005,jacobs:1996, vetter:1995} of RMI.  The required number of grid points needed for a DNS at this high Reynolds number is approximately $\sim 4\times 10^{12}$, which exceeds the capability of today's computational resources.  Simulations using the grids of Table~\ref{tab:DNS} are therefore under resolved with respect to the viscous length scales.  Therefore, the actual diffusion length scales of the simulation will be dependent on the dissipation from the numerics and the model.  Both of which should vanish under grid refinement but will depend heavily on the numerical method and grid spacing.

The large energy containing scales will become increasingly independent of the fine scales associated with the grid as the inertial subrange between the two broadens.  
It is this scale separation and independence of the solution on the fine scale which is probed in a requisite grid convergence study (Figure~\ref{fig:3Dscales}) of an LES calculation. 
Therefore, the energy containing portions of the flow field and global/integral observables such as the mixing width and mixedness will exhibit converging behavior.  However, metrics which are biased to the small scales such as the scalar dissipation rate and enstrophy diverge under grid refinement and show stronger dependencies on the numerical dissipation.  

To explore this grid convergence at high Reynolds numbers, a grid resolution study was conducted for both numerical methods on meshes given in table~\ref{tab:DNS}.  As in the DNS study, the temporal mixing widths and mixedness are plotted for both codes and all resolutions in Figure~\ref{fig:Wles} and Figure~\ref{fig:Mixles}.  Convergence is less pronounced (as compared to the DNS convergence study) and curves diverge with time.  However, for early time ($\tau < 25$) the solutions are nearly indistinguishable at the fine mesh resolution.

\ThreeFig
{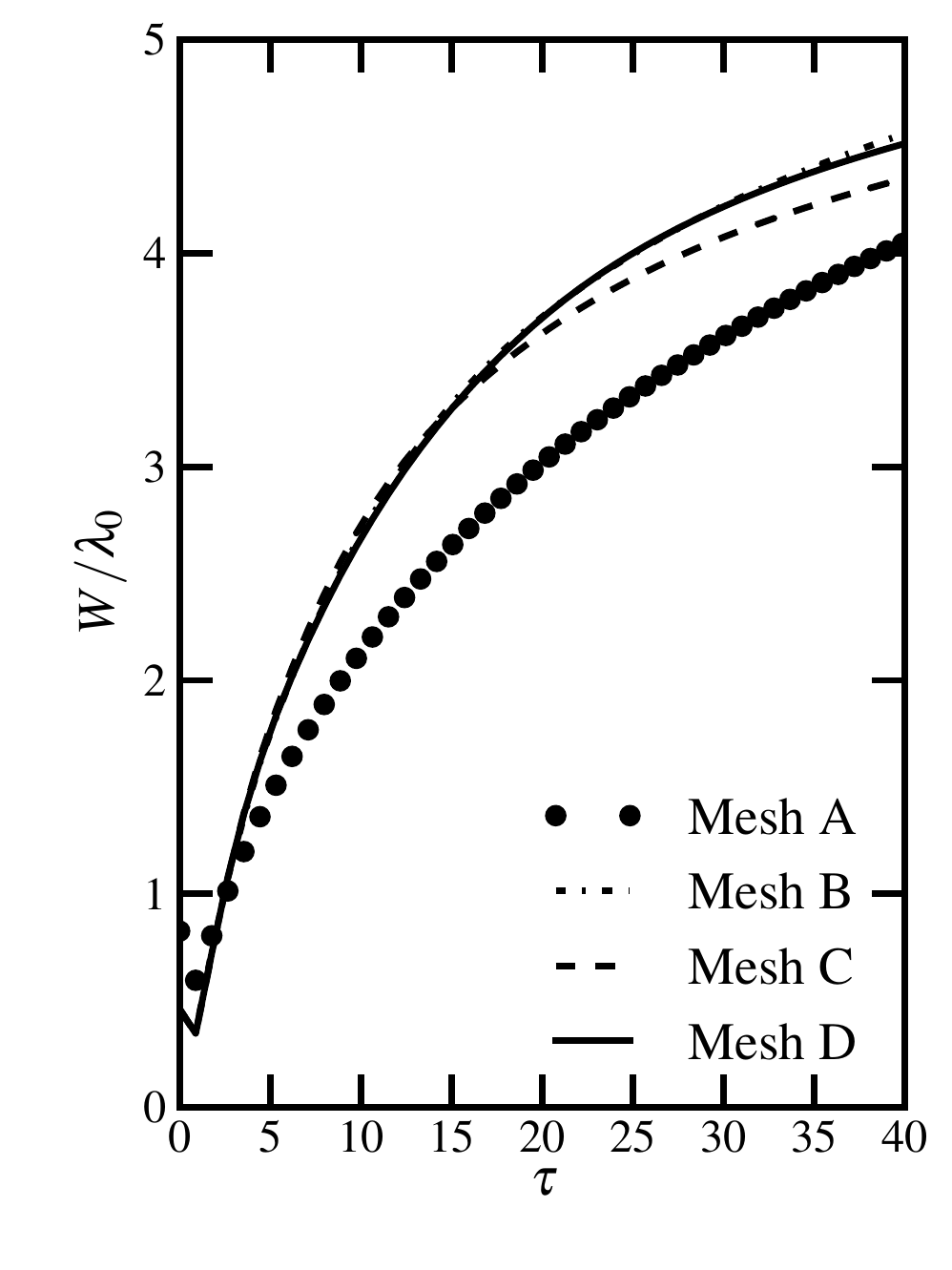}{Miranda}
{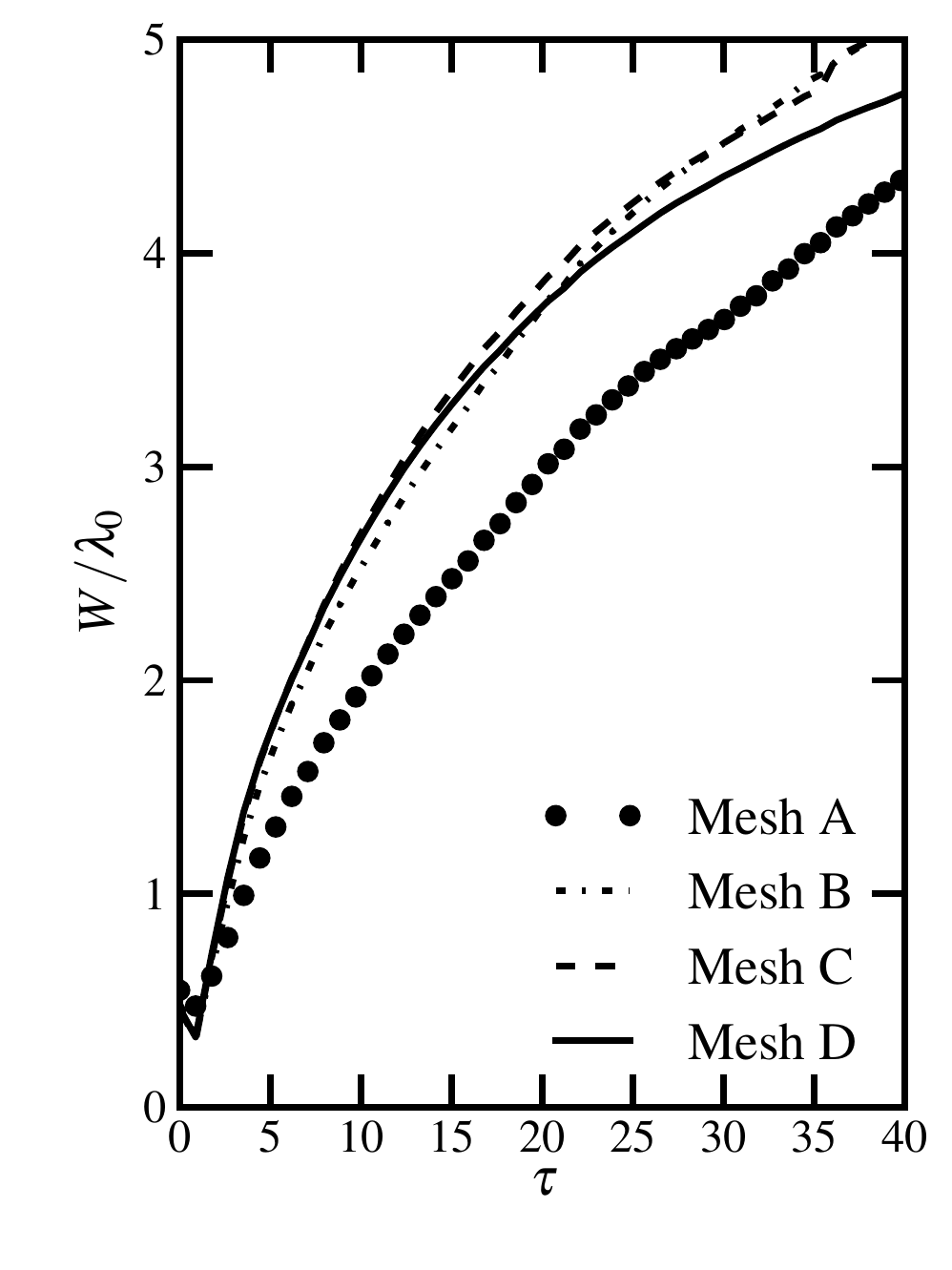}{Ares}
{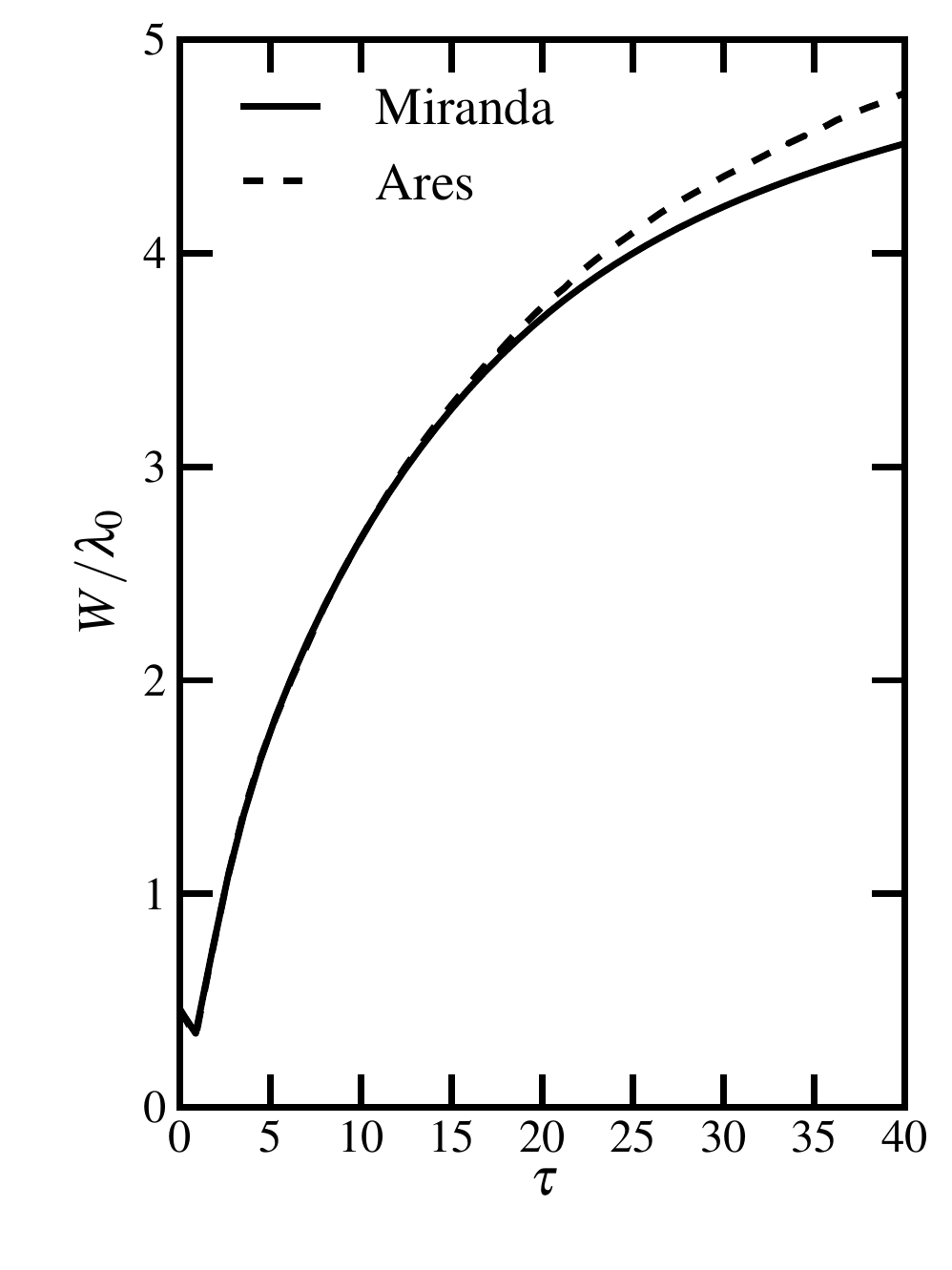}{Mesh D for Miranda \& Ares}
{Non-dimensional mixing width vs. time for meshes A-D at the nominal Reynolds number from Miranda (a) and Ares (b).  Data between codes at the finest resolution are plotted in (c) and show that agreement worsens with time and is at most 5\% different at $\tau=40$.}
{fig:Wles}{.55in}

\ThreeFig
{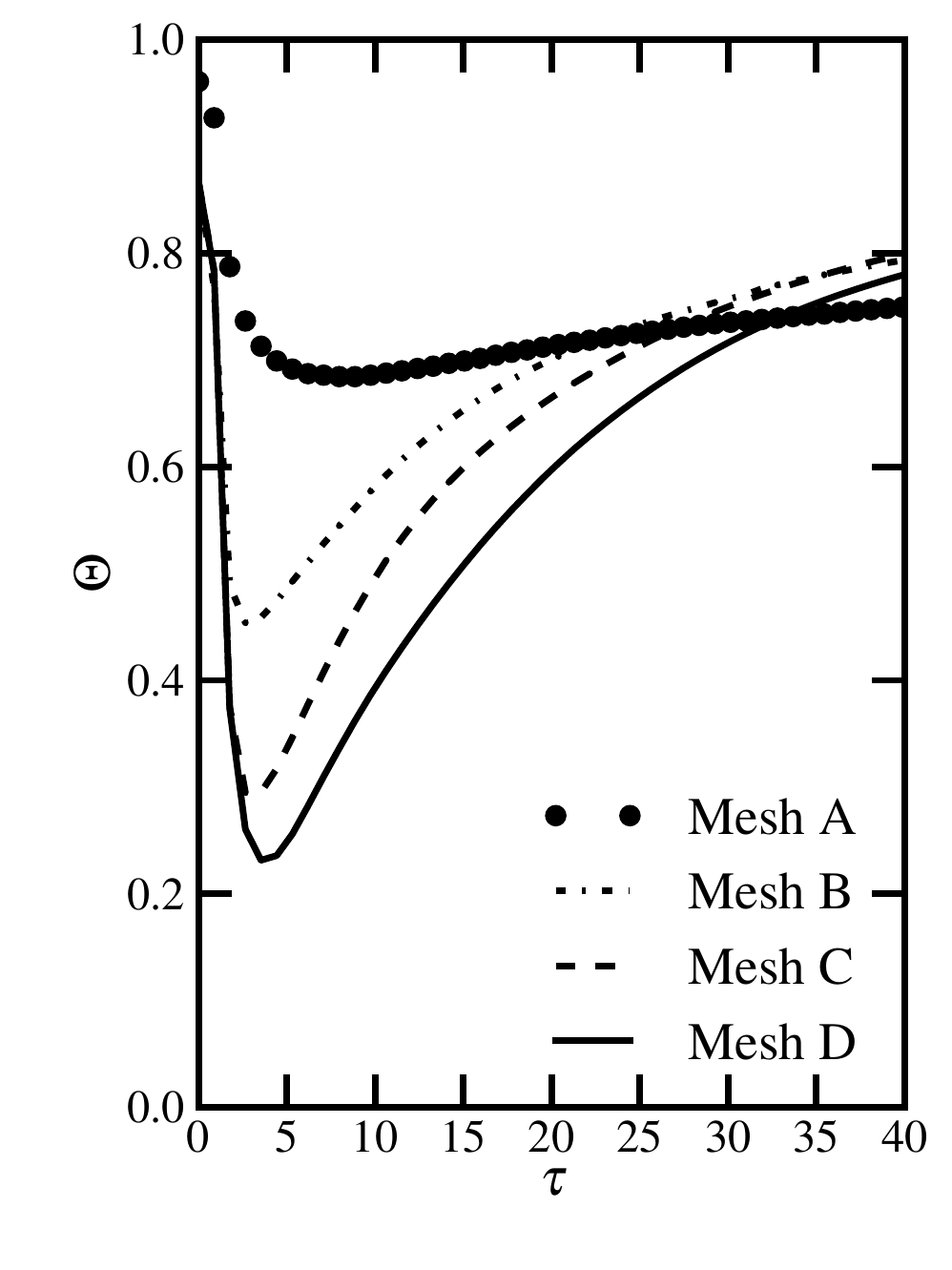}{Miranda}
{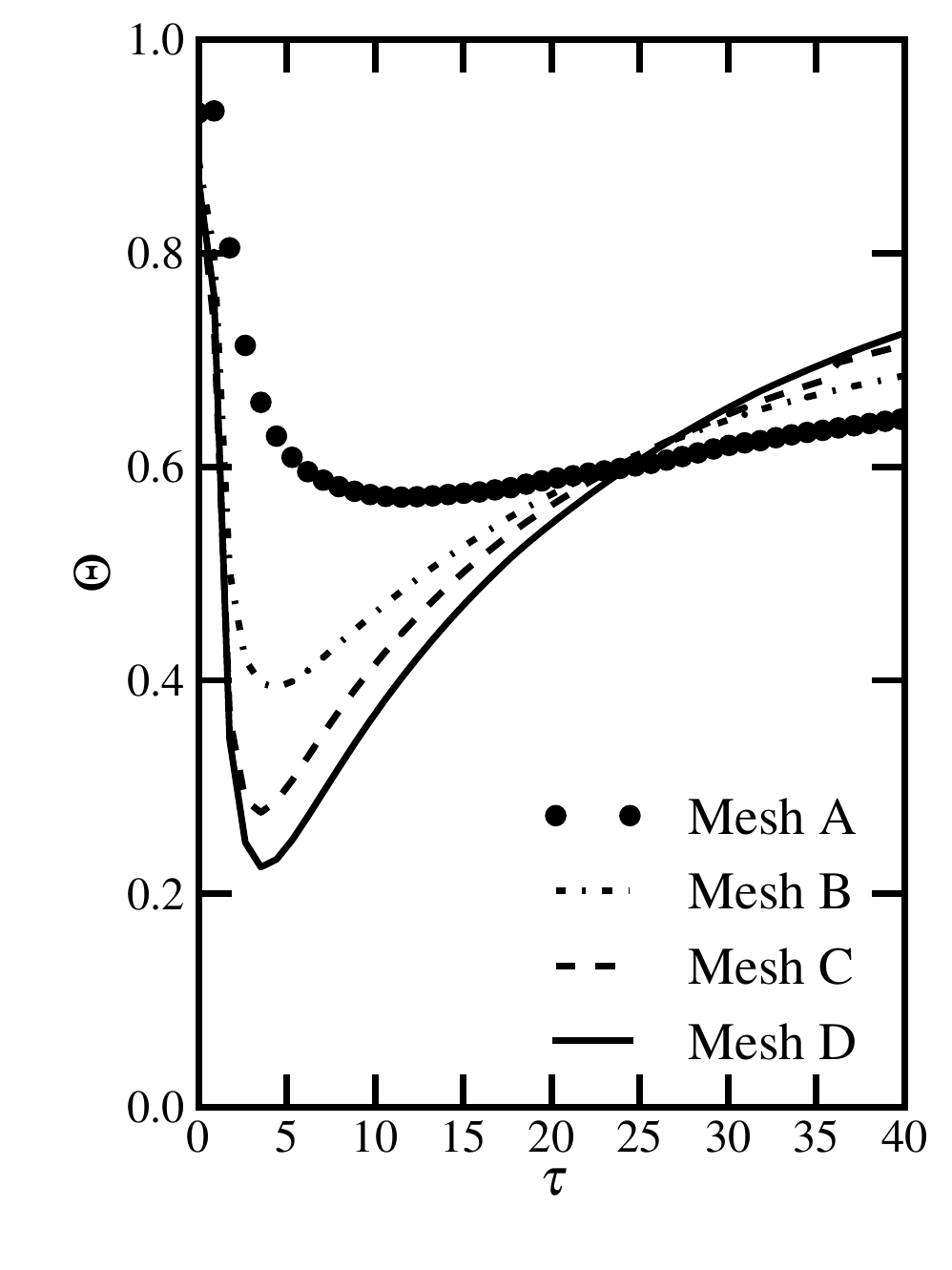}{Ares}
{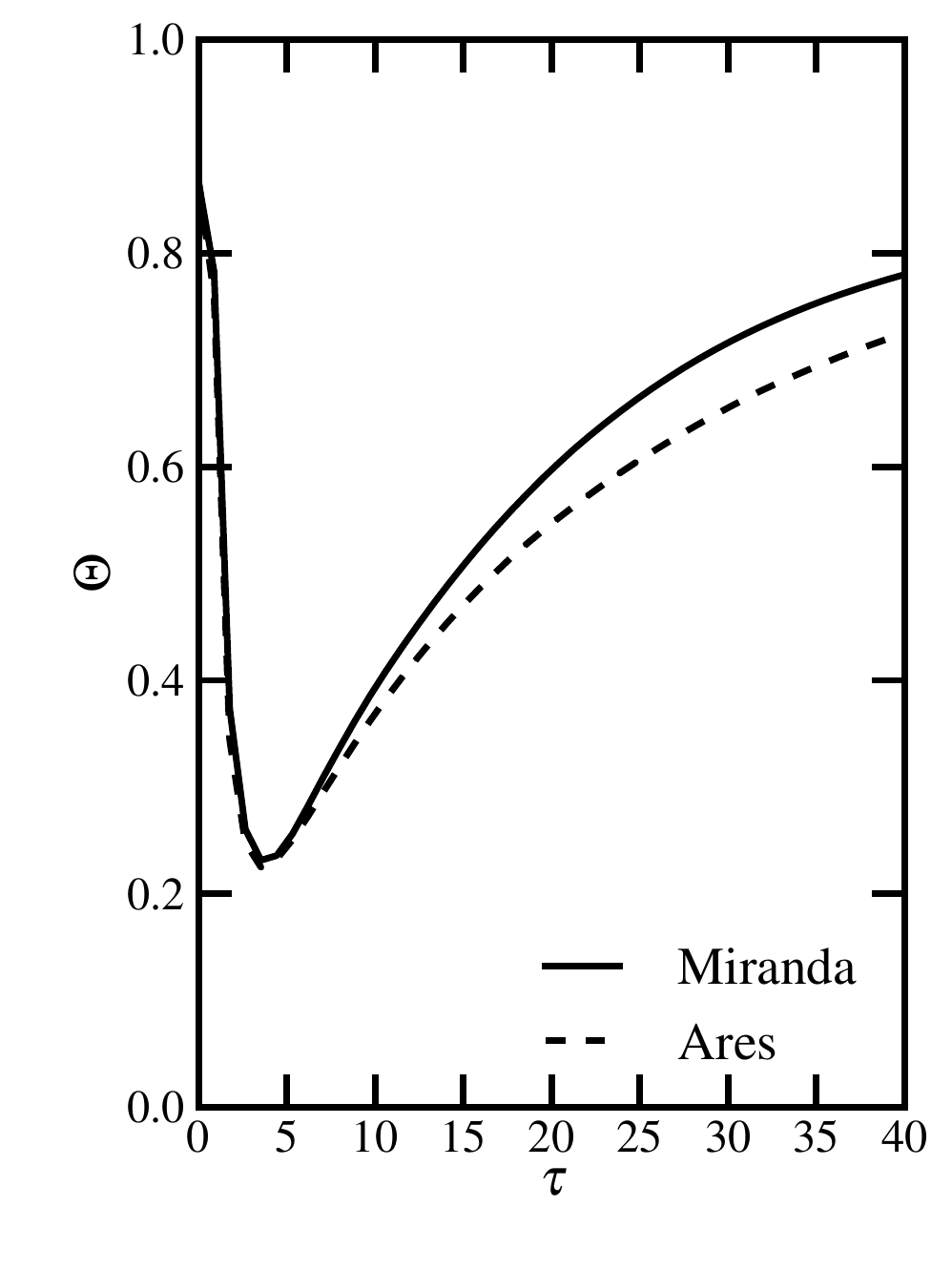}{Mesh D for Miranda \& Ares}
{Mixedness ($\Theta$) vs. time for meshes A-D at the nominal Reynolds number from Miranda (a) and Ares (b).  Data between codes at the finest resolution are plotted in (c) and show that the differences grow with time and are approximately 5\% at $\tau=40$.}
{fig:Mixles}{.52in}

The range of resolved scales can be readily examined by looking at the spectra of velocity and mass fraction fluctuations.  Figure~\ref{fig:UspecLES} shows the power spectra of velocity as a function of the two-dimensional wave number, $k$, computed as was described in the previous section.  The power spectra between codes at the fine resolution are in good agreement for wave number less than 30, after which, they diverge.  The inertial range following the $\sim k^{-5/3}$ spans over a wider range of wave numbers in Miranda than in Ares by approximately a factor of two on mesh D.  At the coarsest calculation (mesh A) the spectra for both Ares and Miranda do not exhibit inertial ranges.  

The mass fraction spectra for the two codes (Figures~\ref{fig:YspecLES}a and~\ref{fig:YspecLES}b) show converged behavior up through wave number 80.  Furthermore, on mesh D, solutions from Miranda and Ares (Figure~\ref{fig:YspecLES}), have equally wide inertial ranges and agree quite well for all wave numbers plotted.

\ThreeFig
{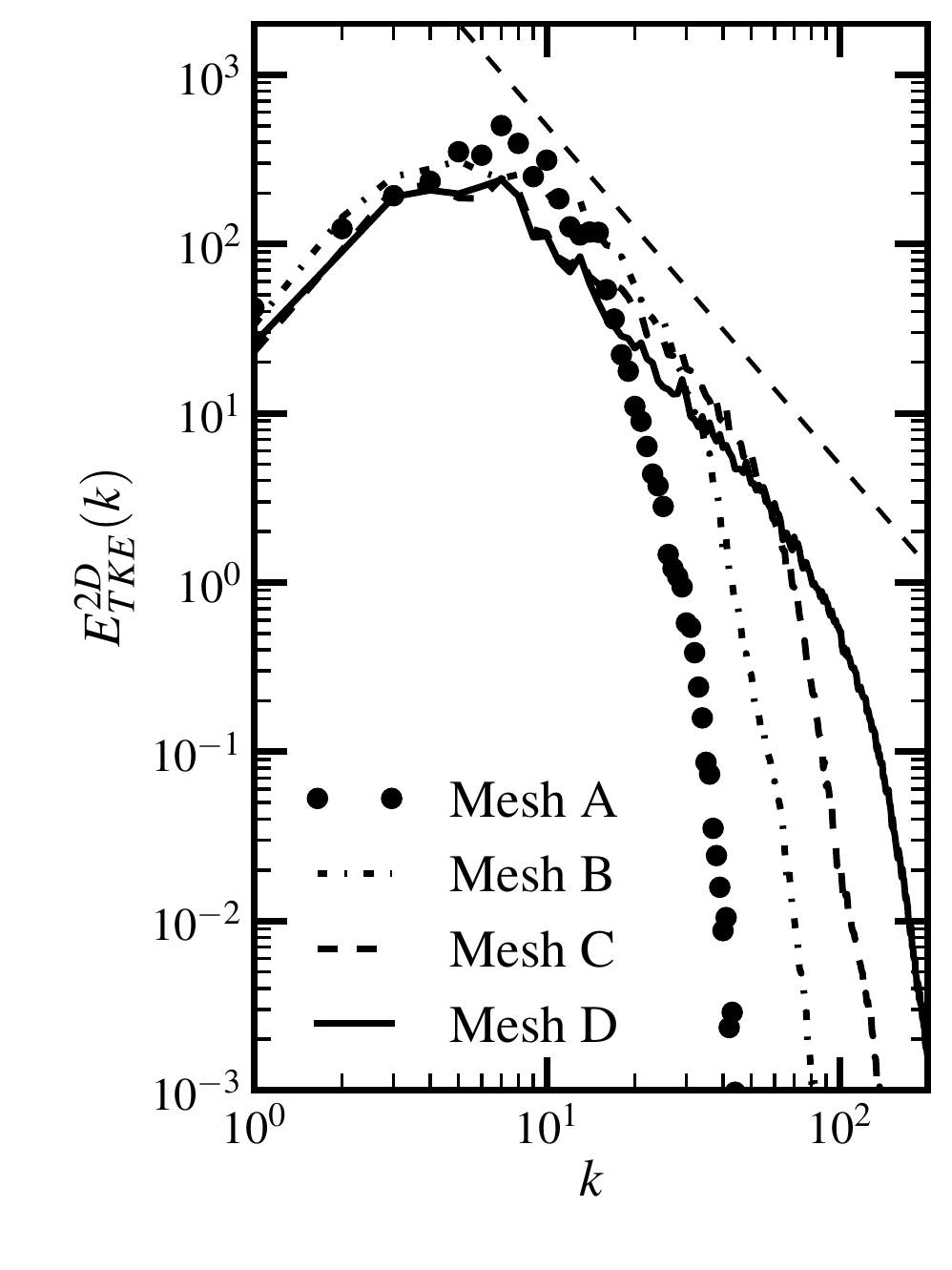}{Miranda}
{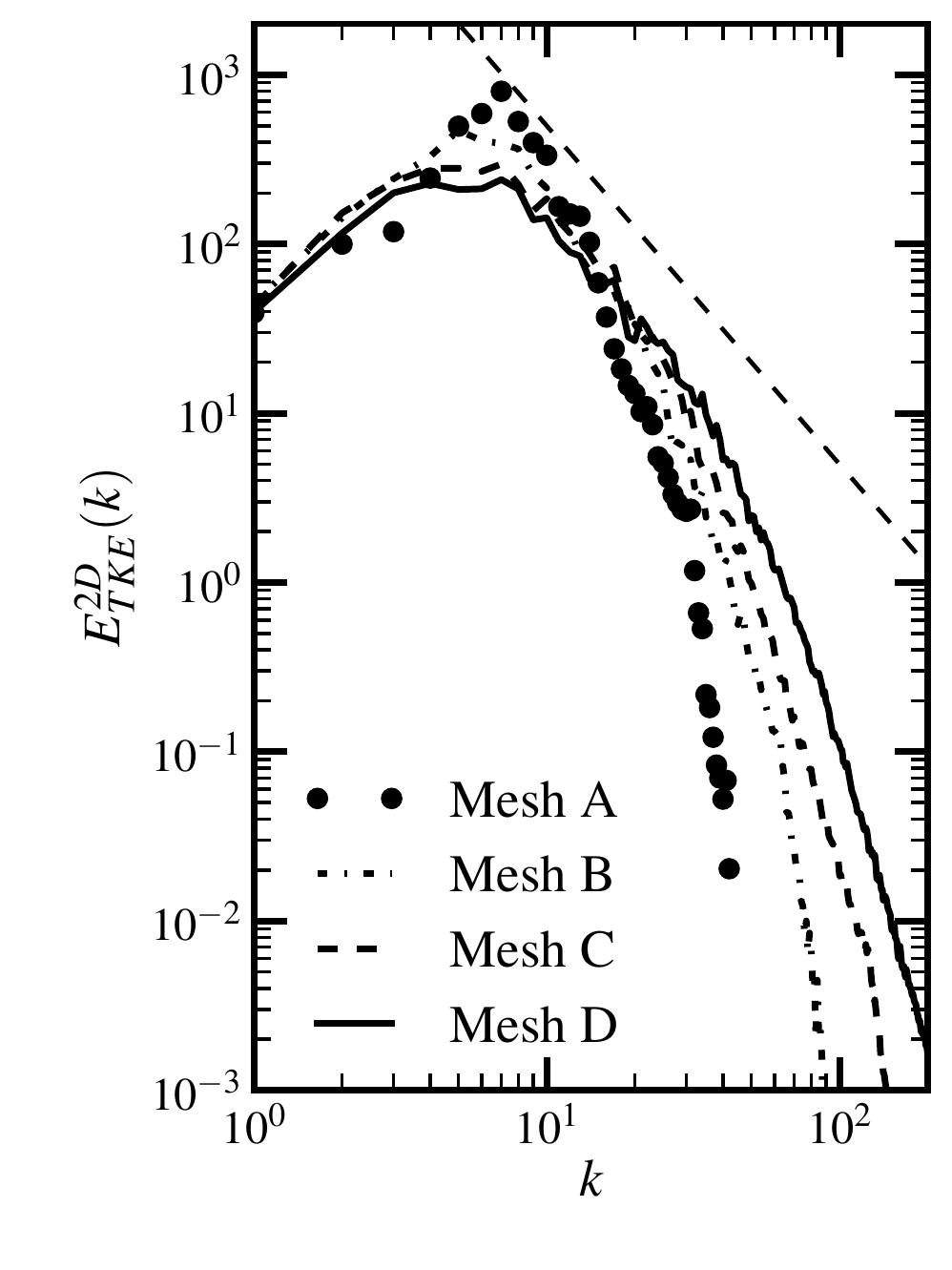}{Ares}
{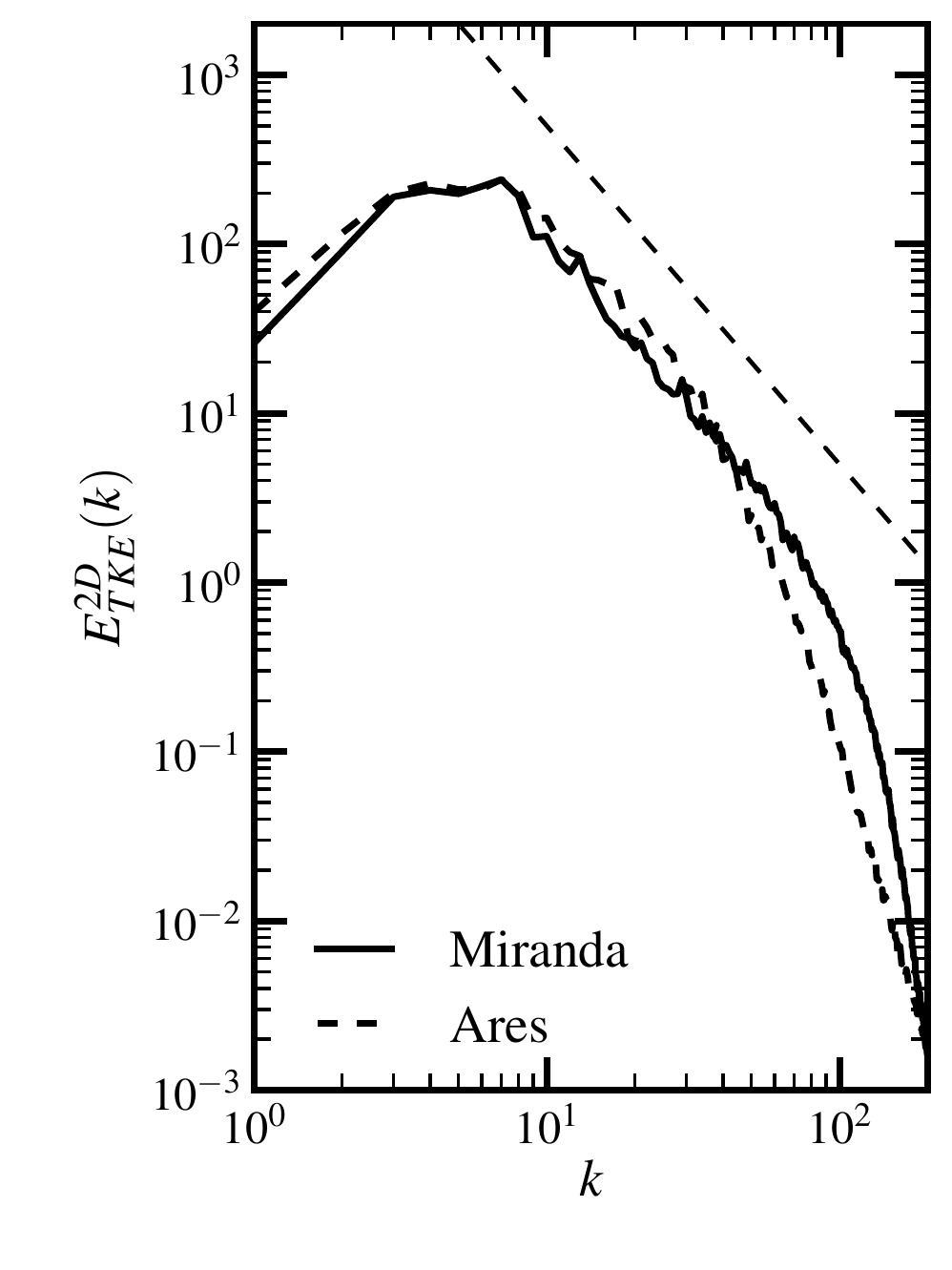}{Mesh D for Miranda \& Ares}
{Power spectra of the fluctuating velocity at $\tau=35$ for Miranda (a) and Ares (b) for meshes A-D at the nominal Reynolds number.  The difference of the spectra for mesh D between the codes (c) number 40.  The $k^{-5/3}$ fiducial is plotted (dashed) and shows a broader inertial subrange as compared to the DNS spectra.}
{fig:UspecLES}{.65in}
	
\ThreeFig
{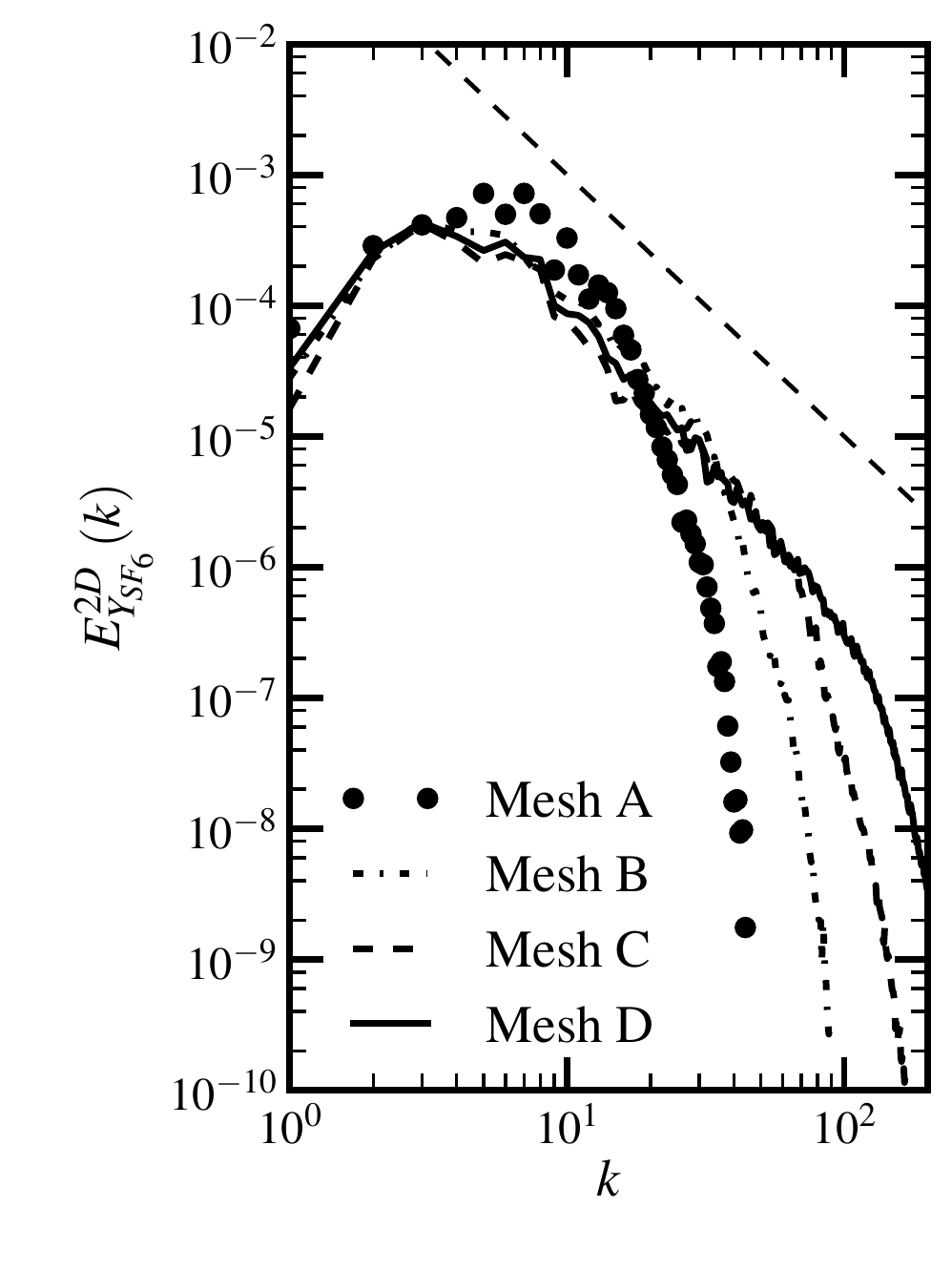}{Miranda}
{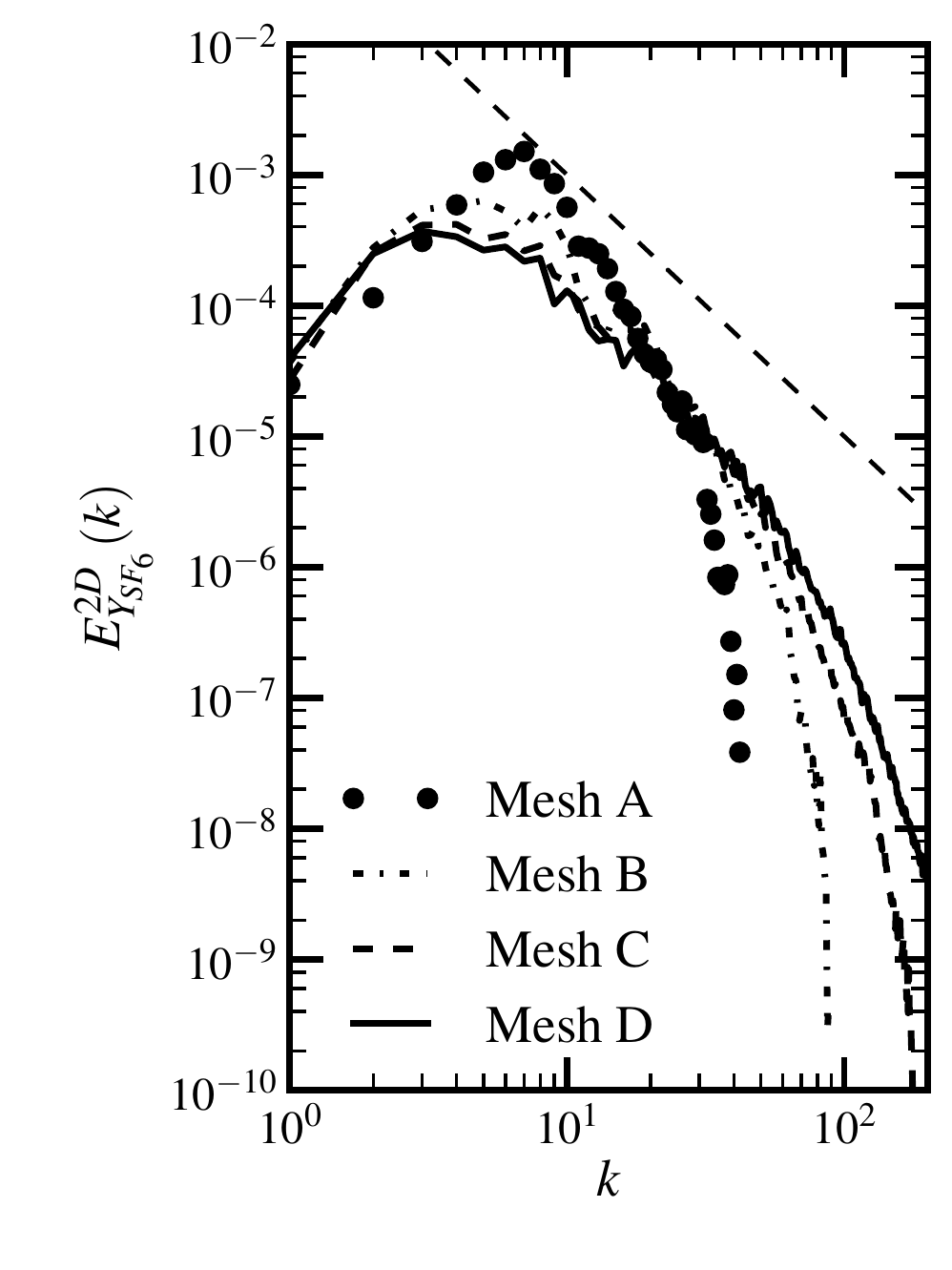}{Ares}
{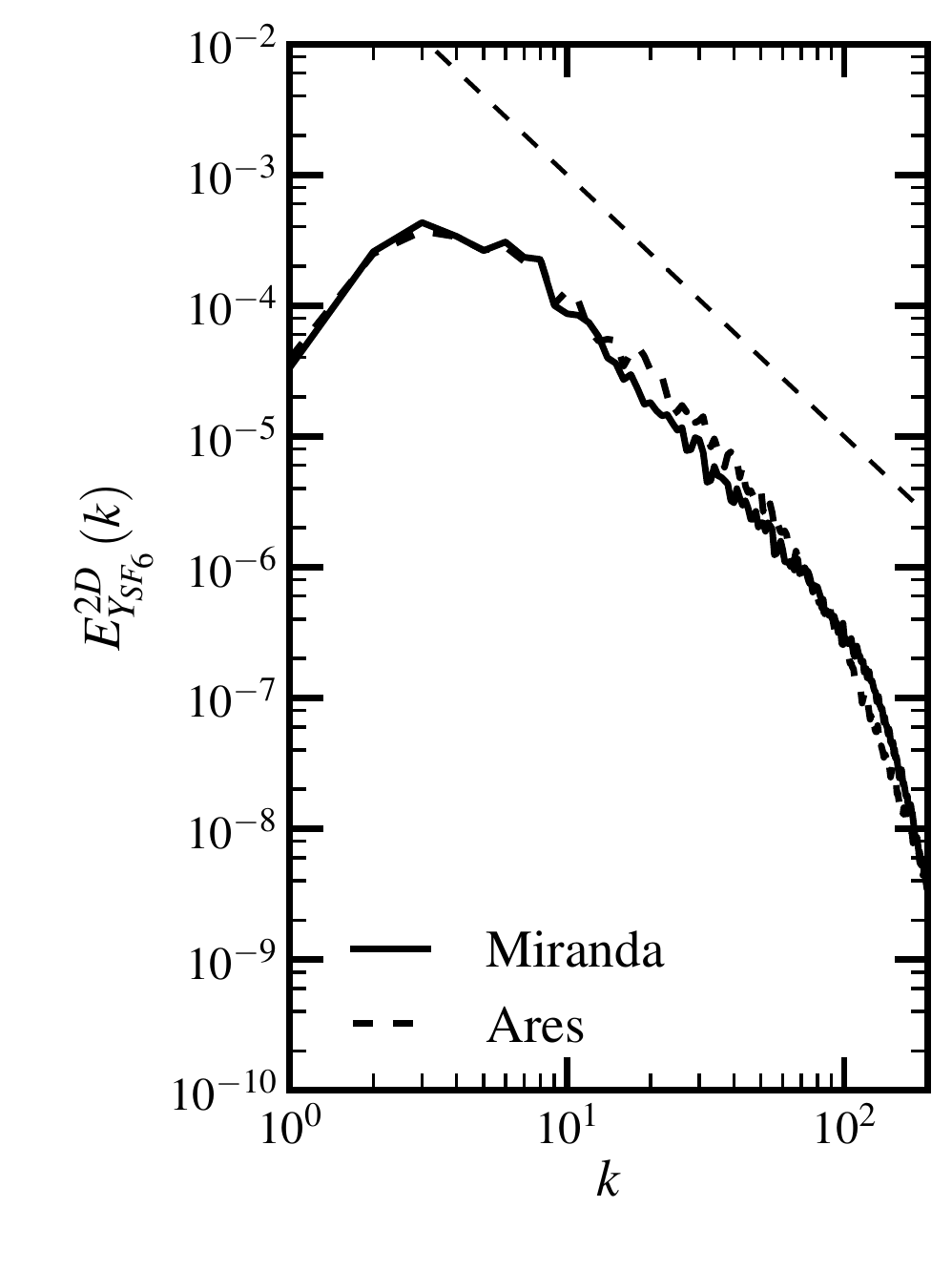}{Mesh D for Miranda \& Ares}
{
Power spectra of the mass fraction at $\tau=35$ for Miranda (a) and Ares (b) for meshes A-D at the nominal Reynolds number.  The difference of the spectra for mesh D between the codes (c) is small over the entire range of plotted wave numbers.  The $k^{-5/3}$ fiducial is plotted (dashed) and shows a that the LES maintains an inertial range before the numerical dissipation effects begin to dominate.
}
{fig:YspecLES}{.65in}

Quantitative measures of dissipation exhibit the largest differences in the under resolved LES calculations.  Since these measures are biased towards gradients of the finest scales (where numerical dissipation is most active), grid and scheme dependence will be most apparent.  The time histories of enstrophy and normalized scalar dissipation are plotted in Figure~\ref{fig:EnstLES} and~\ref{fig:TMRLES}, respectively.  The local maxima of the curves increase in value as the grid is refined.  Values of enstrophy from the mesh C resolution in Miranda are close to those in Ares from mesh D, suggesting that Miranda is capturing finer length scales by roughly a factor of two.  For the scalar dissipation in Figure~\ref{fig:TMRLES}, the disparity is not as large and Ares mesh D data lie somewhere between Miranda mesh C and D data.
	 
\ThreeFig
{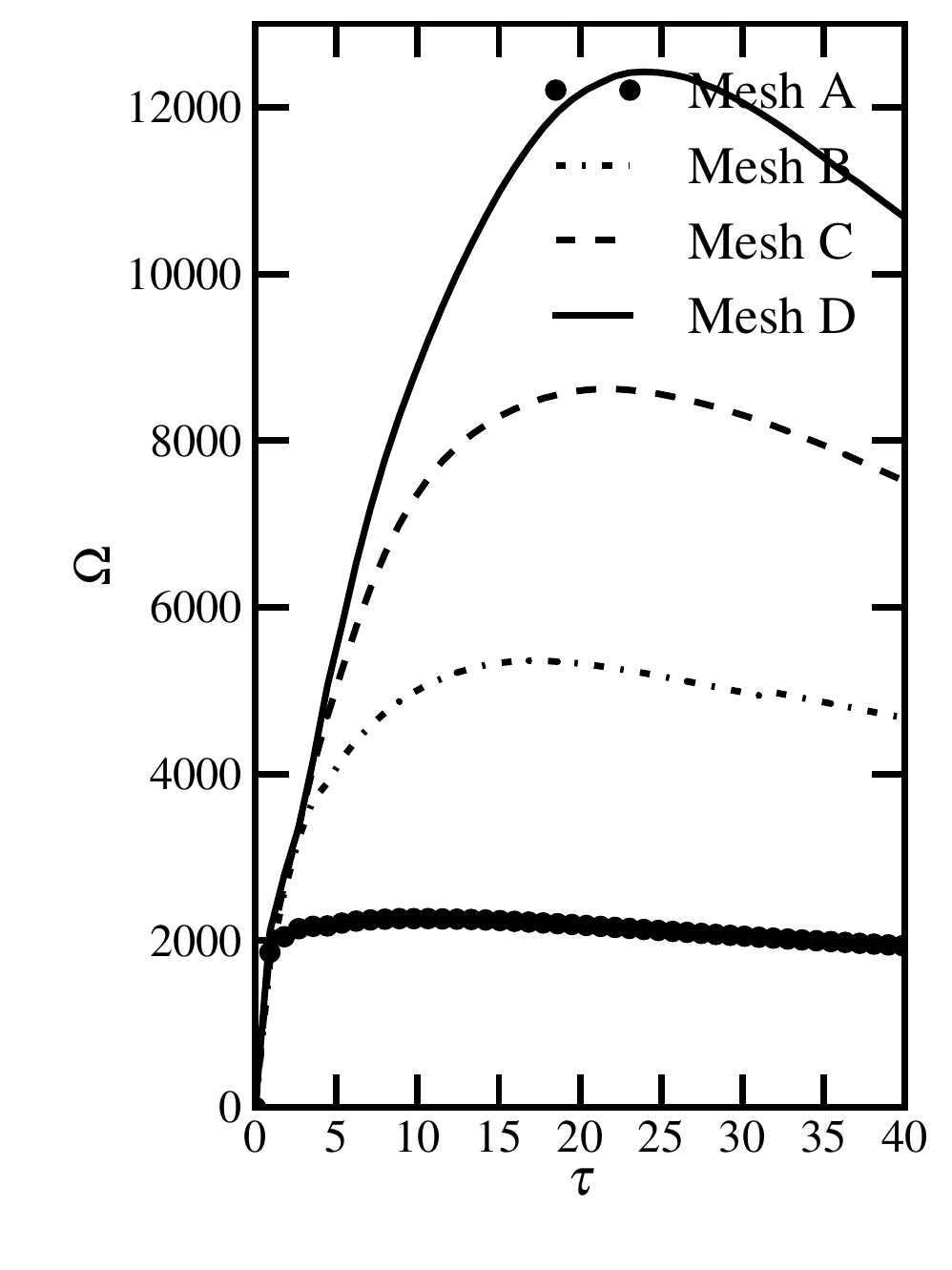}{Miranda}
{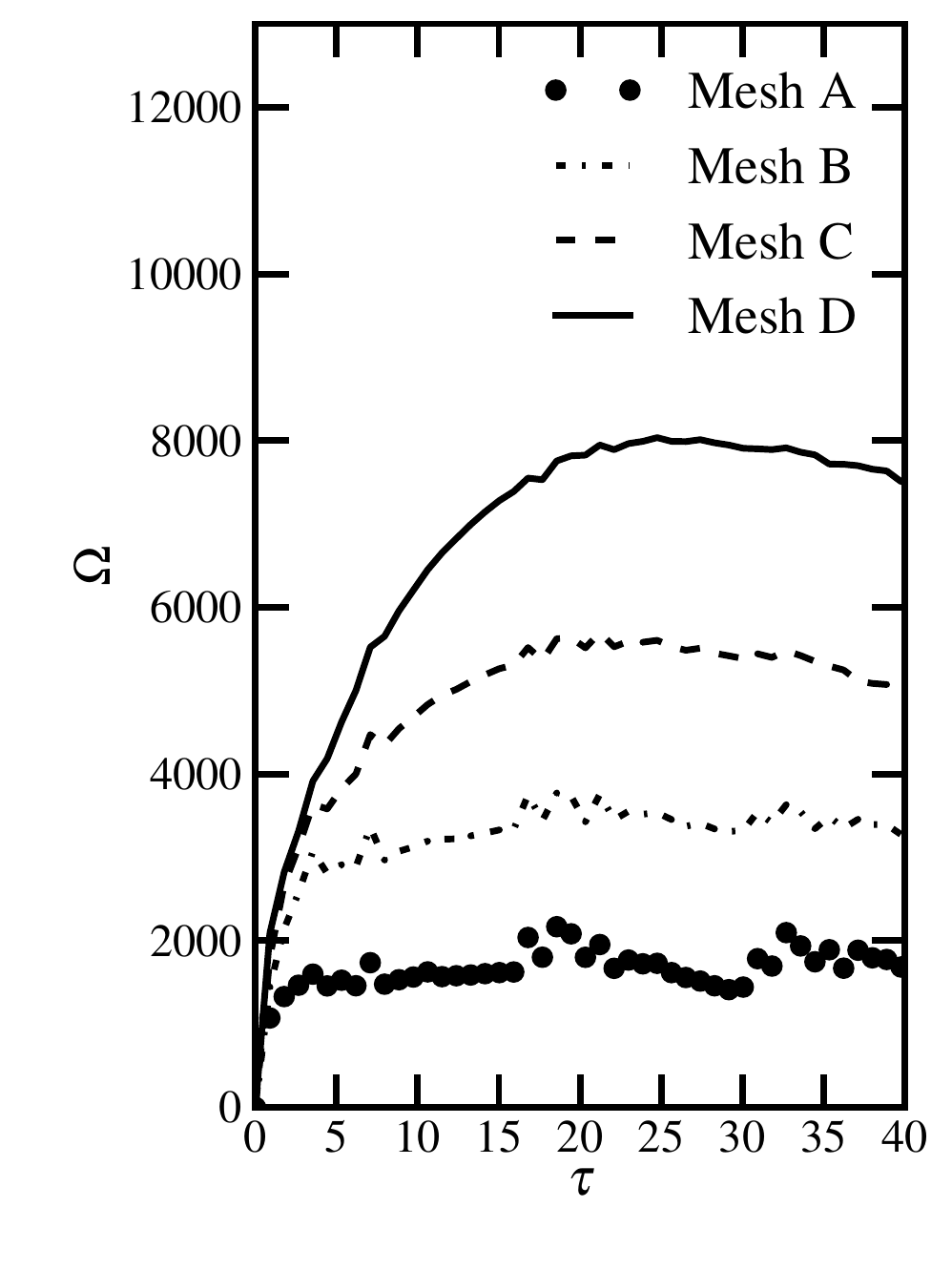}{Ares}
{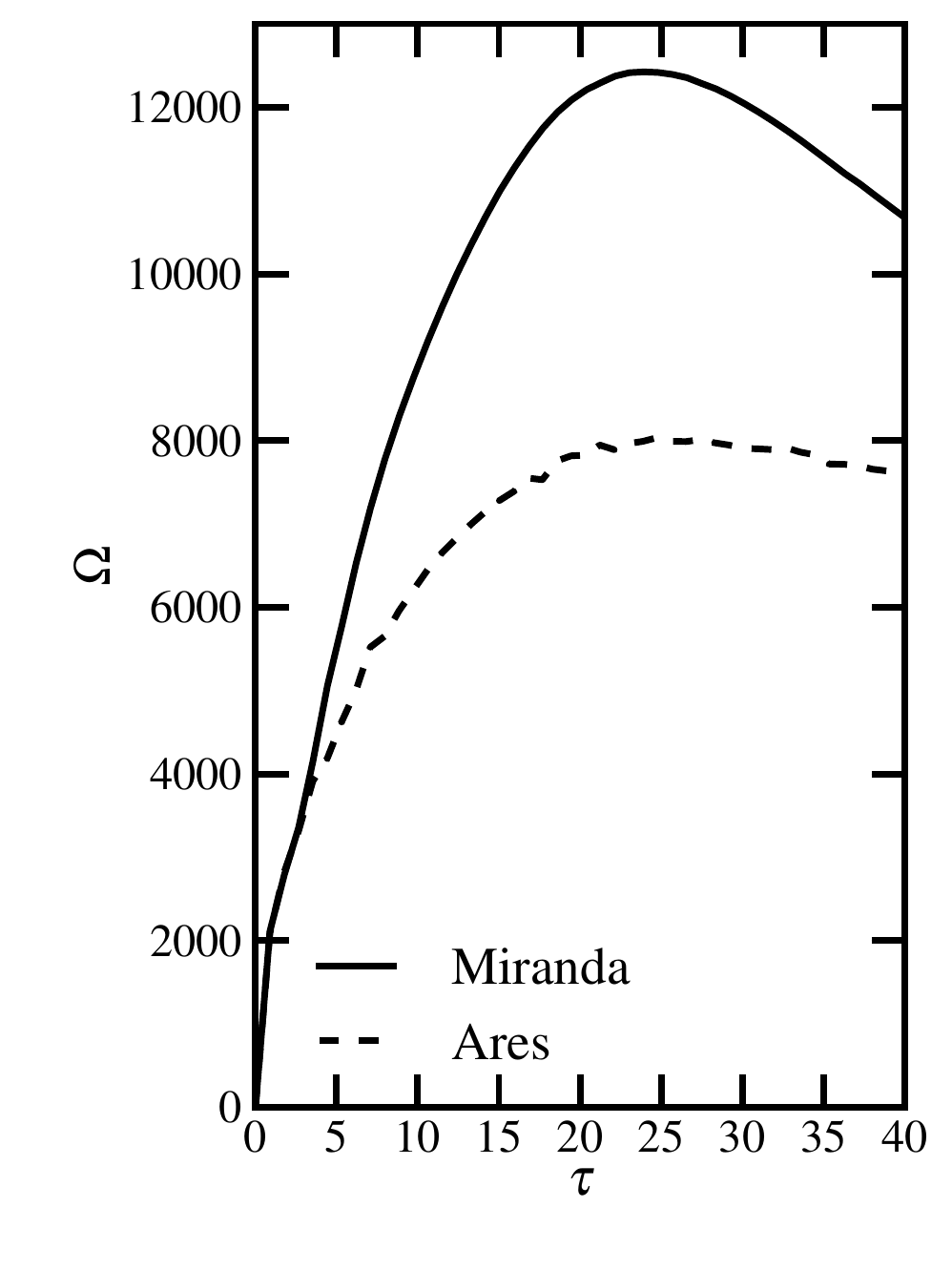}{Mesh D for Miranda \& Ares}
{Time history of the domain integrated enstrophy ($\Omega$, eq.~\ref{eq:enst}) for the nominal Reynolds number LES. The divergent behavior of the data in (a) and (b) suggest that the velocity length scales are proportional to grid spacing.  The comparison of Miranda and Ares (c) on mesh D show the peak enstrophy values of the Miranda calculation on mesh C are approximately equivalent to those of the Ares calculation on mesh D. }
{fig:EnstLES}{.55in}

\ThreeFig
{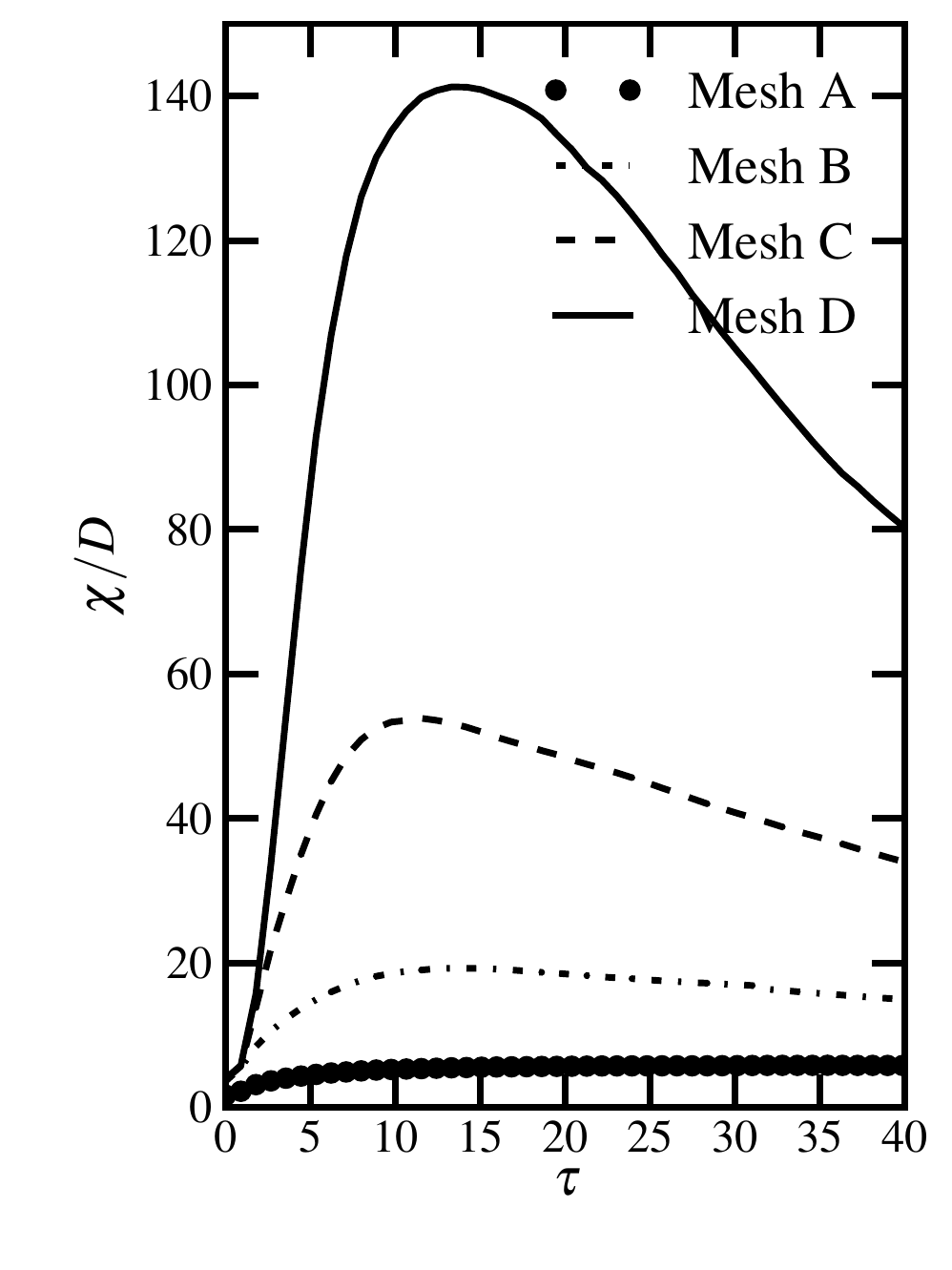}{Miranda}
{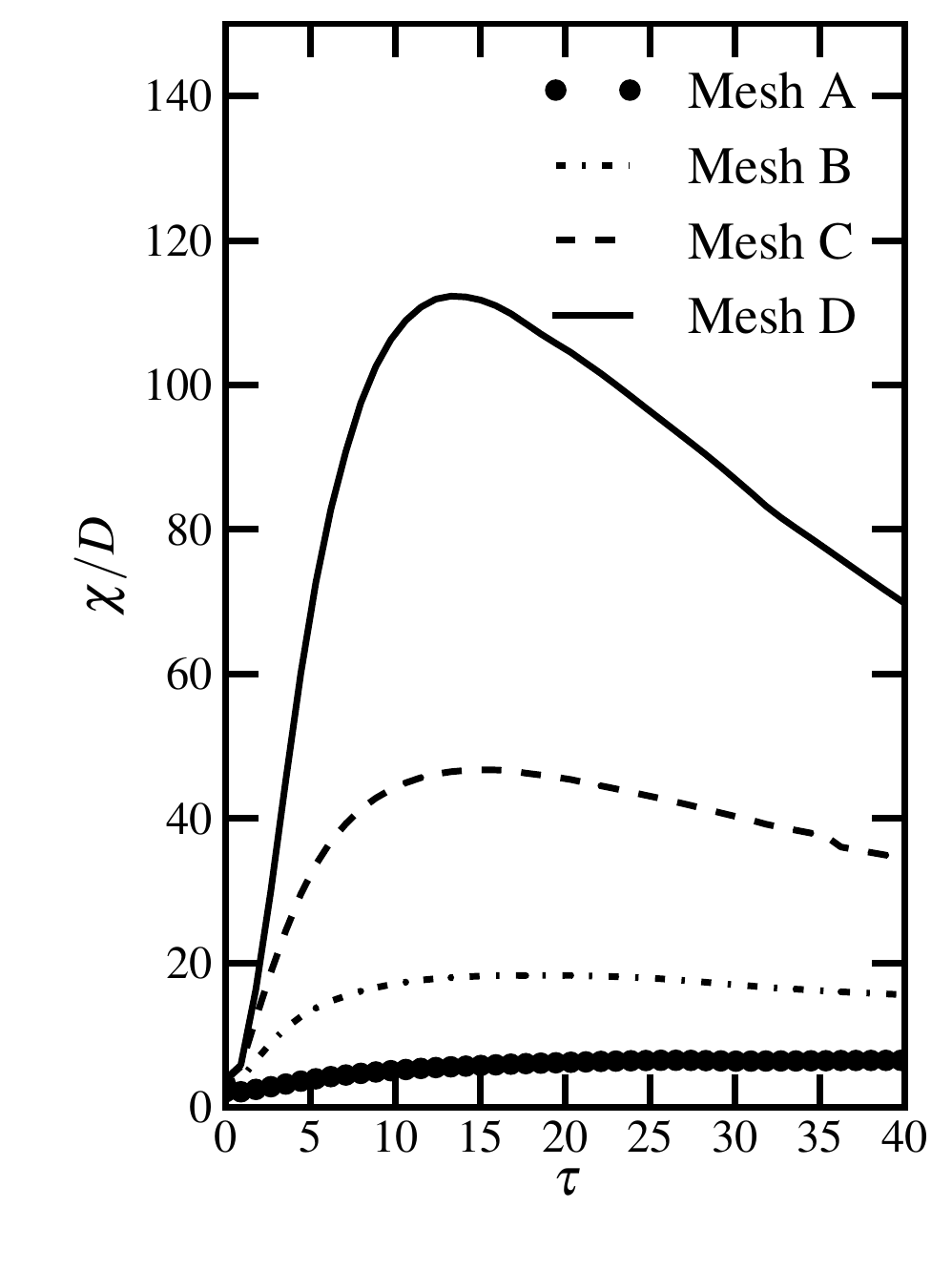}{Ares}
{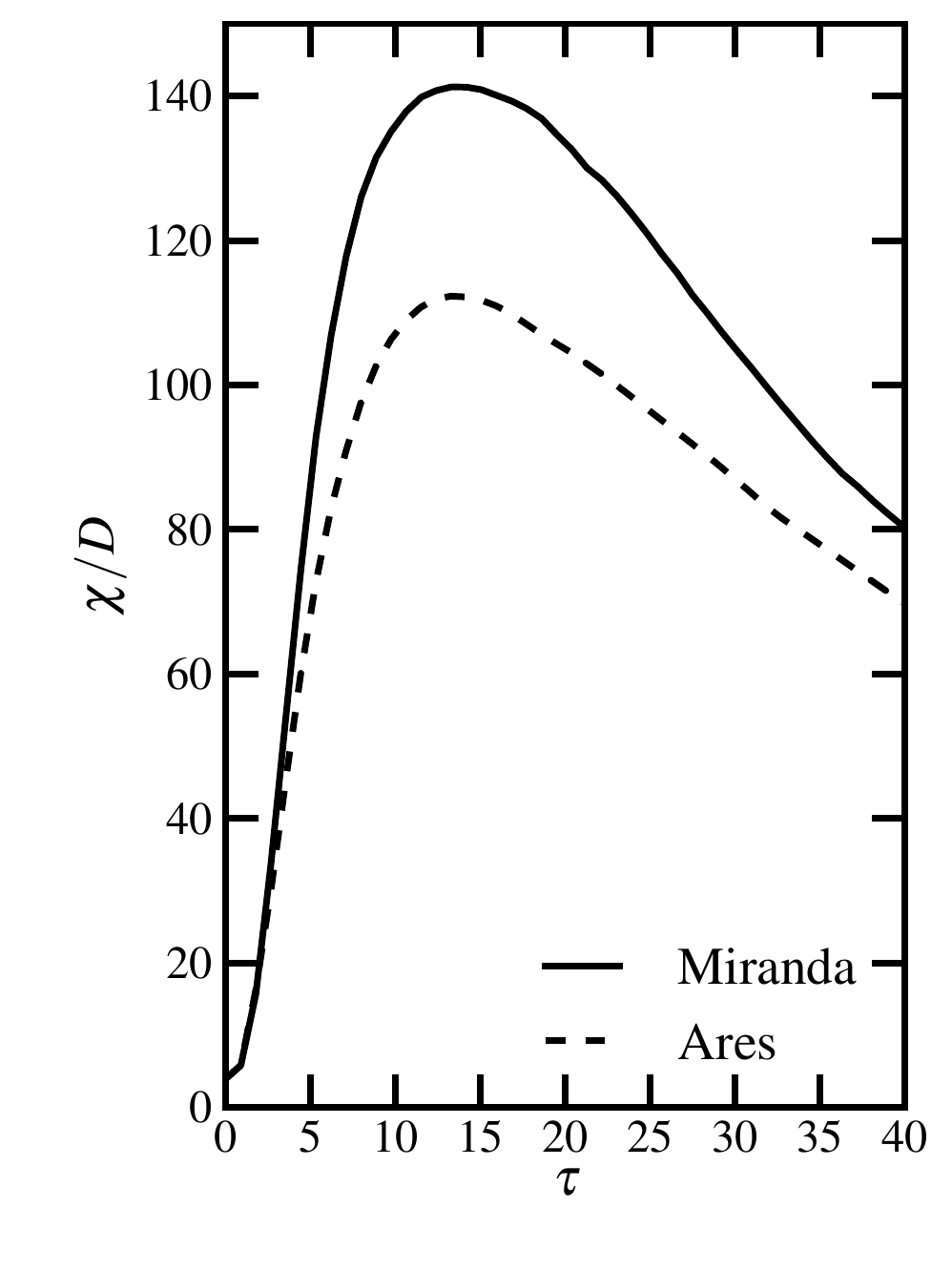}{Mesh D for Miranda \& Ares}
{
Time history of the domain integrated scalar dissipation rate ($\chi/D_{SF_6}$, eq.~\ref{eq:tmr}) for meshes A-D at the nominal Reynolds number from Miranda (a) and Ares (b).  Data between codes at the finest resolution are plotted in (c).
}
{fig:TMRLES}{.55in}

For the high Reynolds numbers calculations, the data clearly suggest that the flow is under resolved.  Although the mean flow field still exhibits dependence on the fine grid scales (Figure~\ref{fig:Wles} and~\ref{fig:Mixles}), the effect is decreasing under grid refinement.  Indeed, as the range of resolved scales grows larger with increased resolution, the effect of the new small scales on the large scales decreases.  This effect can be directly seen in the convergence of the power spectra (Figures~\ref{fig:UspecLES} and~\ref{fig:YspecLES}).  As higher wave number energy is introduced through grid refinement, the effect on the lower wave numbers decreases.  In the limit of infinite scale separation, the large scale solution will approach the Reynolds number independent solution.  Thus, there is a notional connection between grid convergence and Reynolds number independence.   

Conversely, there also exists a connection between grid dependence and Reynolds number dependence.  Grinstein \etal~\cite{grinstein:2011} showed (for the Taylor-Green vortex) comparisons of low Reynolds number calculations and under-resolved high Reynolds number calculations.  They found close correlations between the data, suggesting that poor numerical resolution has a similar effect as large amounts of physical viscosity on a well resolved grid.  Both mechanisms act like a viscosity, damping the fine scales and reducing the length of the inertial range.  In the following section, we seek a general way of comparing arbitrary simulation data which considers grid resolution, Reynolds number and numerical method through formulation of an effective viscosity.

\section{An effective viscosity for assessing the numerical dissipation in LES schemes}
\label{sec:MUeff}

The data presented in the previous sections demonstrate a dependence on Reynolds number, grid resolution and LES method.  The differences arise from the small length scales associated with dissipation.  In this section, an effective viscosity is proposed as an \emph{a posteriori} diagnostic to determine an effective Reynolds number and an effective Kolmogorov length scale of the flow for a given grid size, numerical method and physical Reynolds number.  An analogous effective diffusivity is also proposed, which suggests an effective Batchelor scale and an effective Schmidt number.
 
Given the strong grid dependence in the high wave numbers on the spectra and on profiles of the gradient based quantities, the previous LES in Section~\ref{sec:LES} were poorly resolved with respect to the viscous and diffusion length scales.  
For under resolved calculations, the dissipation provided by the Navier-Stokes terms can be small compared to the dissipation of the SGS model or the numerical discretization.  This has motivated the exclusion of the Navier-Stokes terms entirely in previous ILES studies~\cite{grinstein:2011,thornber:2010, latini:2007} of RMI.  Doing so can reduce the computational cost of the simulation but, used as a general approach, has certain disadvantages.  DNS solutions will be impossible to generate or to approach under grid convergence.  The fine scales of turbulence in an Euler calculation will always scale with those of the grid.  Enstrophy, scalar dissipation rate and other high-order measures of turbulent mixing will never converge.
Furthermore, having never approached the transition between DNS and LES regime, LES schemes which neglect physical transport terms will have less confidence in the assumption of the Reynolds number independence for modeling realistic flows.

A general LES scheme can use any arbitrary set of numerical methods with any arbitrary set of SGS models.  Typically, one selects numerics which balances the overall cost of the flux approximation with adequate resolving power and low numerical dissipation.  SGS models are often selected or developed independently of the numerical scheme and motivated by physical properties of the turbulence.  Some LES approaches combine the two and rely on the natural dissipation of the numerics to act as the SGS model of the scheme.  In all such cases, there exists a non-neglible amount of numerical dissipation which often cannot be directly quantified.  Careful post-processing of the data can reveal the artifacts of the dissipative nature of the scheme when comparisons are made.  Quantities such as enstrophy and scalar dissipation rate are biased toward the high wave numbers and will show greater sensitivity to dissipation compared to conventional measures, such as turbulent kinetic energy (TKE).  

Computing an effective viscosity for LES calculations is instructive in that it allows the net effect of all diffusive processes to be compared on equal terms.  In the absence of an explicit SGS model (as in ILES) previous efforts have shown the utility of an effective viscosity.  Grinstein and Guirguis~\cite{grinstein:1992} compared viscous theory and simulation of two-dimensional shear layer to relate modified equations to an implicit sub-grid scale model.  More recently, Aspden~\etal~provided a method for computing an effective viscosity for incompressible sustained isotropic turbulence.  This viscosity was computed for the entire domain as,

\begin{equation}
	\nu_e = \epsilon/D,
\end{equation}
where
\begin{equation}
	D = \frac{1}{V} \int_V \mathbf{u}\cdot\nabla^2\mathbf{u}\ \mathrm{d}V
\end{equation}
and where $\epsilon$ is the kinetic energy dissipation rate, evaluated directly from the domain time rate of change of kinetic energy.  Aspden showed that $\nu_e$ continuously transitioned between the two extremes; from fully resolved (DNS) where $\nu_f/\nu_e\to1$, to under resolved, quasi-inviscid calculations where $\nu_f/\nu_e\to 0$, where subscript $f$ denotes the physical viscosity.

For compressible turbulence and RMI in particular, we found this form to be insufficient for providing an \emph{a posteriori} approximation of the effective viscosity of the flow.  Firstly, $D$ is not Galilean invariant and will change in magnitude for arbitrary frames of reference as is the case for shock induced mixing.  Secondly, in compressible flow, $\nu$ has thermodynamic dependence and may not be moved outside of the Laplacian of $\bf u$ and therefore the relationship between $\epsilon$ and $D$ will not hold, in general, for a compressible fluid.  Like Aspden, however, we do seek an identical behavior at the limits of DNS and Euler calculations.  

The motion of viscous fluids converts kinetic energy irreversibly to internal energy.  The rate of this conversion due to viscous effects is the dissipation rate ($\epsilon$) and is given~\cite{landau:1959} by

\begin{equation}
	\rho\epsilon = \underline{\tau} : \grad{\bf u} \  .
\end{equation}
Substituting for the stress tensor ($\underline{\tau}$) of a compressible Netwonian fluid, we have

\begin{equation}
	\rho\epsilon = 2\mu {\bf S}^2 + \pth{\beta - \frac{2}{3}\mu}\pth{\divergence {\bf u}}^2  \ .
\end{equation}
SGS models seek to account for sub-grid scale turbulent motion associated primarily with the rotational portion of $S$ and solenoidal portion of the velocity field.  Therefore, if we neglect the purely dilatational term we have

\begin{equation}
	\rho\epsilon = 2\mu {\bf S}^2 \ .
\end{equation}
This is starting point for many SGS models used in the LES community.  
Perhaps the most ubiquitous of which is the Smagorinsky model, which approximates viscous dissipation as

\begin{equation}
	\epsilon = 2 (C_s \Delta x)^2  {\bf S}^3 \ \
\end{equation}
and therefore the SGS viscosity can be written as

\begin{equation}
	\mu_{Smag} = (C_s \Delta x)^2  \rho {\bf S} \ \ .
\end{equation}
Explicit model viscosity will therefore only be dynamically active in regions of the flow where high wave number turbulent energy exists.  In resolved regions the dynamic model will vanish at a rate of $(\Delta x)^2$.

The above overview and description of this particular LES model is not intended to defend nor refute its usage as an LES model.  Rather, its attributes and characteristics are highlighted here only to give context and a starting point for the proposed diagnostic of the present work.  To measure more precisely when the smallest scales of turbulent motion become resolved, an effective viscosity based on the Smagorinsky model and the SGS model of Cook is written as,

\begin{equation}
	\mu^* = C_\mu \rho |  \nabla^2 \mathbf{S}  | \Delta x^4
	\label{eq:muS}
\end{equation}
which is equivalent to Cook's model, with $r=2$ and to Smagorinski where $\bf S$ is replaced with $(\Delta x)^2 \nabla^2 {\bf S}$.  The effect of the Laplacian operator is to amplify the localization of the artificial terms in unresolved regions and to give a convergence rate of $(\Delta x)^4 $ in regions of resolved flow.  Therefore if we write the effective viscosity as

\begin{equation}
	\mu_{\text{eff}} = C_\mu \rho |  \nabla^2 \mathbf{S}  | \Delta x^4 + \mu_f
	\label{eq:mu_eff}
\end{equation}
we have $\mu_f/\mu_\text{eff} \to 1$ for DNS flows and $\mu_f/\mu_\text{eff} \to 0$ for inviscid or highly under revolved calculations.  This form is Galilean invariant, general for compressible flow and can be computed either locally or integrated over some domain.  The coefficient, $C_\mu$ requires closure (which will be discussed below) but is constant for a given numerical method.

\subsubsection{An A Posteriori Analysis of Numerical Dissipation}
\label{sec:apost}

The effective viscosity can be computed at every point in the domain on an existing data set.  For comparison purposes, the derivative operator involved in computing $\bf S$ and in taking the Laplacian should be identical between the two codes.  For the present study, a simple $2^\text{nd}$ order central finite difference method is used for both Miranda and Ares data.  For ease in comparison, a single value for the effective viscosity, $\overline{\mu_\text{eff}}$, is approximated by taking the peak value of the span average of $\mu_e$, written as

\begin{equation}
	\overline{\mu_\text{eff}(t)} = \max\pth{{\barr{\mu_\text{eff}({\bf x},t}}}.
\end{equation}

\begin{figure}
        \centering
        \begin{subfigure}{.5\textwidth}
		\includegraphics[width=3.5in]{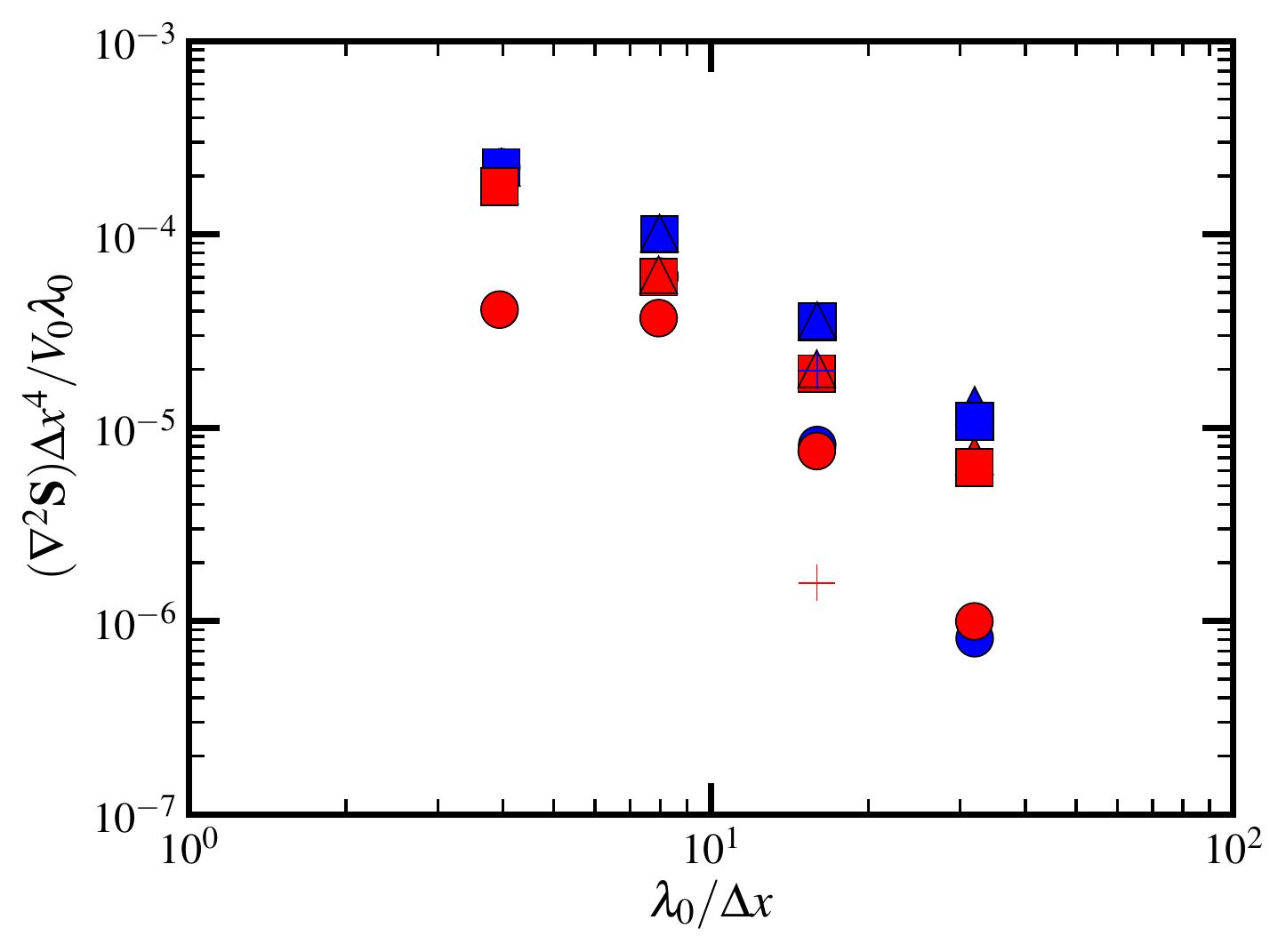}
               	\caption{Inviscid scaling}
        \end{subfigure}%
        \hspace{-.2in}
        \begin{subfigure}{.5\textwidth}
		\includegraphics[width=3.5in]{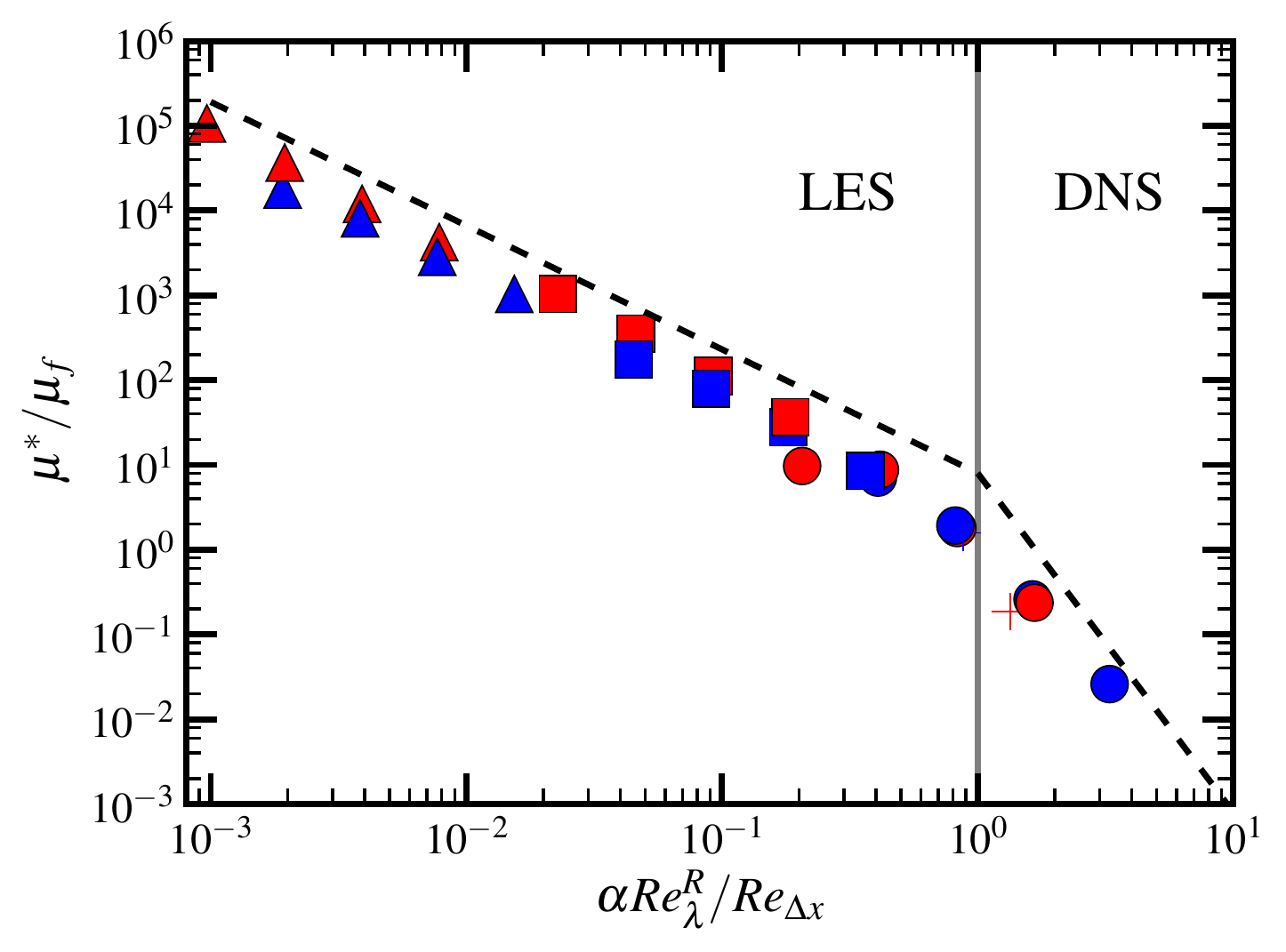}
                	\caption{Viscous scaling}
        \end{subfigure}%
	\caption{\label{fig:lapS}
		Left: Non-dimensional Laplacian of the strain-rate tensor, ${\bf S}$, for all the cases in table \ref{tab:DNS} as a function of inverse grid spacing.  Right: Viscous scaling of the non-physical viscosity as a function of the grid Reynolds number expression.  Blue symbols are data from Miranda and red are from Ares.  The triangle, square and circle symbols correspond to a Reynolds number of $100Re_{\lambda_0}$, $Re_{\lambda_0}$ and $Re_{\lambda_0}/25$, respectively.  The plus symbols reference additional cases described in Table~\ref{tab:Apost}.
	}
\end{figure}

Data for the Laplacian non-dimensionalized by the post shock velocity ($V_0$) and the smallest characteristic wave length of the initial perturbation spectrum ($\lambda_0$) are plotted in Figure~\ref{fig:lapS}a for $\tau = 35$ versus the non-dimensional inverse grid spacing or the number of points per initial wave length.  Data from the two Reynolds numbers at all resolutions are plotted for both Miranda (blue) and Ares (red).  An additional case which used a Reynolds number 100 times larger than the nominal value of Table~\ref{tab:props} was also run and represents the inviscid limit of the flow.

When $\mathbf{S}$ becomes resolved, $\nabla^2 {\bf S}$ will converge and the whole expression in Eq.~\ref{eq:muS} will vanish as $(\Delta x )^4$.  This rapid convergence can be seen in Figure~\ref{fig:lapS} in the circle symbols, which is data from the DNS calculation.  The slope of convergence is clearly steeper than that of the LES calculations (triangles and squares) and indicates that $\bf S$ is nearly converged.
 
For cases where the flow is clearly under-resolved, the magnitude of the effective viscosity (see Figure~\ref{fig:lapS}a) will be proportional to $\Delta x^{-m}$ where $|m| < 4$.
For single shock RMI, both data sets suggest that the value of $m$ is approximately -1.4.  It will be shown later that for LES of high Reynolds number turbulent flows, the value of $m$ is predicted by turbulence theory to be $-4/3$, which is approximately $5\%$ of the measured value.
These convergence slopes are then used to non-dimensionalize the data over all Reynolds numbers.  At the point where the slope becomes $(-4+m)/2$, the approximation is made that the artificial viscosity and the physical viscosity are equivalent or that $\mu^*/\mu_f = 1$.  The degree of freedom used to enforce this constraint gives an explicit value for $C_\mu$, which is dependent on the numerical method of the scheme, but independent of grid spacing and the physical viscosity of the problem.  The values of $C_\mu$ were 8.11 and 63.13 in Miranda and Ares, respectively.  

With $C_\mu$ in hand, the entire expression for $\mu^*$ is known and can be non-dimensionalized by physical viscosity.  The x-axis is also modified to include the effects of both physical viscosity and the grid spacing by computing the quotient $Re_{\lambda_0}^R / Re_{\Delta x}$.  Here, $Re_{\lambda_0}$ is the large scale Reynolds number given by $\rho V_0 \lambda_0 / \mu_f$ and $Re_{\Delta x}$ is the grid Reynolds number given by $\rho V_0 \Delta x / \mu_f$.  The exponent $R$ is given exactly as $R=1+1/m$, which ensures that there is collapse of the data at different physical viscosities.  Note, that if the convergence of $\mu^*$ in the Euler regime gives $m=-1$, then $R=0$ and the data collapse with $1/Re_{\Delta x}$.  

The non-dimensionalization is performed and the data from all the cases are plotted in Figure~\ref{fig:lapS} along with the fiducial slopes for the different convergence rates in each regime.  The x-axis is shifted by a constant such that  $\alpha Re_{\lambda_0}^R / Re_{\Delta x} = 1$ when $\mu^*/\mu_f = 1$ where $\alpha$ is a constant for each code.  For Miranda, $\alpha = 10^n$ with $n=-1.46$ and in Ares, $n=-1.16$.  To the left of this line, the flow is under-resolved and mostly dominated by non-physical dissipation.  To the right, physical viscosity has a large effect on the smallest of length scales and the fourth order convergence indicates DNS levels of resolution.

With this form of the artificial viscosity and after having made the aforementioned non-dimensionalization, one can readily answer two pertinent questions for LES: given the numerics and SGS model of an LES approach, 1) what resolution is needed for a DNS level calculation? 2) what is the effective Reynolds number of an under-resolved LES calculation?
The first asks at which point the viscous scales become numerically resolved.  
The critical point at which this transition occurred (when $\mu^*/\mu_f=1)$ is given as $ Re_{\lambda_0}^R / Re_{\Delta x} = 1/\alpha$, where $\alpha$ was 28.84 in Miranda and 14.46 in Ares.  The ratio between the two ($\alpha_A/\alpha_M$) can be used to compare DNS requirements.  For example, for a given $Re_{\lambda_0}$, if Miranda is predicted to reach a DNS regime at $\Delta x_0$, Ares will reach a DNS regime at $\alpha_A/\alpha_M \Delta x_0$ or $\Delta x_0 / 2.0$.  Additionally, for a constant $\Delta x_0$ for both codes, if Miranda can compute a DNS at $Re_{\lambda_0}$, Ares can compute a DNS at $\pth{\alpha_A/\alpha_M }^{-m} Re_{\lambda_0}$ or $Re_{\lambda_0}/2.64$ using the same grid spacing.  

The second question is relevant to under-resolved LES flows where the effect of physical viscosity may be small and therefore, any Reynolds number which uses that viscosity will have arbitrary significance.  Instead, the effective viscosity (Eq. ~\ref{eq:mu_eff}) can be used to give a more realistic approximation of an effective Reynolds number of the flow.  This Reynolds number will be more indicative of the resolved length scale separation between large production scales and small dissipation scales.  Reynolds number independence and convergence of the large scale flow features will be highly dependent on this Reynolds number.  This effective Reynolds number can be written as

\begin{equation}
	Re_\text{eff} = Re_0 \cdot \pth{\frac{1}{1 + \frac{\mu^*}{\mu_f} }}.
\end{equation}
As $\mu^*/\mu_f$ vanishes with convergence of the DNS solution, the effective Reynolds number will simply be the physical Reynolds number.  For under-resolved LES flows, $\mu^*/\mu_f$ will be arbitrarily large and lead to a substantially lower effective Reynolds number of the flow.

\begin{figure}
	\centering
	\includegraphics[width=3.5in]{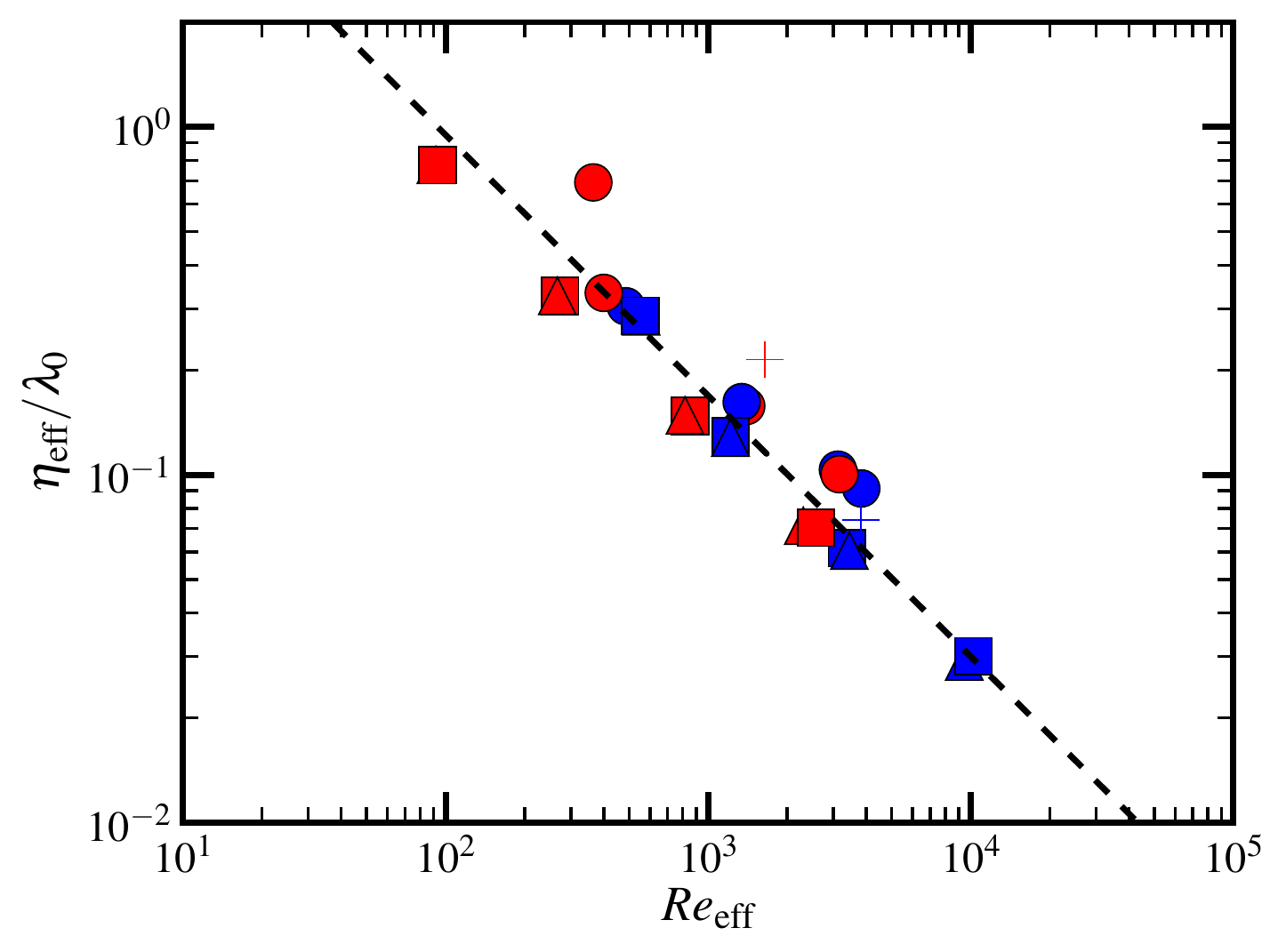}
	\caption{\label{fig:tayRe}
	Non-dimensional Kolmogorov length scale vs. effective Reynolds number at $\tau=30$.	The symbol references are the same as in Figure~\ref{fig:lapS}.  The dashed line is from Kolmogorov theory (Eq.~\ref{eq:Kscale}) and agrees quite well the measured values.}
\end{figure}

To verify that a smaller effective Reynolds number does indeed lead to a smaller range of length scales in the flow, the Kolmogorov length scale is evaluated within the mixing layer.   The Kolmogorov length scale is computed as

\begin{equation}
	\eta_\text{eff} = \pth{\frac{\nu^3}{\epsilon}}^{1/4}
\end{equation}
where $\nu=\mu_\text{eff} / \rho$ is the effective viscosity and where 
$\epsilon = 2 \nu {\mathbf S}^2 $
is being used to approximate the effective dissipation rate.  The effective Kolmogorov length scale plotted in Figure~\ref{fig:tayRe} shows a clear relationship with the effective Reynolds number and a small dependence on the physical Reynolds number of the flow.  Indeed, for sufficiently high Reynolds number, one may assume a balance between the mean turbulent kinetic energy and dissipation rate, $\lambda_0 \sim k^{3/2} / \epsilon$, as suggested from Kolmogorov theory.  Therefore, using the definition of $\eta_\text{eff}$, one may write an approximate scaling of $\eta_\text{eff}$ in terms of the effective Reynolds number as

\begin{equation}
	\frac{\eta_\text{eff}}{\lambda_0} \sim Re_{\text{eff}} ^ {-3/4} .
	\label{eq:Kscale}
\end{equation}
This approximate scaling is plotted in Figure~\ref{fig:tayRe} which shows good agreement with the actual data.
The relationship in Equation~\ref{eq:Kscale} implies a scaling of the effective viscosity with grid spacing.  Earlier, it was reported that $\mu^* \sim \pth{1/ \Delta x}^m$ where $m$ was measured to be $\approx -1.4$.  One can derive an exact value for $m$ using Eq.~\ref{eq:Kscale}, the definition of $Re_\text{eff}$, and the approximation that $\eta \sim \Delta x$ and show that $m=-4/3$.  As the data have indicated, this value and the assumptions needed to derive it, are valid for small vales of $\alpha Re_{\lambda_0}^R / Re_{\Delta x}$, away from the DNS regime.

\subsubsection{Effective Species diffusivity} 

\begin{figure}
        \centering
        \begin{subfigure}{.5\textwidth}
		\includegraphics[width=3.5in]{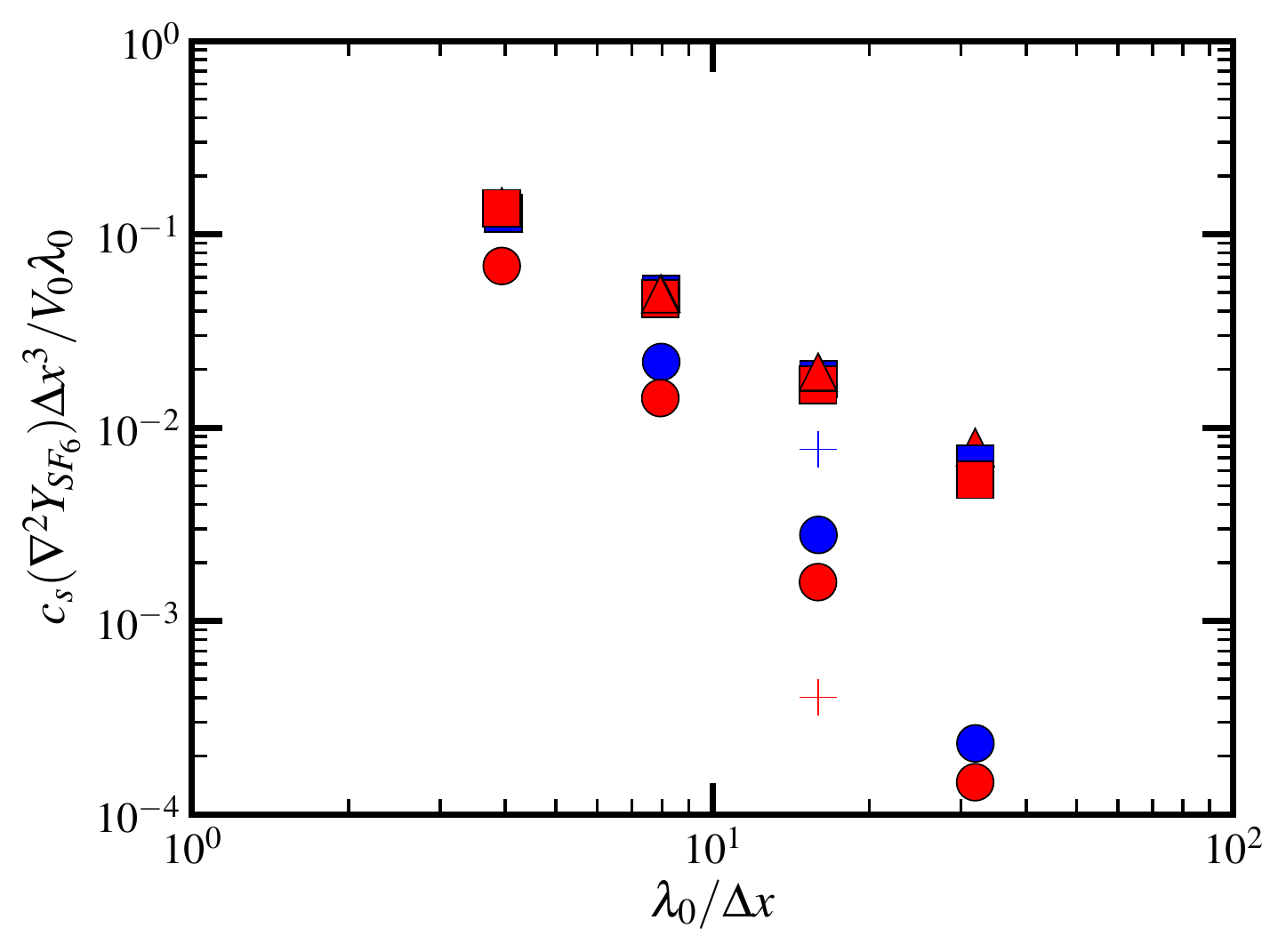}
               	\caption{Inviscid scaling}
        \end{subfigure}%
        \hspace{-.2in}
        \begin{subfigure}{.5\textwidth}
		\includegraphics[width=3.5in]{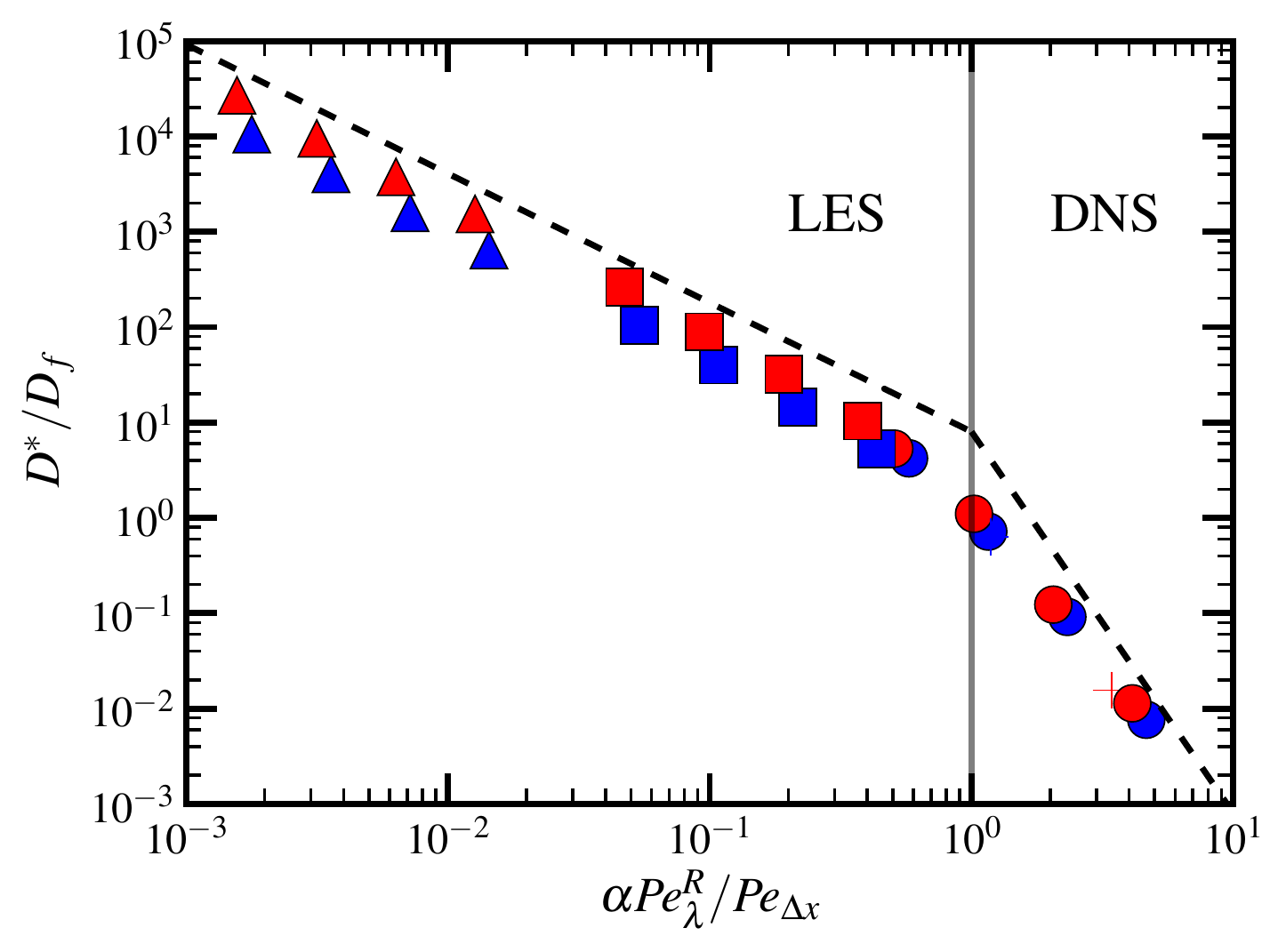}
                	\caption{Diffusive scaling}
        \end{subfigure}%
	\caption{\label{fig:lapY}
		Left: Non-dimensional Laplacian of the magnitude of the scalar gradient, $\sqrt{\nabla Y \cdot\nabla Y}$, for all the cases in table \ref{tab:DNS} as a function of inverse grid spacing.  Right: Viscous scaling of the non-physical diffusivity as a function of the grid P\'eclet number expression.  Blue symbols are data from Miranda and red are from Ares.  The triangle, square and circle symbols correspond to a Reynolds number of $100Re_{\lambda_0}$, $Re_{\lambda_0}$ and $Re_{\lambda_0}/25$, respectively.  The plus symbols reference additional cases described in Table~\ref{tab:Apost}.
	}
\end{figure}

In problems of turbulent multi-component mixing, numerical dissipation will directly affect the diffusive flux of differing materials.  Therefore, the resolved gradients of species mass fraction will largely depend on the numerical scheme, grid resolution and the Reynolds and Schmidt numbers of the flow.  By similar arguments as the effective viscosity, construction of an effective diffusivity can elucidate the differences between methods, resolutions and physical parameters used in LES.  Using the form of the effective viscosity as a template and using $\nabla Y \cdot\nabla Y$ as an indicator for scalar dissipation, the numerical portion is written as

\begin{equation}
	D^* = C_D c_s \left | \nabla^2 \pth{ \sqrt{ \nabla Y \cdot \nabla Y }} \right | \Delta x^4 
\end{equation}
where $c_s$ is the sound speed and $C_D$ is a code dependent coefficient.  The form of $D^*$ follows that of $\mu^*$ where the magnitude of $S$ has been replaced with the magnitude of $\grad Y$ and where $C_\mu \rho$ has been replaced with $C_D c_s$.  For two component flow the $Y$ can be the mass fraction from either gas.  The effective diffusivity is the sum of the numerical and physical portion, written as

\begin{equation}
	D_\text{eff} = D^* + D_f \ .
\end{equation}
Similar to the $\mathbf{S}$ in the effective viscosity expression, as $\grad Y$ becomes resolved in the DNS limit $D_f/D_\text{eff} \to 1$.  For under resolved simulations where the numerical diffusivity dominates, $D_f/D_\text{eff} \to 0$.  Figure~\ref{fig:lapY}a shows the Laplacian of $|\grad Y|$ non-dimensionalized by inviscid mean flow variables and plotted as a function of the number of grid points per $\lambda_0$.  The data show two convergence rates and can be non-dimensionalized in an analogous fashion to $\mu_\text{eff}$, where the P\'eclet number ($Pe_{\lambda_0}=Sc_0 Re_{\lambda_0}$) is the relevant non-dimensional number.  Data indicate that $m=-1.4$ (the same value as $\mu_\text{eff}$) which is the slope of the data from the under resolved calculation.  The coefficient $\alpha$ used to scale the x-axis such that $D^*/D_f = 1$ when $\alpha Pe_{\Delta x}^R / Pe_{\lambda_0} = 1$ is $10^{1.77}$ in Miranda and $10^{1.71}$ in Ares.  Again, by construction, $R=1+1/m$, which is constant for all cases and codes.  This gives coefficients $C_D$ of .039 and .097 for Miranda and Ares, respectively.  Figure~\ref{fig:lapY} shows the non-dimensional numerical diffusion in the under resolved and resolved regions.  The fiducial slopes indicate where the flow is becoming resolved on the grid.  Ares (red) data are shifted slightly to the left of the Miranda data, indicating that Miranda solutions reach DNS levels of convergence at a slightly coarser resolution than Ares.  Therefore, for a given grid resolution and physical Reynolds number, one would expect higher values of $Pe_{\lambda_0}$ and smaller scalar length scales in Miranda.

\begin{figure}
	\centering
	\includegraphics[width=3.5in]{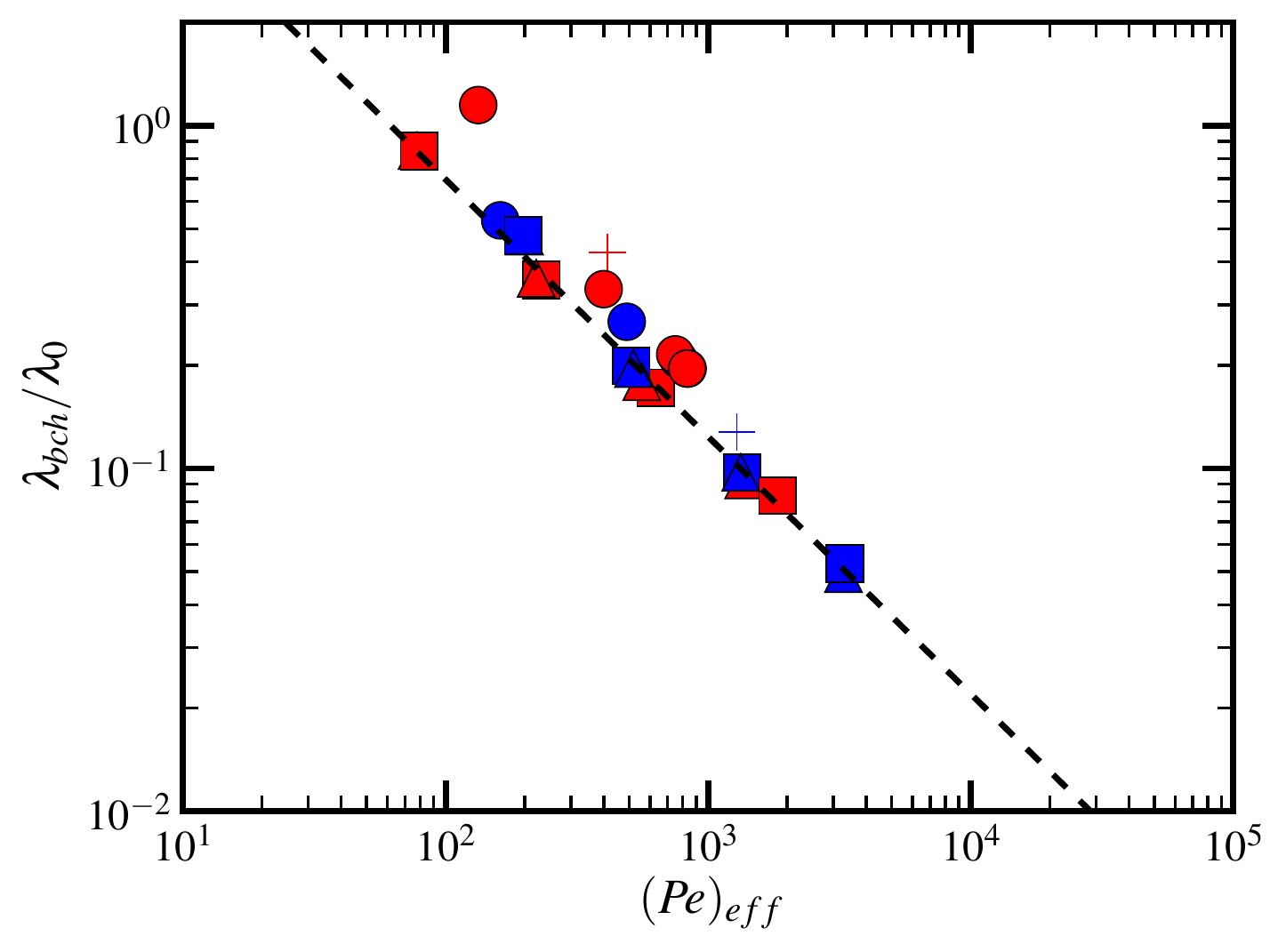}
	\caption{\label{fig:lamYReSc}
	Non-dimensional Batchelor length scale vs. effective P\'eclet number at $\tau=30$.  The symbol references are the same as in Figure~\ref{fig:lapY}.  The dashed line is the scaling for $\lambda_{bch}$ as predicted by Kolmogorov theory (Eq.~\ref{eq:Bscale}) which shows good agreement with the data.
	}
\end{figure}

\begin{figure}
	\centering
	\includegraphics[width=3.5in]{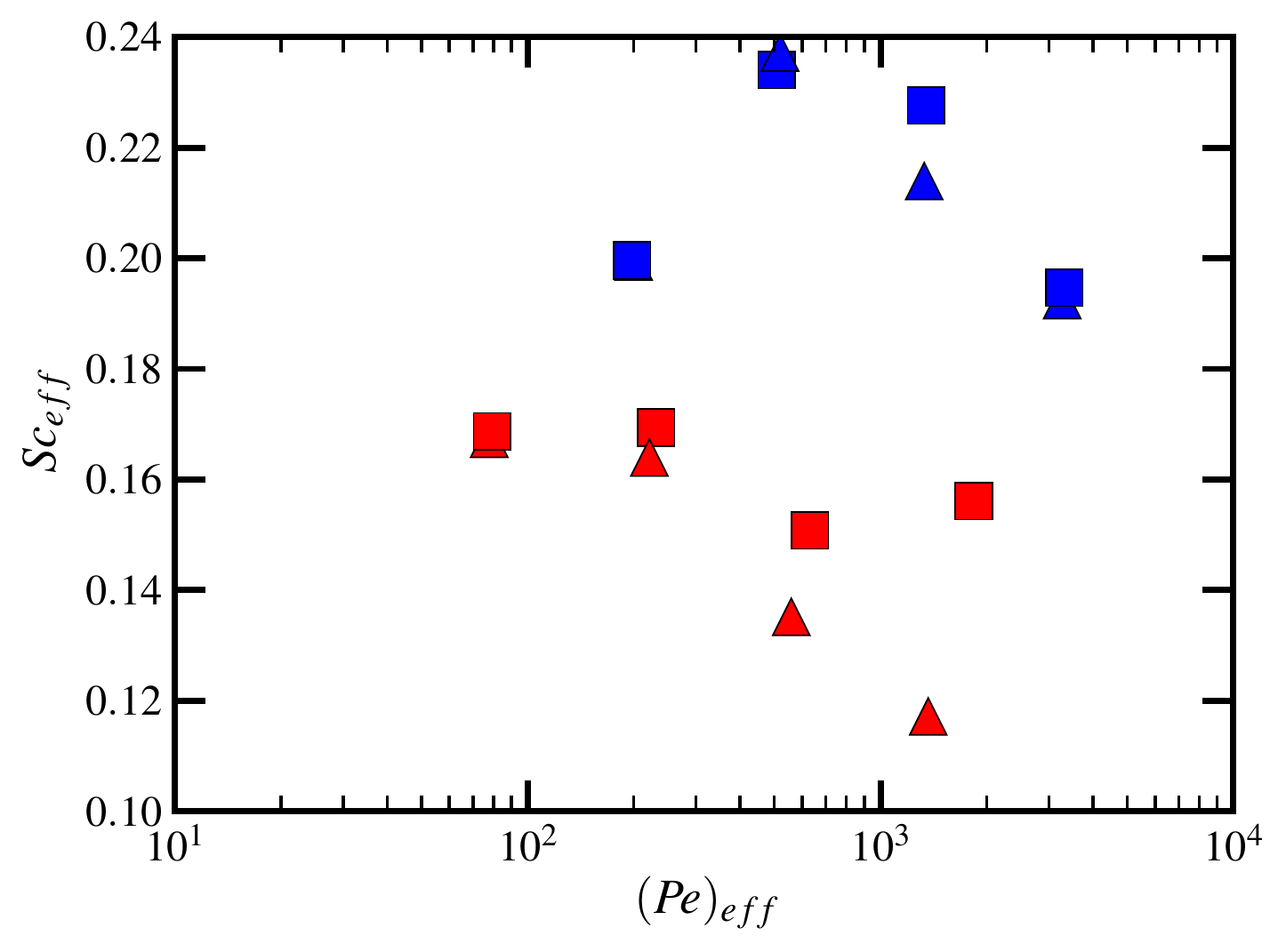}
	\caption{\label{fig:ScReSc}
	Effective Schmidt number vs. effective P\'eclet number at $\tau=30$ in the mixing layer.  The physical Schmidt number was between .18 and 1.11 (see Table~\ref{tab:props}) on the heavy and light side, respectively.  The symbol references are the same as in Figure~\ref{fig:lapY}.
	}
\end{figure}

Similar to the Kolmogorov length scale, the Batchelor scale describes the smallest length scales in the scalar gradient that can exist before diffusion dominates.  This length scale can be related to the Kolmogorov scale as

\begin{equation}
	\lambda_{bch} = \frac{\eta }{ Sc_\text{eff}^{1/2}} \ 
\end{equation}
where the effective Schmidt number is defined as 
\begin{equation}
	Sc_\text{eff} = \frac{\mu_\text{eff} }{\rho D_\text{eff} } \ .
\end{equation}
The non-dimensional Batchelor scale is plotted in Figure~\ref{fig:lamYReSc} and shows an exponential relationship with $Pe_\text{eff}$.  
Indeed, Kolmogorov theory also suggest a scaling of the Batchelor scale and P\'eclet number as

\begin{equation}
	\frac{\lambda_{bch}}{\lambda_0} \sim \pth{Pe_{\text{eff}}} ^ {-3/4} ,
	\label{eq:Bscale}
\end{equation}
which is plotted as dashed line in Figure~\ref{fig:lamYReSc} and shows good agreement with the actual data.  Similar to the artificial viscosity, it can be shown that the artificial species diffusivity scales as $D^* \sim \pth{1/\Delta x}^m$.  Measured data and theory predict the value of $m$ to be, respectively, $-1.4$ and $-4/3$, identical to the values associated with $\mu^*$ in the LES regime. 
The effective Schmidt number is also plotted vs. $Pe_\text{eff}$ in Figure~\ref{fig:ScReSc} and shows that Miranda data have a slightly higher Schmidt number than Ares.

\section{Discussion of LES requirements}
\label{sec:discuss}
LES results and the effective viscosity/diffusivity suggest that dissipation from the numerical method, grid resolution, and physical properties affect the small scales of motion.  The above framework enables all three sources of dissipation to be assessed directly by examination of the large data set.  As one might expect, the low order code produced larger amounts of effective viscosity than the higher order code.  The difference between the two can be quantified as the equivalent $\Delta x_{lo}$ needed in the low order code (Ares), to have an equal amount of effective viscosity as the high order code (Miranda) at grid spacing $\Delta x_{ho}$.  
The ratio between mesh spacing when $\mu^*_{lo}/\mu^*_{ho} = 1$ is defined as $N\equiv \Delta x_{ho} / \Delta x_{lo}$.  It was observed that the value of $N$ was dependent on the level of resolution of the physical viscous scales.  Near the DNS limit, it was previously shown that $N = (\alpha_{ho}/\alpha_{lo})$.   Away from the DNS regime $N$ was larger as evidenced by the poor collapse between codes in Figure~\ref{fig:lapS}b for small values of $\alpha Re^{R}_{\lambda} / Re_{\Delta x}$.  The upper bound for $N$ can be approximated as $N= \pth{ C_{\mu,ho} / C_{\mu,lo} }^{1/m} $ which assumes that the $\nabla^2 {\bf S}$ is the same between codes for a given case.  Therefore, the equivalent grid spacing can be expressed as

\begin{equation}
	\frac{\alpha_{ho}}{\alpha_{lo}} \le N \le \pth{\frac{C_{\mu,ho}}{C_{\mu,lo}}}^{1/m} \ .
	\label{eq:Eres}
\end{equation}
It is also important to note that for three-dimensional time dependent simulations, the additional cost of running a calculation at a finer grid spacing scales approximately with $N^4$ and will be less if using AMR.  The predicted bound of $N$ for the viscous terms at $\tau=30$ was $2.0 \le N \le 4.3$ which is consistent with the time histories of enstrophy and in the spectra of the velocity field.  For example, in the velocity spectra and enstrophy plots (Figures~\ref{fig:UspecLES} and~\ref{fig:EnstLES}), the data from the mesh D Ares calculation lies in between data from mesh B and C from the Miranda calculation.  These Miranda data are 2 and 4 times as coarse which is consistent with the  predicted bounds on $N$ evaluated from Equation~\ref{eq:Eres}.  

The resolution difference of the scalar field was less pronounced than in the velocity field and equation~\ref{eq:Eres} (where $\mu$ is replaced with $D$) gives $1.15\le N\le 1.92$.  Here, the mass fraction spectra and scalar dissipation (Figures~\ref{fig:YspecLES} and~\ref{fig:TMRLES}) show that the data from the scalar dissipation rate and the spectra of the density are slightly less than a factor of two different between Ares and Miranda, which again is consistent with the $N$ from Equation~\ref{eq:Eres}.

The value of $N$ for the viscous and diffusive scales are supported by the measured effective Kolmogorov and Batchelor length scales.  
Both the Kolmogorov and Batchelor length scales represent the smallest length scales of turbulent motion, where fluctuations are dissipated by the viscosity of diffusivity.  
The lower numerical dissipation in Miranda leads to smaller values of these inner diffusive scales and therefore, a broader inertial range of turbulent fluctuations (see Figure~\ref{fig:UspecLES}).  As stated earlier, it is this range which must be sufficiently large as to produce large scale LES results which are grid independent and which approximate a real flow in the Reynolds number independent regime.  From the present data, it was observed that sufficient scale separation occurred at and above $Re_\text{eff} = 2500$, which was represented on grid C and D in Miranda and grid D in Ares.  Such information could be used in approximating the resolution requirements for a given scheme and Reynolds number if one wanted to compute either a DNS solution or grid independent LES solution.  We note that a grid independent LES calculation of a Reynolds number dependent flow must be a DNS if the flow is truly grid independent.  Furthermore, as was implicit in the analysis of Aspden~\etal, the maximum Reynolds number that a given mesh can capture at DNS resolution must always be less than the effective Reynolds number of an Euler calculation on that same mesh.

As an \apos test of the above analysis to approximate the level of resolution of the simulated flow, two additional simulations were run.  One in Miranda using $Re_{\lambda_0}/10$ at mesh C resolution and one in Ares using $Re_{\lambda_0}/50$ at mesh C resolution.  For the viscous terms, the relative resolution metric, $\alpha Re^R_{\lambda_0}/Re_{\Delta x}$, was 0.87 in Miranda and 1.34 in Ares.  The Miranda case falls in the under-resolved regime (since $\alpha Re^R_{\lambda_0}/Re_{\Delta x} < 1$) and the predicted value by a fit from data in Figure~\ref{fig:lapS}b of $\mu^*/\mu_f$ is 1.59.  The actual measured value for $\mu^*/\mu_f$ was 1.57 which is strikingly close to the predicted value considering the analysis is in logarithmic space.  These data are also plotted in Figure~\ref{fig:lapS}b as cross symbols and show good agreement with the other non-dimensionalized data.  This same assessment is made for the Ares data and for both viscous and diffusive terms.  The results are summarized in Table~\ref{tab:Apost} and plotted as cross symbols in Figures~\ref{fig:lapS}b and~\ref{fig:lapY}b.  

\begin{table}[!h]
\begin{center}
	\begin{tabular}{|c |c |c  c  |c |c c|}
	\hline 
	Case & $\alpha Re^R_{\lambda_0}/Re_{\Delta x}$  &  \multicolumn{2}{c|}{$\mu^*/\mu_f$}  &  $\alpha Pe^R_\lambda / Pe_{\Delta x}$ & \multicolumn{2}{c|}{$D^*/D_f$}  \\
	~& ~  &  \scriptsize{measured} & \scriptsize{predicted}  &  ~ & \scriptsize{measured} & \scriptsize{predicted}  \\
	\hline 
	$Re_{\lambda_0} / 10$ (Miranda) & 0.87 & 1.57 & 1.59 & 1.30 & 0.64 & 0.65  \\
	$Re_{\lambda_0} / 25$ (Ares)  & 1.34 & 0.18 & 0.45 & 3.74 & 0.016 & 0.022 \\
	\hline
	\end{tabular}
\end{center}
	\caption{Summary of an \apos test of the analysis in Section~\ref{sec:apost} using two independent calculations in Ares and Miranda.  The analysis predicts the observed dissipation measures ($\mu^*/\mu_f$ and $D^*/D_f$) quite well, when compared to the collapsed data in Figures~\ref{fig:lapS}b and~\ref{fig:lapY}b.
	\label{tab:Apost}
	}
\end{table}

It is certainly not feasible to conduct the full analysis contained in this study for every LES problem one encounters.  However, once the coefficients are determined, one can expect~\cite{cook:2007:POF} them to be fairly universal across a broad range of turbulent flows.  Therefore, on new problems where little resolution requirement information is known, one can quite easily compute $\mu_{f}/\mu_\text{eff}$ as a scalar quantity in the flow field and determine which regions are under resolved or where $\mu_{f}/\mu_\text{eff} << 1$.  Such an indicator can be useful for flagging regions which are to undergo adaptive mesh refinement (AMR) or indicate where numerical errors arising from non-physical dissipation are expected to be most pronounced.  Furthermore, since it has been shown that $\mu^*/\mu_f \sim \pth{1/\Delta x}^m$ in the under-resolved LES regime, a reasonable prediction in the grid spacing needed to reach the minimal DNS requirement ($\mu^* = \mu_f$) can be provided by the expression,

\begin{equation}
	\Delta x_{DNS} \approx \frac{\Delta x_{LES}}{\pth{\mu^*/\mu_f}_{LES}^{1/m}}
\end{equation}
where the ``LES'' subscripts make reference to value from an under-resolved LES calculation.

\section{Conclusion}
\label{sec:conclusion}

We have investigated the effects of numerical method, grid resolution and Reynolds number on the Richtmyer-Meshkov instability through a suite of LES and DNS calculation in the Ares and Miranda codes.  Four mesh resolutions were used between the two codes in the simulation of the RMI using five different Reynolds numbers.  Large scale integral quantities such as mixing layer width and integral mixedness were compared and showed close agreement under refinement.  Frequency dependent terms demonstrated dependence on the mesh, scheme and Reynolds number of the flow.  Gradient based terms which were related to dissipation rates also showed large dependence on the difference sources of dissipation.  The results confirm the expected behavior, that the high-order method captures and a broader range of length scales and has better convergence than the low-order method.  Although this finding is not particularly novel, the fidelity of the simulation database is novel, and therefore, has been interrogated to establish a new framework for LES comparisons.

A simple form for an effective viscosity and diffusivity were proposed and applied \emph{a posteriori} to the data and which indicate the cumulative amount of dissipation in the flow field.  
The effective viscosity and diffusivity scalings collapse all the data between codes, resolutions and physical Reynolds numbers in one common framework which indicates the breadth of the dynamic range of scales supported in a particular LES calculation.  
An effective Reynolds number was also constructed which indicated that grid independence occurs at $Re_\text{eff} > 2500$ and that the smallest viscous and diffusive scales supported on the grid are proportional to, respectively, the effective Reynolds number and P\'eclet number to the -3/4 power.  
The effective viscosity and diffusivity can be used to determine regions of under resolved flow and make predictions of the level of resolution needed to either produce a DNS result or an LES solution which is grid independent.  The predictive capability of the framework was assessed for two additional, independent calculations which showed excellent collapsed onto the original data.

\section*{Acknowledgements}
This work was performed under the auspices of the U.S. Department of Energy by Lawrence Livermore National Laboratory under contract number DE-AC52-07NA27344.  The authors wish to thank A. Cook, W. Cabot, O. Schilling and B. Morgan for many valuable discussions and for help in running the codes.

\bibliographystyle{agsm}

\bibliography{BJO_THESIS}

\end{document}